\documentclass[sigconf,nonacm]{acmart}
\usepackage[T1]{fontenc}
\usepackage{aecompl}

\usepackage{amssymb}
\usepackage{pifont}
\usepackage{bm}
\usepackage{bbm}
\usepackage{amsmath}
\usepackage{amsthm}
\usepackage{amssymb}
\usepackage{multirow}
\usepackage{subfigure}
\usepackage{algorithm}
\usepackage{algpseudocode}
\usepackage{siunitx}
\usepackage{color}
\usepackage{colortbl}
\usepackage{titletoc}  
\renewcommand\footnotetextcopyrightpermission[1]{}
\newcommand{\hrefassumption}[1]{\hyperref[#1]{Ass.~\ref*{#1}}}
\newcommand{\hreffigure}[1]{\hyperref[#1]{Figure~\ref*{#1}}}
\newcommand{\hreffigurea}[1]{\hyperref[#1]{Figure~\ref*{#1}a}}
\newcommand{\hreffigureb}[1]{\hyperref[#1]{Figure~\ref*{#1}b}}
\newcommand{\hreffigurec}[1]{\hyperref[#1]{Figure~\ref*{#1}c}}
\newcommand{\hreffigured}[1]{\hyperref[#1]{Figure~\ref*{#1}d}}
\newcommand{\hrefappendix}[1]{\hyperref[#1]{Appx.~\ref*{#1}}}
\newcommand{\hrefequation}[1]{\hyperref[#1]{Eq.~\ref*{#1}}}
\newcommand{\hrefknowledge}[1]{\hyperref[#1]{Kno.~\ref*{#1}}}
\newcommand{\hrefdefinition}[1]{\hyperref[#1]{Def.~\ref*{#1}}}
\newcommand{\hreftable}[1]{\hyperref[#1]{Table~\ref*{#1}}}
\newcommand{\hrefsection}[1]{\hyperref[#1]{Section.~\ref*{#1}}}
\newcommand{\hrefalgorithm}[1]{\hyperref[#1]{Algorithm.~\ref*{#1}}}
\newcommand{\INPUT}{\item[\textbf{Input:}]}
\newcommand{\OUTPUT}{\item[\textbf{Output:}]}

\theoremstyle{plain}
\newtheorem{theorem}{Theorem}[section]

\theoremstyle{definition}
\newtheorem{definition}[theorem]{Definition}
\newtheorem{assumption}[theorem]{Assumption}
\newtheorem{knowledge}[theorem]{Knowledge}
\theoremstyle{remark}

\AtBeginDocument{%
  \providecommand\BibTeX{{%
    \normalfont B\kern-0.5em{\scshape i\kern-0.25em b}\kern-0.8em\TeX}}}

\begin{document}

\title{Deep Hierarchical Knowledge Loss for Fault Intensity Diagnosis}

\author{Yu Sha\textsuperscript{1,5,\textdagger}, Shuiping Gou\textsuperscript{3}, Bo Liu\textsuperscript{3}, Haofan Lu\textsuperscript{3}, Ningtao Liu\textsuperscript{4}, Jiahui Fu\textsuperscript{5}, Horst Stoecker\textsuperscript{5,6,7}, \\Domagoj Vnucec\textsuperscript{8}, Nadine Wetzstein\textsuperscript{8}, Andreas Widl\textsuperscript{8} and Kai Zhou\textsuperscript{1,2,5,\textdaggerdbl}}
\affiliation{
\institution{\textsuperscript{1}School of Science and Engineering, The Chinese University of Hong Kong, Shenzhen 518172, China\\
             \textsuperscript{2}School of Artificial Intelligence, The Chinese University of Hong Kong, Shenzhen 518172, China\\
             \textsuperscript{3}School of Artificial Intelligence, Xidian University, Xi'an 710126, China\\
             \textsuperscript{4}School of Computer, Luoyang Institute of Science and Technology, Luoyang 471023, China\\
             \textsuperscript{5}Frankfurt Institute for Advanced Studies, Frankfurt am Main 60438, Germany\\
             \textsuperscript{6}Institut f{\"u}r Theoretische Physik, Goethe Universit{\"a}t Frankfurt, Frankfurt am Main 60438, Germany\\
             \textsuperscript{7}GSI Helmholtzzentrum f{\"u}r Schwerionenforschung GmbH, Darmstadt 64291, Germany\\
             \textsuperscript{8}SAMSON AG, Frankfurt am Main 60314, Germany}
\country{}}

\renewcommand{\shortauthors}{Yu Sha et al.}

\begin{abstract}
Fault intensity diagnosis (FID) plays a pivotal role in intelligent manufacturing while neglecting dependencies among target classes hinders its practical deployment. This paper introduces a novel and general framework with deep hierarchical knowledge loss (DHK) to achieve hierarchical consistent representation and prediction. We develop a novel \textbf{hierarchical tree loss} to enable a holistic mapping of same-attribute classes, leveraging tree-based positive and negative hierarchical knowledge constraints. We further design a \textbf{focal hierarchical tree loss} to enhance its extensibility and devise two adaptive weighting schemes based on tree height. In addition, we propose a \textbf{group tree triplet loss} with hierarchical dynamic margin by incorporating hierarchical group concepts and tree distance to model boundary structural knowledge across classes. The joint two losses significantly improve the recognition of subtle faults. Extensive experiments are performed on four real-world datasets from various industrial domains (three cavitation datasets from SAMSON AG and one publicly available dataset) for FID, all showing superior results and outperforming recent state-of-the-art FID methods.
\end{abstract}

\keywords{Cavitation Intensity Diagnosis; Acoustic Signals; Hierarchical Knowledge Loss; Hierarchical classification and Representation Learning}

\maketitle
\renewcommand{\thefootnote}{}
\footnotetext{\textsuperscript{\textdagger} Work performed while at Xidian University.}
\footnotetext{\textsuperscript{\textdaggerdbl} Corresponding author: Kai Zhou (zhoukai@cuhk.edu.cn)}
\section{Introduction}
\label{sec: intro}
Deep learning has achieved excellent advancement in numerous signal processing, like fault detection, fault location and fault intensity diagnosis (FID) \cite{lu2016deep}. In this paper, we focus on FID for cavitation or other faults in complex industrial mechanical systems, which has become an active research topic during recent years. The core of FID is to achieve non-destructive and finer fault recognition of acoustic or vibration signals from the target machine with the assistance of artificial intelligence methods, which is regarded as a critical technology for intelligent manufacturing and the fourth industrial revolution \cite{javaid2022artificial}.
\begin{figure}
    \centering
    \includegraphics[width=0.4\textwidth,height=40mm]{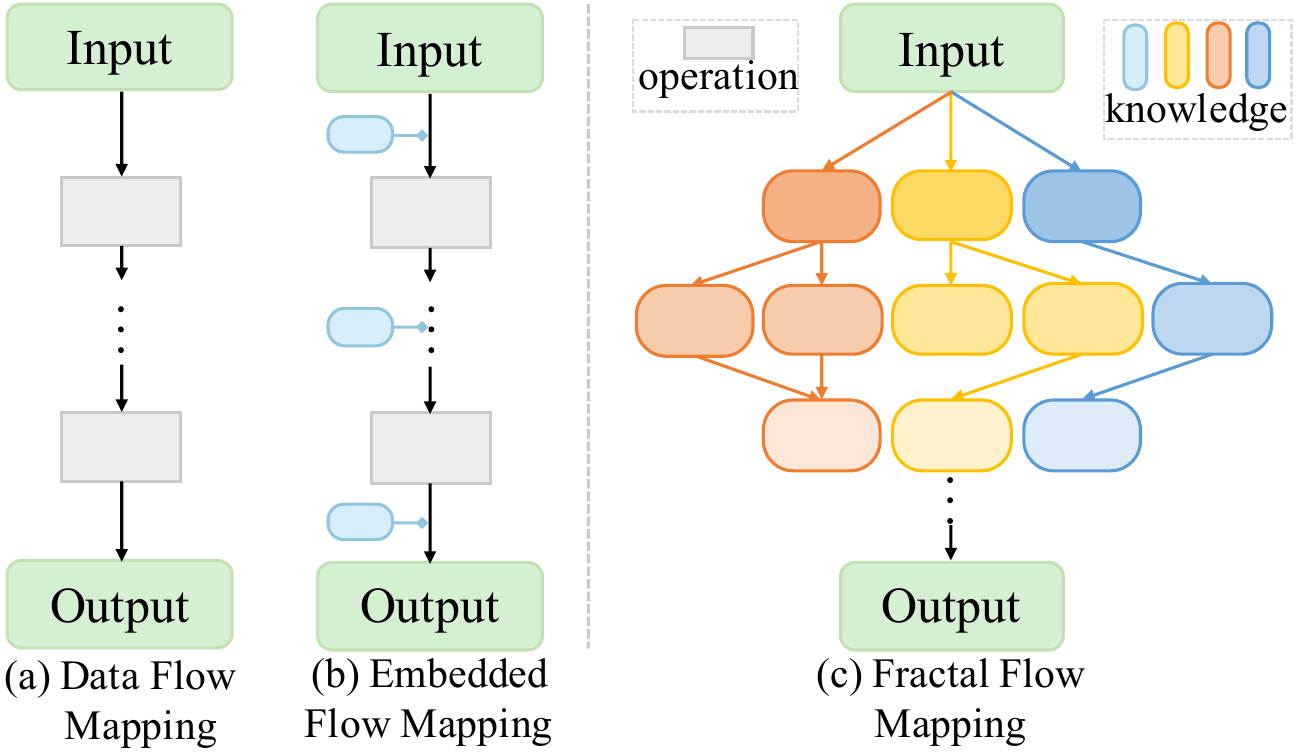}
    \caption{Schematic diagrams for visualising different mappings. Each rectangular box represents an operation, which is a filter in the CNN. Different coloured ellipse boxes of the same colour scheme indicate knowledge associations.}
    \label{fig: DifferentThoughts}
    \vspace{-18pt}
\end{figure}

Formally, FID data consists of multiple physical state signals, with each signal capturing an entire dynamic physical process from the initiation to the termination of a fault in a complex system. Since physical states transition progressively from one to another, there is a natural inherent inter-correlation among these physical signals. This specific correlation is regarded as hierarchical knowledge across measurement events, which can be systematically expressed via a hierarchical knowledge tree \cite{zhang2024representation,yang2024hypformer}. In order to improve the performance of equipment health monitoring in complex industrial systems, the model should exploit the hierarchical dependencies within signals, i.e. optimize the fault recognition and diagnosis process by hierarchical knowledge.

Previous FID research has focused on data-driven convolutional neural (CNN) \cite{sha2022multi,li2023attention,mohammad2023one,li2024deep,zhang2024cavitation,song2024optimized} and transformer \cite{yu2023adaptive,eldele2023self,cui2024self,xiao2024bayesian,rao2024feature} methods. More recently, efforts have been made to design knowledge-embedded data-driven hybrid methods \cite{li2024predicting,sun2024target,guo2024causal,sha2024hierarchical}. The knowledge-embedded data-driven hybrid methods can further optimize the performance and robustness of FID, promoting the deployment and adoption of FID technology in real-world industrial applications.

Although data-driven CNN and Transformers methods have their own advantages, their common limitation is that they rely on typical input-output data flow mapping (DFM), failing to consider hierarchical dependencies among target classes, see \hreffigurea{fig: DifferentThoughts}. To address this issue, a hybrid-driven approach of knowledge-embedded data flow mapping (EFM) is proposed, as shown in \hreffigureb{fig: DifferentThoughts}, which forcibly embeds knowledge into data flow introduces undesired noise and suffers from serious knowledge cross-domain issues \cite{an2024ddcdr}. Therefore, we suggest fusing hierarchical knowledge among classes into the objective function of the DFM, i.e. fractal flow mapping (FFM), see \hreffigurec{fig: DifferentThoughts}. Each node of the FFM represents a specific fault state and edges denote the hierarchical relationships across nodes \cite{zhao2019cambricon}. The FFM not only allows the simultaneous propagation of knowledge and data during backpropagation, but also naturally guides the model to learn hierarchical knowledge by optimizing the objective function \cite{hu2020machine}. This ensures consistency within the predictions and class hierarchical structure, allowing more efficient handling of complex hierarchical data and improving the balance and stability of the model.

To overcome these challenges and effectively apply in real-world scenarios, we propose a general framework with \underline{d}eep \underline{h}ierarchical \underline{k}nowledge loss for FID, called DHK. The core of DHK is converting DFM to FFM by incorporating hierarchical knowledge constraints among same-attribute classes and boundary structural knowledge of different classes to naturally guide the model to learn hierarchical knowledge. More specifically, there is a natural and complementary relationship between child nodes and ancestor nodes of a hierarchical tree $\mathcal{T}$, i.e. positive and negative $\mathcal{T}$ knowledge. Therefore, we propose a hierarchical tree loss ${\mathcal{L}^{HT}}$ by modifying the classical binary cross-entropy loss based on hierarchical positive and negative knowledge, which ensures the consistency of predictions across various hierarchies. We also further develop the focal hierarchical tree loss ${\mathcal{L}^{FHT}}$ to enhance its applicability. In addition, we design two adaptive weighting strategies based on $\mathcal{T}$ height, i.e. normalized height weight and proportional height weight, to dynamically achieve a balanced treatment of various hierarchical classes. To exploit boundary structural knowledge among different classes within the same hierarchy, we introduce a group tree triplet loss ${\mathcal{L}^{GTT}}$ with hierarchical dynamic margin by integrating the group concept and tree distance. The ${\mathcal{L}^{GTT}}$ can precisely establish the dynamic boundary relations between different classes and enable the model to more accurately learn the boundary structures across classes. The above combination of ${\mathcal{L}^{FHT}}$ and ${\mathcal{L}^{GTT}}$ constitutes our proposed DHK.


Our contributions can be summarized as:
\begin{itemize}
    \item A novel and general hierarchical fault intensity diagnosis framework with deep hierarchical knowledge loss is proposed, which can be applied to any representation learning approaches.
    \item We develop a focal hierarchical tree loss to ensure consistency between prediction results and hierarchical structure and further design two adaptive weighting schemes based on hierarchical tree height to maintain the balance of different hierarchies.
    \item We propose a group tree triplet loss with hierarchical dynamic margin, which can capture the boundary structural relationships between different classes within the same hierarchy and effectively improve class boundary learning.
    \item Our DHK outperforms state-of-the-art methods in experiments conducted on four real-world datasets from different industrial domains (three cavitation datasets provided by SAMSON AG and one public dataset). Moreover, ablation studies further demonstrate the effectiveness of our proposed DHK for industrial fault intensity diagnosis. 
\end{itemize}
\section{Preliminaries}
\label{sec: Preliminaries}
\subsection{Cavitation Event Intensity Diagnosis}
\label{subsec: Cavitation Event Intensity Diagnosis}
Cavitation defines the process of localized bubbles or vapour cavities forming, expanding and collapsing within a liquid \cite{plesset1977bubble}. In pipe systems, acoustic sensors are used to record cavitation of different flow conditions as continuous acoustic waveforms. Different intensities of cavitation are defined in \hrefappendix{app: Cavitation Intensity Definition}. Each recorded acoustic signal captures the entire physical process from the beginning to the end of the event related to the specific flow state. In our experiments, cavitation intensity diagnosis distinguishes incipient cavitation, constant cavitation, choked flow cavitation and non-cavitation ($cf.$ \hrefappendix{app: Cavitation Event Intensity Diagnosis}). Any intensity of cavitation indicates the potential issues or faults in system operation. Therefore, intelligent industrial systems need to effectively and precisely identify different intensities of cavitation to implement appropriate responses.

\subsection{Typical Fault Intensity Diagnosis}
\label{subsec: Typical Fault Intensity Diagnosis}
Typical Fault Intensity Diagnosis (FID) is regarded as a classification task. Given a signal dataset ${\mathcal{X}}\subseteq {\mathbb{R}^{M \times N}}$ and corresponding labels ${\mathcal{Y}} \subseteq {\mathbb{R}^{C \times N}}$ with $N$ streams, $M$ measurements for each stream and $C$ fault classes. The transformed dataset ${\mathcal{\tilde X}} \subseteq {\mathbb{R}^{T \times F \times 3}}$ is obtained by applying sliding window ($cf.$ \hrefappendix{app: Acoustic Signals Augmentation}) and Short-Time Fourier Transform ($cf.$ \hrefappendix{app: Time-Frequency Transform}). Assume each sample $(\tilde x \in {\mathcal{\tilde X}},\tilde y \in {\mathcal{\tilde Y}})$ is identically and independently sampled from the joint distribution $p(\tilde x,\tilde y)$. The classifier $f(\theta ):\mathcal{\tilde {X}} \to {\mathbb{R}^C}$ with learnable parameters $\theta$ predicts $p(\tilde {y}|\tilde{x};\theta)$ and is trained to maximize the log-likelihood $\log P(\tilde{X},\tilde{Y};\theta )$, i.e. ${\theta ^*} = \mathop {\arg \max } \log P(\tilde{X},\tilde{Y};\theta )$. This training process results in the categorical cross-entropy (CCE) loss function:
\begin{equation}
\label{eq: CCE Loss}
{\mathcal{L}^{CCE}} =  \sum\limits_{i = 1}^{kN} {\sum\limits_{c = 1}^C} \bm{-}\mathbbm{1}[{\tilde{y}_i} = c]\log p({\tilde{y}_i}|{\tilde{x}_i};\theta ),
\end{equation}
where $\mathbbm{1}[\cdot]$ is an indicator function and $k=\frac{\left(M-w\right)}{{s}}$ is the number of sub-sequences with width ${w}$ and step size ${s}$.
\section{Method}
\label{sec: Method}
Our goal is to enhance traditional fault intensity diagnosis methods by incorporating hierarchical information between classes to achieve hierarchical consistent representation and prediction. Given this goal, we develop a \underline{D}eep \underline{H}ierarchical \underline{K}nowledge loss (DHK) consisting of a hierarchical tree loss with two novel adaptive weighting schemes and a group tree triplet loss with a hierarchical dynamic margin.

\subsection{Hierarchical Fault Intensity Diagnosis}
\label{subsec: Hierarchical Fault Intensity Diagnosis}
In contrast to typical FID methods that treat classes as disjoint labels, hierarchical FID considers potential hierarchical dependencies between classes and represent these relationship using a hierarchical tree $\mathcal{T} = (\mathcal{V},\mathcal{E})$. Each node $v \in {\mathcal{V}}$ represents a real target class and each edge $(u,v) \in {\mathcal{E}}$ encodes a decomposition relationship between class $u\in {\mathcal{V}}$ and class $v$, where the parent node $u$ is a generalized conceptual superclass of the child node $v$, e.g. $(u,v) = (\mathrm{cavitation},\mathrm{incipient\;cavitation})$. In addition, node $u' \in \mathcal{V}$ and node $u$ are on the same level and node $v' \in \mathcal{V}$ is a child of node $u'$, e.g. $(u',v') = (\mathrm{health},\mathrm{turbulent})$. We also assume $(v,v) \in {\mathcal{E}}$, i.e. each class as both a subclass and a superclass of itself. The root node ${v^r \in \mathcal{V}}$ denotes the broadest class and the leaf nodes ${\mathcal{V}_C} = \{ {v_c}\} _{c = 1}^C \in {\mathcal{V}}$ denote the finest classes, i.e. the true precision labels of the dataset, e.g. ${\mathcal{V}_C} = \{\mathrm{incipient},\mathrm{constant}, \mathrm{choked\;flow}\cdots \}$ for cavitation parsing.

To convert typical FID into hierarchical FID, we simply need to map the hierarchy of the whole tree $\mathcal{T}$ to the corresponding classes, i.e. the classification task transformed into a multi-label classification task. Specifically, we replace a $\mathrm{softmax}$ classifier head with a $\mathrm{sigmoid}$ classifier head, resulting in $\bm{S} = \mathrm{sigmoid}(f(\tilde {x};\theta )) \in {[0,1]^{T \times F \times |\mathcal{V}|}}$ w.r.t. the entire class hierarchy $\mathcal{V}$. Meanwhile, the $\mathcal{L}^{CCE}$ ($cf$. \hrefequation{eq: CCE Loss}) is modified to a binary cross-entropy (BCE) loss, i.e. given a predicted score vector $\bm{s} = {[{s_v}]_{v \in \mathcal{V}}} \in {[0,1]^{|\mathcal{V}|}}$ and a ground truth one-hot label $\tilde{\bm{y}} = {[{{\tilde y}_v}]_{v \in \mathcal{V}}} \in {\{0,1\}^{|\mathcal{V}|}}$, the optimization of the BCE loss is as follows:
\begin{equation}
\label{eq: BCE Loss}
{\mathcal{L}^{BCE}} = \sum\limits_{v \in \mathcal{V}} { - {{\tilde y}_v}\log ({s_v}) - (1 - {{\tilde y}_v})\log (1 - {s_v})}.
\end{equation}
 During inference, each event is associated with a root-to-leaf path in the hierarchical class tree $\mathcal{T}$, as follows:
 \begin{equation}
 \label{eq: BCE Inference}
\{ v_i^*\} _{i = 1}^{\left| \mathcal{P} \right|} = \mathop {\arg \max }\limits_{\mathcal{P} \subseteq \mathcal{T}} \sum\limits_{{v_i} \in \mathcal{P}} {{s_{{v_i}}}},
 \end{equation}
 where $\mathcal{P} = \{ {v_1}, \cdots ,{v_{c}}\}\subseteq \mathcal{T}$ refers to a feasible root-to-leaf path of $\mathcal{T}$, i.e. ${v_1} = {v^r}$, ${v_{c}} \in {\mathcal{V}_{C}}$ and $\forall {v_i},{v_{i + 1}} \in \mathcal{P}$, we get $({v_i},{v_{i + 1}}) \in \mathcal{E}$. \hrefequation{eq: BCE Inference} ensures the consistency between the predicted scores and the class hierarchy during inference, allowing us to derive \hrefassumption{ass: hierarchical classification}.

\begin{assumption}
\label{ass: hierarchical classification}
(Hierarchical Inference). \textit{Given a hierarchical tree $\mathcal{T}$ and an input $\tilde {x}$, the prediction of the path depends only on the current node ${{v_i}}$, i.e. $p({v_1} \to  \cdots  \to {v_{c}}|\tilde x) = \prod\nolimits_{i = 1}^{|\mathcal{P}|} {p({v_i}|\tilde x)}$.}
\end{assumption}

The assumption implies that the predicted conditional probability of each node ${{v_i}}$ of the hierarchical path depends solely on the input $\tilde {x}$ and each step is conditionally independent. Furthermore, hierarchical inference degenerates to typical inference when $\mathcal{T}$ has one hierarchy, i.e. $\mathcal{V} = {\mathcal{V}_{C}}$ (proof in \hrefappendix{app: Hierarchical Inference}).  

\begin{figure*}
    \centering
    \includegraphics[width=\textwidth,height=40mm]{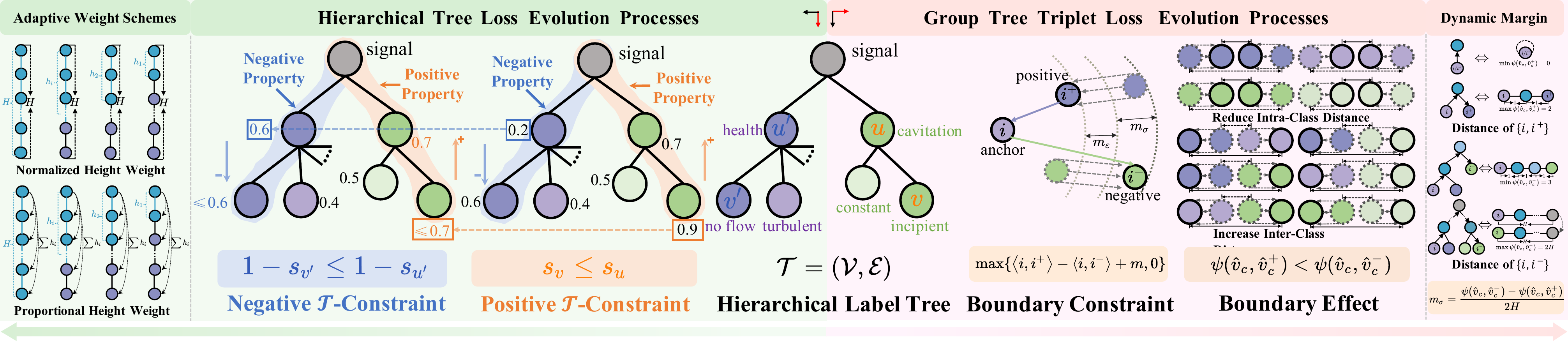}
    \caption{Evolution schematic of deep hierarchical knowledge loss (DHK), comprising hierarchical tree loss ${\mathcal{L}^{HT}}$ with two adaptive weighting schemes and group tree triplet loss ${\mathcal{L}^{GTT}}$ with hierarchical dynamic margin.}
    \label{fig: framework}
\end{figure*}

\subsection{Hierarchical Tree Loss}
\label{subsec: Hierarchical Tree Loss}
Since the $\mathcal{L}^{BCE}$ ($cf.$ \hrefequation{eq: BCE Loss}) is calculated independently for each class, without introducing any hierarchical relationships among classes. To address this issue, the $\mathcal{L}^{BCE}$ can be refined by integrating the following two types of hierarchical knowledge.

\begin{knowledge}
\label{know: Positive Hierarchical Knowledge}
(Positive $\mathcal{T}$-Knowledge). \textit{For each event, if its class ${v}$ is labelled positive, all its parent node $u$ of $\mathcal{T}$ should labelled as positive, i.e. ${v} \in {\mathcal{V}_{pos}} \Rightarrow {u}\in {\mathcal{V}_{pos}}$}.
\end{knowledge}

\begin{knowledge}
\label{know: Negative Hierarchical Knowledge}
(Negative $\mathcal{T}$-Knowledge). \textit{For each event, if its class ${u'}$ is labelled negative, all its child node ${v'}$ of $\mathcal{T}$ should labelled as negative, i.e. ${u'} \in {\mathcal{V}_{neg}} \Rightarrow {{v'}}\in {\mathcal{V}_{neg}}$}.
\end{knowledge}
The above knowledge is complementary, considering the hierarchical tree $\mathcal{T}$ from both bottom-to-top ($cf.$ \hrefknowledge{know: Positive Hierarchical Knowledge}) and top-to-bottom ($cf.$ \hrefknowledge{know: Negative Hierarchical Knowledge}) perspectives, respectively. Based on the above knowledge, we can derive two hierarchical constraints for each event prediction.

\begin{definition}
\label{def: Positive Hierarchical Constraint}
(Positive $\mathcal{T}$-Constraint). \textit{Given an event, if its class ${v}$ is labelled positive and ${u}$ is the parent node of ${v}$, then it holds that ${s_v} \le {s_u}$.}
\end{definition}

\begin{definition}
\label{def: Negative Hierarchical Constraint}
(Negative $\mathcal{T}$-Constraint). \textit{Given an event, if its class ${u'}$ is labelled negative and ${v'}$ is the child node of ${u'}$, then it holds that $1 - {s_{v'}} \le 1 - {s_{u'}}$.}
\end{definition}
Positive $\mathcal{T}$-constraint and negative $\mathcal{T}$-constraint complement each other and their main difference is the directional perspective. With the premises of positive $\mathcal{T}$-constraint and negative $\mathcal{T}$-constraint, positive $\mathcal{T}$-knowledge and negative $\mathcal{T}$-knowledge can always be guaranteed, respectively. Based on the above two constraints, the updated prediction vector $\bm{\hat s} = {[{{\hat s}_v}]_{v \in \mathcal{V}}} \in {[0,1]^{|\mathcal{V}|}}$ is expressed as follows:
\begin{equation}
\label{eq: updated prediction score}
\left\{\begin{aligned}
{{\hat s}_v} &= \mathop{\min}\limits_{u \in {\mathcal{V}_A}} ({s_u}) \;\;\;\;\;\;\;\;\;\;\;\;\;\;\;\;\;\;{{\tilde y}_v} = 1, \\
1 - {{\hat s}_v} &= \mathop{\min}\limits_{u \in {\mathcal{V}_R}} (1 - {s_u}) \;\;\;\;\;\;\;\;\;\;\;\;{{\tilde y}_v} = 0,
\end{aligned}\right.
\end{equation}
where ${\mathcal{V}_A}$ and ${\mathcal{V}_R}$ represent the set of superclasses and subclasses of node $v$, respectively. Based on \hrefequation{eq: BCE Loss} and \hrefequation{eq: updated prediction score}, the \textbf{hierarchical tree loss} (HT) can be obtained:
\begin{equation}
\label{eq: HT Loss}
\begin{array}{l}
{\mathcal{L}^{HT}} = \sum\limits_{v \in \mathcal{V}} { - {{\tilde y}_v}\log (\mathop {\min }\limits_{u \in {\mathcal{V}_A}} ({s_u}))}  - \;(1 - {{\tilde y}_v})\log (1 - \mathop {\max }\limits_{u \in {\mathcal{V}_R}} ({s_u})).
\end{array}
\end{equation}
The ${\mathcal{L}^{HT}}$ strictly adheres to two hierarchical constraints and successfully incorporates class hierarchy information compared to the ${\mathcal{L}^{BCE}}$. The derivation, convergence and differentiability of \hrefequation{eq: HT Loss} are detailed in \hrefappendix{app: HT Loss}-\ref{app: Differentiability HT Loss}, respectively. 

\noindent\textbf{Adaptive Weight Schemes.} In order to ensure the model can effectively learn across different hierarchies and achieve hierarchical consistency. Based on the height of the hierarchical tree $\mathcal{T}$, we design two novel adaptive weighting schemes for ${\mathcal{L}^{HT}}$, named normalized height weight (NHW) and proportional height weight (PHW), as follows:
\begin{equation}
\label{eq: Adaptive Weight Schemes}
\mathrm{NHW}:\frac{{{h_i}}}{H} \times {\mathcal{L}^{HT}}, \;\;\;\;\;\;\;\mathrm{PHW}:\frac{{{h_i}}}{{\sum\nolimits_i {{h_i}}}}\times {\mathcal{L}^{HT}},
\end{equation}
where ${h_i}$ is an $i$-th hierarchy height and $H$ is the total height of $\mathcal{T}$. Both weighting schemes can dynamically adjust weights based on the height of different hierarchical trees. The NHW focuses more on the fine-grained nodes at the bottom hierarchy, while the PHW exhibits more adaptability to avoid excessive focus on nodes at any particular hierarchy.

\noindent\textbf{Focal Hierarchical Tree Loss.} Motivated by \cite{ross2017focal}, we introduce a modulating factor in ${\mathcal{L}^{HT}}$ to focus on hard-to-classify samples, as follows:
\begin{equation}
\label{eq: FHT Loss}
\begin{array}{l}
{\mathcal{L}^{FHT}} = \sum\limits_{v \in \mathcal{V}} { - {{\tilde y}_v}{{(1 - \mathop {\min }\limits_{u \in {\mathcal{V}_A}} ({s_u}))}^\gamma}\log (\mathop {\min }\limits_{u \in {\mathcal{V}_A}} ({s_u}))} \\
\;\;\;\;\;\;\;\;\; - \;(1 - {{\tilde y}_v}){(\mathop {\max }\limits_{u \in {\mathcal{V}_R}} ({s_u}))^\gamma }\log (1 - \mathop {\max }\limits_{u \in {\mathcal{V}_R}} ({s_u})),
\end{array}
\end{equation}
where $\gamma  \in [0,5]$ is a tunable focusing factor to control the down-weighting speed of easily distinguished class samples. When $\gamma  = 0$, ${\mathcal{L}^{FHT}}$ degrades to ${\mathcal{L}^{HT}}$. The NWH and PHW are also suitable for ${\mathcal{L}^{FHT}}$, i.e. ${{({h_i}} \mathord{\left/{\vphantom {{({h_i}} {H) \times L\;}}} \right.\kern-\nulldelimiterspace} {H) \times {\mathcal{L}^{FHT}}\;}}$ and $({{{h_i}} \mathord{\left/{\vphantom {{{h_i}} {\sum\nolimits_i {{h_i}} }}} \right.\kern-\nulldelimiterspace} {\sum\nolimits_i {{h_i}} }}) \times {\mathcal{L}^{FHT}}\;$. 

\subsection{Group Tree Triplet Loss}
\label{subsec: Group Tree Triplet Loss}
To achieve a comprehensive holistic mapping between event and hierarchical classes, we embed the inherent knowledge among different hierarchies of $\mathcal{T}$ as constraints into the BCE loss. In addition, the same hierarchy of $\mathcal{T}$ also provides rich structural information across different class concepts. Therefore, we aim to exploit this structured knowledge to further improve the performance of the model.

\noindent\textbf{Boundary Structural Knowledge.} The boundary structured knowledge is defined by the tree distance $\psi ( \cdot , \cdot )$. Specifically, for any pair of nodes $(u,v)$, let $\psi (u,v)$ represent their distance in $\mathcal{T}$, i.e. the sum of the path lengths from node $u$ and node $v$ to their nearest common ancestor. In fact, $\psi (u,v)$ is an inter-node similarity measure in $\mathcal{T}$, which is a non-negative and symmetric function, i.e. $\psi (u,v) = \psi (v,u)$, $\psi (u,v) \ge 0$ and 
$u = v \Leftrightarrow \psi (u,u) = 0$. Moreover, $\psi (u,v)$ also satisfies the triangle inequality, i.e. $\forall u,v,w \in \mathcal{T}$, $\psi (u,w) \le \psi (u,v) + \psi (v,w)$ (proof $cf.$ \hrefappendix{app: triangle inequality}). This distance metric effectively captures the semantic similarity among nodes of $\mathcal{T}$ and reflects the progressive abstraction and correlation of concepts within the hierarchical structure.

\noindent\textbf{Hierarchical Tree Group.} Inspired by the concept of a ‘bag’ in \cite{maron1997framework,jang2024multiple}, we find that hierarchical tree $\mathcal{T}$ also inherently contains a similar notion, named hierarchical $\mathcal{T}$-group ($cf.$ \hrefdefinition{def: Hierarchical Tree Group}). Through introducing the hierarchical tree group, the model can better capture the semantic differences and hierarchical connections among various classes.

\begin{definition}
\label{def: Hierarchical Tree Group}
(Hierarchical $\mathcal{T}$-Group). \textit{Given a hierarchical tree $\mathcal{T}$, each leaf node ${v_c}$ is defined as an independent concept group ${g_c}$ and its parent node represents a supergroup ${g_s}$ with semantic associations.}
\end{definition}

\begin{definition}
\label{def: Sample Selection Strategy}
(Sampling Strategy). \textit{The anchor, positive and negative samples $\{ i,{i^ + },{i^ - }\}$ are selected only from the set of samples corresponding to the true leaf nodes $\{ {{\hat v}_c},\hat v_c^ + ,\hat v_c^ - \}$, ensuring ${g_c}(i) = {g_s}({i^ + })$ and ${g_c}(i) \ne {g_s}({i^ - })$.}
\end{definition}
Based on the above description, we propose a novel \textbf{group tree triplet loss (GTT)} based on \cite{schroff2015facenet}. This loss is optimized over a set of event triplets $\{ i,{i^ + },{i^ - }\}$, with $i$, ${i^ - }$ and ${i^ + }$ are the anchor, positive and negative samples, respectively. $\{ i,{i^ + },{i^ - }\}$ are sampled from the entire training batch according to sample selection strategy ($cf.$ \hrefdefinition{def: Sample Selection Strategy}) and allows $\psi ({{\hat v}_c},\hat v_c^ + ) < \psi ({{\hat v}_c},\hat v_c^ - )$. In our GTT, the positive sample is semantically similar to the anchor than the negative sample. In contrast to the classical triplet loss without hierarchical groups, our GTT loss selects the anchor and positive samples from the same supergroup classes, while the anchor and negative samples come from different supergroup classes, i.e. ${g_c}(i) = {g_s}({i^ + })$ and ${g_c}(i) \ne {g_s}({i^ - })$. Given a triplet $\{ i,{i^ + },{i^ - }\}$, the GTT loss is defined as:
\begin{equation}
\label{eq: GTT Loss}
{\mathcal{L}^{GTT}} = \frac{1}{N}\sum\nolimits_{i = 1}^N {\max \{ \left\langle {i,{i^ + }} \right\rangle  - \left\langle {i,{i^ - }} \right\rangle  + m,0\},}     
\end{equation}
where $\left\langle { \cdot , \cdot } \right\rangle$ is a similarity distance function and $m$ denotes a separation margin.

\noindent\textbf{Hierarchical Dynamic Margin.} To address the shortcomings of fixed margin in typical triplet loss, we design a hierarchical dynamic margin strategy based on tree distance, as follows:
\begin{equation}
\label{eq: Hierarchical Dynamic Margin}
\begin{aligned}
m &= {m_\varepsilon } + 0.5{m_\sigma }, \;\;\;\;\;\;\;m_\sigma  = \frac{{\psi ({{\hat v}_c},\hat v_c^ - ) - \psi ({{\hat v}_c},\hat v_c^ + )}}{2H},
\end{aligned}
\end{equation}
where ${m_\varepsilon} = 0.15$ is used to tolerate intra-class variance. ${m_\sigma } \in (0,1]$ is the hierarchical dynamic penalty margin and its boundary proof in \hrefappendix{app: Boundary}.

\subsection{Training Objective}
\label{subsec: Training Objective}
Our model (framework $cf.$ \hrefappendix{app: Framework of Model}) is trained by minimizing the joint loss of the focal hierarchical tree loss ${\mathcal{L}^{FHT}}$ ($cf.$ \hrefequation{eq: FHT Loss}) and the group tree triplet loss ${\mathcal{L}^{GTT}}$ ($cf.$ \hrefequation{eq: GTT Loss}): 
\begin{equation}
\label{eq: Training Objective}
\mathcal{L} = \frac{{{h_i}}}{{\sum\nolimits_i {{h_i}} }} \times {\mathcal{L}^{FHT}} + \alpha {\mathcal{L}^{GTT}},
\end{equation}
where $\alpha  \in [0,0.5]$ is a magnitude scaling factor. The pseudocode of DHK see \hrefalgorithm{alg: deep hierarchical knowledge loss}.
 \begin{algorithm}[htbp]
   \caption{DHK loss for Fault Intensity Recognition}
   \label{alg: deep hierarchical knowledge loss}
   \begin{algorithmic}
   \INPUT original signal dataset ${\mathcal{X}}\subseteq {\mathcal{R}^{M \times N}}$, the corresponding label ${\mathcal{Y}} \subseteq {\mathbb{R}^{C \times N}}$, tunable focusing factor $\gamma  \in [0,5]$ and magnitude scaling factor $\alpha  \in [0,0.5]$
    \OUTPUT hierarchical predicted score $\bm{s} = {[{s_v}]_{v \in \mathcal{V}}} \in {[0,1]^{|\mathcal{V}|}}$
   \For{epoch $= 0,1,\ldots,N$}
   \State \textbf{Acoustic Signals Pre-processing:}
   \State ${\mathcal{X}} = \{ {X_i},i = 1,2, \ldots ,N\} \Rightarrow {{\mathcal{\tilde X}}} \subseteq {\mathbb{R}^{T \times F \times 3}}$ 
   \State \textbf{Feature Representation Learning:}
   \State ${{\mathcal F}_{\mathrm{FL}}}({{\mathcal{\tilde X}}},\boldsymbol{\theta}) \Rightarrow \boldsymbol{F}\subseteq {{\mathbb{R}}^D}$ \hfill$\rhd$ e.g. CNNs, Transformer
   \State \textbf{Hierarchical Learning:}
   \State Create hierarchical tree:
   \State ${\mathcal{Y}}  \Rightarrow {\mathcal{T}} \subseteq {{\mathbb{R}}^{C \times H}}$ $\hfill\rhd$ Manual, Unsupervised Learning
   \State Build the focal hierarchical tree loss function: \hfill$\rhd$ \hrefequation{eq: FHT Loss}
   \State ${\mathcal{L}^{FHT}} = \sum\limits_{v \in \mathcal{V}} { - {{\tilde y}_v}{{(1 - \mathop {\min }\limits_{u \in {\mathcal{V}_A}} ({s_u}))}^\gamma}\log (\mathop {\min }\limits_{u \in {\mathcal{V}_A}} ({s_u}))}$ 
    \State \;\;\;\;\;\;\;\;\;\;\;\;\;$ - \;(1 - {{\tilde y}_v}){(\mathop {\max }\limits_{u \in {\mathcal{V}_R}} ({s_u}))^\gamma }\log (1 - \mathop {\max }\limits_{u \in {\mathcal{V}_R}} ({s_u}))$ 
   \State Define two weighting schemes:
   \State $\mathrm{NHW}:\frac{{{h_i}}}{H}, \;\;\;\;\;\;\;\mathrm{PHW}:\frac{{{h_i}}}{{\sum\nolimits_i {{h_i}}}}$\hfill$\rhd$ \hrefequation{eq: Adaptive Weight Schemes}
   \State Build the group tree triplet loss function:
   \State ${\mathcal{L}^{GTT}} = \frac{1}{N}\sum\nolimits_{i = 1}^N {\max \{ \left\langle {i,{i^ + }} \right\rangle  - \left\langle {i,{i^ - }} \right\rangle  + m,0\}} $ \hfill$\rhd$ \hrefequation{eq: GTT Loss}
   \State $m = {m_\varepsilon } + 0.5{m_\sigma} \;\;\;\;{m_\sigma = \frac{{\psi ({{\hat v}_c},\hat v_c^ - ) - \psi ({{\hat v}_c},\hat v_c^ + )}}{2H}}$ \hfill$\rhd$ \hrefequation{eq: GTT Loss}
   \State Obtain prediction score $\bm{s}$:
   \State $\bm{s} = \mathrm{sigmoid}({{\mathcal F}_{\mathrm{FL}}}({{\mathcal{\tilde X}}},\boldsymbol{\theta})) \in {[0,1]^{T \times F \times |\mathcal{V}|}}$ 
   \State Update parameters of DHK by minimizing $\mathcal{L}$:
   \State min $\frac{{{h_i}}}{{\sum\nolimits_i {{h_i}} }} \times {\mathcal{L}^{FHT}} + \alpha {\mathcal{L}^{GTT}}$ 
   \State Save parameters $\boldsymbol{\theta}$ of DHK in current epoch
   \EndFor
\end{algorithmic}
\end{algorithm}

\section{Experiments}
\label{sec: Experiments}
\subsection{Evaluation Metrics}
To evaluate the DHK for fault intensity diagnosis, we use the same evaluation metrics as in traditional fault intensity diagnosis. Therefore, we select Accuracy (Acc), Precision (Pre), Recall (Rec) and F1-score (F1) as the evaluation metrics following previous studies ($cf.$ \hrefappendix{app: Evaluation Metrics}).

\subsection{Implementation Details}
\label{subsection: Implementation Details}
Our proposed DHK uses convolutional neural networks (CNNs) \cite{he2016deep,simonyan2014very,huang2017densely,liu2022convnet}, lightweight neural networks (LNNs) \cite{sandler2018mobilenetv2,howard2019searching,ma2018shufflenet} and transformers \cite{dosovitskiy2020image,liu2021swin,min2022peripheral,li2023uniformer} as feature extractors, respectively. 
The input T-F domain spectrogram is horizontally and vertically flipped, rotated by \SI{180}{\degree}. The parameters $\gamma$, ${m_\varepsilon}$, $\alpha$ are set to 2, 0.15 and 0.1, respectively. During training, the DHK uses Adam as the optimizer with $({\beta _1},{\beta _2}) = (0.9,0.999)$ and an epsilon of ${10^{ - 8}}$. The initial learning rate of the Adam is set to ${10^{ - 3}}$ and dynamically adjusted by the cosine annealing warm with periodic restarts at every ${T_{cur}} = 20$ epochs and a restart period ${T_{mult}} = 1$. The DHK is trained for 100 epochs with a batch size of 64. During inference, our method strictly adheres to \hrefequation{eq: BCE Inference}.

\subsection{Datasets}
\label{subsection: datasets}
\noindent\textbf{Cavitation Datasets.} The SAMSON AG provides three real-world cavitation datasets, called \textbf{Cavitation-Short}, \textbf{Cavitation-Long} and \textbf{Cavitation-Noise} (real noise). The cavitation acoustic signals are acquired from various valves under different upstream and downstream with varying valve openings in a professional laboratory. The cavitation consists of incipient cavitation, constant cavitation and choked flow cavitation. The non-cavitation includes turbulent flow and no flow. All recorded acoustic signals are sampled at \SI{1562.5}{kHz}. \textbf{Cavitation-Short} consists of a total of 356 samples with each sample lasting \SI{3}{s}. \textbf{Cavitation-Long} and \textbf{Cavitation-Noise} contain 806 and 160 acoustic signals and each signal has a duration of \SI{25}{s} ($cf.$ \hrefappendix{app: Datasets}).

\noindent\textbf{PUB Dataset.} The PUB dataset \cite{lessmeier2016condition} from Paderborn University consists of bearing vibration signals, including three states: Inner Ring (IR) damage, Outer Ring (OR) damage and healthy. Signals are sampled at \SI{64}{kHz} for \SI{4}{s}. The dataset is organized into three hierarchies: bearing diagnosis (Hierarchy I), damage type diagnosis (Hierarchy II), and IR/OR intensity diagnosis (Hierarchy III-IR/III-OR). For this dataset, $80\%$ of data is used for training and $20\%$ for testing ($cf.$ \hrefappendix{app: Datasets}).

\subsection{Experimental Results}

\begin{figure}[htbp]
\centering
\subfigure[ResNet34]{
\begin{minipage}[t]{0.333\linewidth}
\centering
\includegraphics[width=\textwidth,height=25mm]{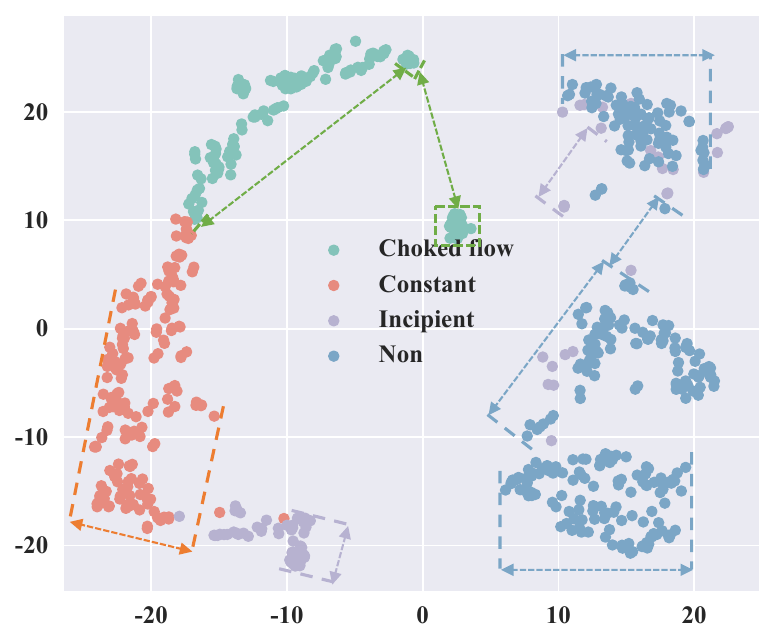}
\end{minipage}%
}%
\subfigure[HKG+ResNet34]{
\begin{minipage}[t]{0.333\linewidth}
\centering
\includegraphics[width=\textwidth,height=25mm]{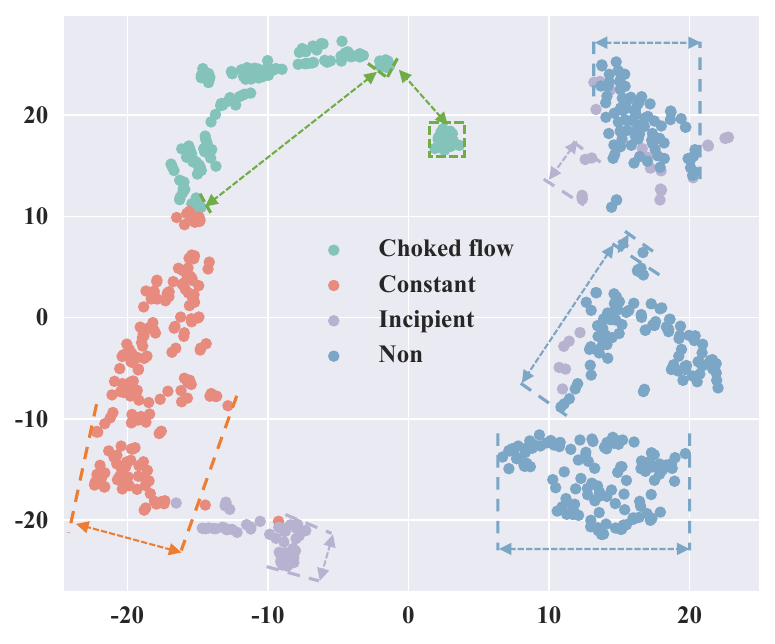}
\end{minipage}%
}%
\subfigure[DHK+ResNet34]{
\begin{minipage}[t]{0.333\linewidth}
\centering
\includegraphics[width=\textwidth,height=25mm]{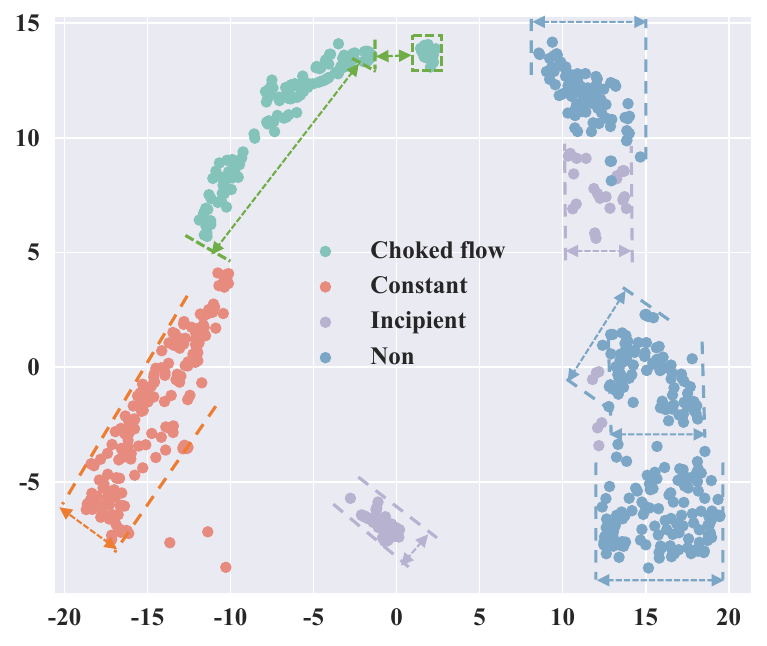}
\end{minipage}%
}%
\centering
\caption{Visualisation of the learned deep feature distribution of vanilla ResNet34, HKG+ResNet34 and DHK+ResNet34 via t-SNE on Cavitation-Short.}
\label{app: examples features distribution}
\end{figure}

\noindent\textbf{Results on Cavitation Datasets.} \hreffigure{app: examples features distribution} shows examples of cavitation deep feature distributions with and without the guidance of hierarchical knowledge. It indicates that hierarchical knowledge can better constrain and guide cavitation deep features. As seen in \hreftable{tab: Different Cavitation Results}, DHK shows superior results under different representation learning backbones across various cavitation datasets. 

\begin{table}[htbp!]
\caption{Results of different backbones on three real-world cavitation datasets averaged over three runs. We compare different representation learning methods including CNNs, LNNs and Transformers. According to our previous research \cite{sha2022multi}, the sliding window with a window size of 466944 and a step size of 466944.}
\label{tab: Different Cavitation Results}
\scriptsize
\setlength{\tabcolsep}{0.15mm}{
\begin{tabular}{l|c|cccc|cccc|cccc}
\toprule
\multicolumn{2}{c}{Dataset}      & \multicolumn{4}{c}{Cavitation-Short}  & \multicolumn{4}{c}{Cavitation-Long}  & \multicolumn{4}{c}{Cavitation-Noise}    \\ 
\midrule
\multicolumn{1}{c|}{Methods}     & \multicolumn{1}{c|}{\begin{tabular}[c]{@{}c@{}}Image \\ size\end{tabular}} & \multicolumn{1}{c}{Acc} & \multicolumn{1}{c}{Pre} & \multicolumn{1}{c}{Rec} & \multicolumn{1}{c|}{F1} 
                                 & \multicolumn{1}{c}{Acc} & \multicolumn{1}{c}{Pre} & \multicolumn{1}{c}{Rec} & \multicolumn{1}{c|}{F1} 
                                 & \multicolumn{1}{c}{Acc} & \multicolumn{1}{c}{Pre} & \multicolumn{1}{c}{Rec} & \multicolumn{1}{c}{F1} \\ 
\midrule
ResNet18                     &${256^2}$   &87.57  &90.25  &77.16  &77.95    &90.83  &86.31  &90.75  &88.21    &98.94  &98.94  &98.94  &98.94   \\
ResNet34                     &${256^2}$   &88.57  &91.47  &76.22  &74.00    &91.88  &87.60  &91.86  &89.42    &98.81  &98.82  &98.81  &98.81   \\
ResNet50                     &${256^2}$   &84.43  &86.97  &71.70  &66.83    &91.45  &87.09  &91.25  &88.89    &98.19  &98.20  &98.19  &98.19   \\
VGG11                        &${256^2}$   &84.29  &62.31  &70.10  &65.95    &80.78  &74.46  &81.11  &76.69    &97.81  &97.82  &97.81  &97.81   \\
VGG13                        &${256^2}$   &80.71  &58.96  &66.85  &62.60    &81.56  &75.21  &82.00  &77.50    &98.19  &98.19  &98.19  &98.19    \\
VGG16                        &${256^2}$   &79.86  &57.77  &66.53  &61.25    &83.38  &77.30  &83.14  &79.44    &98.44  &98.44  &98.44  &98.44    \\
DenseNet121                    &${256^2}$   &84.86  &87.36  &72.01  &67.21    &91.16  &86.73  &91.24  &88.66    &97.94  &97.94  &97.94  &97.94    \\
DenseNet161                    &${256^2}$   &85.00  &89.49  &71.96  &71.14    &91.84  &87.84  &92.07  &89.67    &98.25  &98.25  &98.25  &98.25    \\
DenseNet169                    &${256^2}$   &87.00  &64.39  &73.04  &68.28    &92.08  &87.78  &92.14  &89.65    &98.69  &98.69  &98.69  &98.69    \\     
ConvNeXt-T                     &${224^2}$   &88.14  &85.42  &88.75  &86.73    &91.80  &87.62  &91.60  &89.36    &97.56  &97.57  &97.56  &97.56   \\
ConvNeXt-S                     &${224^2}$   &88.29  &85.60  &88.17  &86.54    &91.97  &87.81  &92.02  &89.63    &97.88  &97.88  &97.88  &97.87    \\
ConvNeXt-B                     &${224^2}$   &88.57  &85.99  &89.00  &87.09    &93.00  &89.12  &92.58  &90.67    &98.13  &98.13  &98.13  &98.13    \\
\rowcolor[HTML]{EFEFEF}
\textbf{DHK+ResNet18}        &${256^2}$       
&\textbf{92.57}              &\textbf{90.67}               &\textbf{91.95}             &\textbf{91.27}    
&\textbf{93.10}              &\textbf{89.36}               &\textbf{93.20}             &\textbf{91.04}    
&\textbf{99.31}              &\textbf{99.31}               &\textbf{99.31}             &\textbf{99.31} \\

\rowcolor[HTML]{EFEFEF}
\textbf{DHK+ResNet34}        &${256^2}$       
&\underline{\textbf{93.57}}  &\underline{\textbf{92.27}}  &\underline{\textbf{93.79}}  &\underline{\textbf{92.94}}    
&\textbf{93.97}              &\textbf{90.39}              &\textbf{93.85}              &\textbf{91.94}    
&\textbf{99.25}              &\textbf{99.25}              &\textbf{99.25}              &\textbf{99.25} \\

\rowcolor[HTML]{EFEFEF}
\textbf{DHK+VGG11}           &${256^2}$       
&\textbf{89.71}              &\textbf{86.84}              &\textbf{89.72}              &\textbf{88.04}    
&\textbf{84.18}              &\textbf{78.29}              &\textbf{84.13}              &\textbf{80.48}   
&\textbf{98.81}              &\textbf{98.82}              &\textbf{98.81}              &\textbf{98.81} \\

\rowcolor[HTML]{EFEFEF}
\textbf{DHK+DeNet169}       &${256^2}$       
&\textbf{90.14}              &\textbf{87.92}              &\textbf{90.22}              &\textbf{88.94}    
&\textbf{94.07}              &\textbf{90.59}              &\textbf{94.16}              &\textbf{92.18}  
&\textbf{98.94}              &\textbf{98.94}              &\textbf{98.94}              &\textbf{98.94} \\

\rowcolor[HTML]{EFEFEF}
\textbf{DHK+CoNeXt-B}        &${224^2}$       
&\textbf{92.14}              &\textbf{90.71}              &\textbf{91.80}              &\textbf{91.23}    
&\underline{\textbf{94.35}}  &\underline{\textbf{91.03}}  &\underline{\textbf{94.30}}  &\underline{\textbf{92.52}}    
&\underline{\textbf{99.81}}  &\underline{\textbf{99.81}}  &\underline{\textbf{99.81}}  &\underline{\textbf{99.81}} \\

\midrule
MoNetv2                     &${256^2}$       &84.00  &62.15  &69.98  &65.82    &88.30  &82.99  &88.08  &85.04    &94.13  &94.13  &94.13  &94.13    \\
MoNetv3S                    &${256^2}$       &76.71  &55.36  &63.38  &58.82    &85.38  &79.45  &85.56  &81.72    &92.38  &92.44  &92.38  &92.37    \\
MoNetv3L                    &${256^2}$       &77.43  &59.02  &61.39  &58.10    &86.87  &81.14  &86.43  &83.24    &93.44  &93.49  &93.44  &93.44    \\
ShuNet-0.5                   &${256^2}$       &74.14  &58.67  &57.20  &53.04    &83.27  &77.15  &83.39  &79.36    &90.06  &90.06  &90.06  &90.06    \\
ShuNet-1.0                   &${256^2}$       &74.86  &63.13  &57.98  &55.51    &83.00  &76.85  &83.06  &79.04    &90.56  &90.60  &90.56  &90.57    \\
ShuNet-1.5                   &${256^2}$       &84.86  &63.12  &71.55  &67.07    &88.89  &83.68  &89.06  &85.84    &90.94  &90.94  &90.94  &90.92    \\
ShuNet-2.0                   &${256^2}$       &73.29  &64.63  &58.35  &55.91    &82.53  &76.11  &81.94  &78.17    &89.94  &89.95  &89.94  &89.94    \\

\rowcolor[HTML]{EFEFEF}
\textbf{DHK+MbNetv2}        &${256^2}$       
&\textbf{88.57}              &\textbf{85.44}              &\textbf{88.17}               &\textbf{86.54}    
&\textbf{90.37}              &\textbf{85.59}              &\textbf{90.21}               &\textbf{87.52}    
&\underline{\textbf{97.38}}  &\underline{\textbf{97.39}}  &\underline{\textbf{97.38}}   &\underline{\textbf{97.37}} \\

\rowcolor[HTML]{EFEFEF}
\textbf{DHK+ShNet-1.5}      &${256^2}$       
&\underline{\textbf{88.71}}  &\underline{\textbf{86.30}}  &\underline{\textbf{88.77}}   &\underline{\textbf{87.38}}    
&\underline{\textbf{90.83}}  &\underline{\textbf{86.31}}  &\underline{\textbf{90.75}}   &\underline{\textbf{88.21}}    
&\textbf{94.56}              &\textbf{94.57}              &\textbf{94.56}               &\textbf{94.56} \\
\midrule

ViT-T                        &${224^2}$       &86.86  &89.32  &76.81  &77.77    &91.23  &86.85  &91.09  &88.69    &96.19  &96.19  &96.19  &96.19   \\
ViT-S                        &${224^2}$       &87.43  &90.08  &77.02  &77.79    &91.93  &87.84  &92.08  &89.68    &96.50  &96.50  &96.50  &96.50   \\
ViT-B                        &${224^2}$       &87.86  &90.81  &75.17  &72.39    &92.65  &88.77  &92.46  &90.40    &97.25  &97.25  &97.25  &97.25   \\
Swin-T                       &${224^2}$       &87.71  &90.63  &75.03  &72.24    &91.64  &87.35  &91.59  &89.16    &97.38  &97.40  &97.38  &97.38   \\
Swin-S                       &${224^2}$       &87.86  &90.81  &75.17  &72.39    &91.90  &87.72  &91.43  &89.35    &97.63  &97.63  &97.63  &97.63   \\
Swin-B                       &${384^2}$       &88.43  &91.39  &75.90  &73.41    &92.88  &88.91  &93.09  &90.73    &99.19  &99.19  &99.19  &99.19   \\  
PerViT-T                     &${224^2}$       &88.43  &85.42  &88.36  &86.64    &92.64  &88.68  &92.81  &90.46    &97.63  &97.63  &97.63  &97.63   \\
PerViT-B                     &${224^2}$       &88.71  &85.86  &89.36  &87.21    &93.03  &89.08  &93.30  &90.90    &98.25  &98.25  &98.25  &98.25   \\
UniFormer-S                    &${224^2}$       &88.57  &86.01  &89.56  &87.45    &92.82  &88.82  &93.12  &90.69    &97.94  &97.94  &97.94  &97.94   \\
UniFormer-B                    &${224^2}$       &88.86  &86.23  &88.97  &87.42    &93.56  &89.92  &93.70  &91.59    &98.31  &98.31  &98.31  &98.91 \\

\rowcolor[HTML]{EFEFEF}
\textbf{DHK+ViT-B}             &${224^2}$       
&\textbf{91.43}              &\textbf{89.26}  &\textbf{91.51}  &\textbf{90.18}    
&\textbf{94.01}              &\textbf{90.60}  &\textbf{93.84}  &\textbf{92.06}    
&\textbf{99.06}              &\textbf{99.07}  &\textbf{99.06}  &\textbf{99.06} \\

\rowcolor[HTML]{EFEFEF}
\textbf{DHK+Swin-B}          &${384^2}$       
&\textbf{91.86}              &\textbf{89.87}  &\textbf{92.23}  &\textbf{90.94}    
&\textbf{94.19}              &\textbf{90.88}  &\textbf{94.43}  &\textbf{92.45}    
&\textbf{99.75}              &\textbf{99.75}  &\textbf{99.75}  &\textbf{99.75} \\

\rowcolor[HTML]{EFEFEF}
\textbf{DHK+PerViT-B}        &${224^2}$       
&\textbf{92.43}              &\textbf{90.42}              &\textbf{92.84}  &\textbf{91.50}    
&\textbf{94.81}              &\underline{\textbf{92.14}}  &\textbf{94.70}  &\underline{\textbf{93.33}}    
&\textbf{99.88}              &\textbf{99.88}              &\textbf{99.88}  &\textbf{99.88} \\

\rowcolor[HTML]{EFEFEF}
\textbf{DHK+UFmer-B}       &${224^2}$       
&\underline{\textbf{92.86}}  &\underline{\textbf{91.75}}  &\underline{\textbf{93.05}}  &\underline{\textbf{92.37}}    
&\underline{\textbf{94.92}}  &\textbf{91.82}              &\underline{\textbf{94.79}}  &\textbf{93.17}    
&\underline{\textbf{99.94}}  &\underline{\textbf{99.94}}  &\underline{\textbf{99.94}}  &\underline{\textbf{99.94}} \\

\bottomrule
\end{tabular}}
\end{table}

In detail, (1) \textbf{Cavitation-Short:} The DHK achieves an accuracy of over \textbf{88}$\%$ across all backbones, surpassing the respective baseline. The DHK improves upon ResNet34 by \textbf{5}$\%$, \textbf{0.8}$\%$, \textbf{17.57}$\%$ and \textbf{18.94}$\%$ across various metrics. (2) \textbf{Cavitation-Long:} The accuracy of DHK with DenseNet169, ConvNeXt-B, ViT, Swin-B, PerViT and UniFormer-B as backbones all surpasses \textbf{94}$\%$, showing improvements of \textbf{1.99}$\%$, \textbf{1.35}$\%$, \textbf{1.36}$\%$, \textbf{1.31}$\%$, \textbf{1.78}$\%$ and \textbf{1.36}$\%$ over their respective baselines. The DHK also achieves the highest precision, recall and F1-score of \textbf{92.14}$\%$, \textbf{94.79}$\%$ and \textbf{93.33}$\%$, respectively. (3) \textbf{Cavitation-Noise:} The DHK achieves accuracies of \textbf{99.81}$\%$, \textbf{97.38}$\%$ and \textbf{99.94}$\%$ on ResNet18, MobileNetv2 and UniFormer-B, with improvements of \textbf{1.68}$\%$, \textbf{3.25}$\%$ and \textbf{1.63}$\%$ over the corresponding models. There are several reasons why our approach delivers excellent results: First, while this dataset has real background noise compared to other cavitation datasets, the STFT can filter out most of noise \cite{sha2024hierarchical}. Second, this dataset is balanced across all classes ($cf.$ \hreftable{tab: CavitationDatasets-FlowStatus}). Third, this dataset is obtained with the same valve stroke and upstream pressure ($cf.$\hreftable{tab: CavitationDatasets-operation}). 
\begin{table}[htbp!]
\caption{Results of different comparison methods on three real-world cavitation datasets averaged over three runs. Non-hierarchical and hierarchical methods from fault intensity diagnosis are used as comparative methods.}
\label{tab: Compared Methods Cavitation Results}
\scriptsize
\setlength{\tabcolsep}{0.15mm}{
\begin{tabular}{l|c|cccc|cccc|cccc}
\toprule
\multicolumn{2}{c}{Dataset}      & \multicolumn{4}{c}{Cavitation-Short}  & \multicolumn{4}{c}{Cavitation-Long}  & \multicolumn{4}{c}{Cavitation-Noise}    \\ 
\midrule
\multicolumn{1}{c|}{Methods}    & \multicolumn{1}{c|}{\begin{tabular}[c]{@{}c@{}}Image \\ size\end{tabular}} & \multicolumn{1}{c}{Acc} & \multicolumn{1}{c}{Pre} & \multicolumn{1}{c}{Rec} & \multicolumn{1}{c|}{F1} 
                                & \multicolumn{1}{c}{Acc} & \multicolumn{1}{c}{Pre} & \multicolumn{1}{c}{Rec} & \multicolumn{1}{c|}{F1} 
                                & \multicolumn{1}{c}{Acc} & \multicolumn{1}{c}{Pre} & \multicolumn{1}{c}{Rec} & \multicolumn{1}{c}{F1} \\ 
\midrule
LiftingNet               &${256^2}$       & 85.29 & 82.75 & 75.89 & 73.95     & 88.43 & 86.31 & 87.97 & 87.05      & 95.86 & 94.62 & 95.84 & 95.20   \\
MIPLCNet                 &${256^2}$       & 86.57 & 83.38 & 77.71 & 75.00     & 89.14 & 87.06 & 88.18 & 87.55      & 96.57 & 95.78 & 96.61 & 96.18   \\
ResNet-APReLU            &${256^2}$       & 86.86 & 84.04 & 77.89 & 75.56     & 90.71 & 88.83 & 90.91 & 89.70      & 97.29 & 96.77 & 97.49 & 97.12  \\
LSTM-RDRN                &${256^2}$       & 87.71 & 85.42 & 78.50 & 76.72     & 91.14 & 89.36 & 91.55 & 90.32      & 98.71 & 98.40 & 98.53 & 98.46  \\
BCNN                     &${256^2}$       & 81.71 & 78.63 & 72.25 & 70.03     & 85.71 & 82.42 & 85.09 & 83.56      & 92.14 & 89.75 & 92.71 & 91.03  \\                                            
HKG-ResNet34             &${256^2}$       & 89.71 & 90.54 & 78.02 & 76.74     & 92.44 & 88.55 & 92.21 & 90.16      & 99.06 & 99.06 & 99.06 & 99.06  \\
HKG-Swin-B               &${384^2}$       & 89.57 & 92.35 & 77.88 & 76.66     & 93.18 & 89.58 & 93.40 & 91.27      & 99.63 & 99.63 & 99.63 & 99.62   \\
\rowcolor[HTML]{EFEFEF}
\textbf{DHK+ResNet34}        &${256^2}$       
&\underline{\textbf{93.57}}  &\underline{\textbf{92.27}}  &\underline{\textbf{93.79}}  &\underline{\textbf{92.94}}    
&\textbf{93.97}              &\textbf{90.39}              &\textbf{93.85}              &\textbf{91.94}    
&\textbf{99.25}              &\textbf{99.25}              &\textbf{99.25}              &\textbf{99.25} \\

\rowcolor[HTML]{EFEFEF}
\textbf{DHK+CoNeXt-B}        &${224^2}$       
&\textbf{92.14}              &\textbf{90.71}              &\textbf{91.80}              &\textbf{91.23}    
&\textbf{94.35}  &\textbf{91.03}  &\textbf{94.30}  &\textbf{92.52}    
&\textbf{99.81}  &\textbf{99.81}  &\textbf{99.81}  &\textbf{99.81} \\

\rowcolor[HTML]{EFEFEF}
\textbf{DHK+Swin-B}          &${384^2}$       
&\textbf{91.86}              &\textbf{89.87}  &\textbf{92.23}  &\textbf{90.94}    
&\textbf{94.19}              &\textbf{90.88}  &\textbf{94.43}  &\textbf{92.45}    
&\textbf{99.75}              &\textbf{99.75}  &\textbf{99.75}  &\textbf{99.75} \\

\rowcolor[HTML]{EFEFEF}
\textbf{DHK+UFmer-B}       &${224^2}$       
&\textbf{92.86}  &\textbf{91.75}  &\textbf{93.05}  &\textbf{92.37}   
&\underline{\textbf{94.92}}  &\underline{\textbf{91.82}}              &\underline{\textbf{94.79}}  &\underline{\textbf{93.17}}    
&\underline{\textbf{99.94}}  &\underline{\textbf{99.94}}  &\underline{\textbf{99.94}}  &\underline{\textbf{99.94}} \\
\bottomrule
\end{tabular}}
\end{table}

In addition, we also report the accuracy of each cavitation state on three real-world cavitation datasets, see \hreftable{app: Different Cavitation Fine Results}. It can be seen that our method can significantly improve the performance of each cavitation state, especially the incipient cavitation state. Specifically, DHK+Uniformer-B outperforms UniFormer-B by \textbf{7.5}$\%$, \textbf{0.3}$\%$ and \textbf{2.25}$\%$ for the incipient cavitation state on three different cavitation datasets, respectively. 
\begin{table}[htbp]
\centering
\caption{Accuracy of various fine states on three real-world cavitation datasets, encompassing non-cavitation (non), choked flow cavitation (cho), constant cavitation (con), and incipient cavitation (inc).  We show different representation learning methods including CNNs, LNNs and Transformers. All results are averaged over three runs.}
\label{app: Different Cavitation Fine Results}
\scriptsize
\setlength{\tabcolsep}{0.3mm}{
\begin{tabular}{l|cccc|cccc|cccc}
\toprule
\multicolumn{1}{c}{Dataset} & \multicolumn{4}{c}{Cavitation-Short}  & \multicolumn{4}{c}{Cavitation-Long} & \multicolumn{4}{c}{Cavitation-Noise}    \\ 
\midrule
\multicolumn{1}{c|}{Methods}                    & \multicolumn{1}{c}{cho}     & \multicolumn{1}{c}{con}     & \multicolumn{1}{c}{inc}    & \multicolumn{1}{c|}{non} 
                                                & \multicolumn{1}{c}{cho}     & \multicolumn{1}{c}{con}     & \multicolumn{1}{c}{inc}    & \multicolumn{1}{c|}{non} 
                                                & \multicolumn{1}{c}{cho}     & \multicolumn{1}{c}{con}     & \multicolumn{1}{c}{inc}    & \multicolumn{1}{c}{non} \\ 
\midrule
ResNet34                                        &\underline{100.0}            &96.11                        &8.75                         &\underline{100.0}        
                                                &92.21                        &92.24 	                    &92.17                        &90.84         
                                                &98.50                        &98.75 	                    &99.00 	                      &99.00  \\
                                                
VGG11                                           &85.71 	                      &95.00                        &0.00                         &99.67                    
                                                &81.04                        &80.33 	                    &81.93                        &81.14         
                                                &97.75                        &98.00 	                    &97.25 	                      &98.25  \\
                                                
DenseNet169                                     &92.14 	                      &\underline{100.0}            &0.00                         &\underline{100.0}        
                                                &92.45                        &91.96 	                    &92.17                        &91.99         
                                                &98.50                        &98.25 	                    &98.50 	                      &99.50  \\
                                                
ConvNeXt-B                                      &89.29 	                      &90.56                        &86.25                        &87.33                    
                                                &92.33                        &93.14                        &91.06                        &93.80         
                                                &98.25                        &98.00 	                    &98.00 	                      &98.25   \\
                                              
HKG-ResNet34                                    &\underline{100.0} 	          &98.33                        &13.75                        &\underline{100.0}                
                                                &91.81                        &92.82                        &91.87                        &92.35         
                                                &99.50                        &98.50 	                    &99.50 	                      &98.75   \\
                                                
HKG-VGG11                                       &93.57 	                      &87.22                        &25.00                        &97.33                    
                                                &83.25                        &82.78                        &83.73                        &82.47         
                                                &98.25                        &98.50 	                    &99.00 	                      &98.75   \\
                                                
HKG-DenseNet169                                 &93.57 	                      &\underline{100.0}            &7.50                         &\underline{100.0}                
                                                &92.81                        &92.71                        &91.47                        &92.92         
                                                &98.75                        &99.50 	                    &99.75 	                      &98.75   \\                                                
                                                
\rowcolor[HTML]{EFEFEF}\textbf{DHK-ResNet34}    &\textbf{90.00}               &\textbf{93.33}               &\underline{\textbf{97.50}}   &\textbf{94.33}       
                                                &\textbf{94.10}               &\textbf{94.05}               &\textbf{93.37}               &\textbf{93.89}        
                                                &\textbf{99.25} 	          &\underline{\textbf{100.0}}   &\textbf{99.00}               &\textbf{98.75}    \\
                                                
\rowcolor[HTML]{EFEFEF}\textbf{DHK-VGG11}       &\textbf{94.29}               &\textbf{87.78}               &\textbf{87.50}               &\textbf{89.33}         
                                                &\textbf{84.90}               &\textbf{83.97}               &\textbf{83.33}               &\textbf{84.31}      
                                                &\textbf{99.25} 	          &\textbf{99.25}               &\textbf{98.00}               &\textbf{98.75}   \\
                                                
\rowcolor[HTML]{EFEFEF}\textbf{DHK-DenseNet169} &\textbf{89.29}               &\textbf{90.00}               &\textbf{91.25}               &\textbf{90.33}     
                                                &\underline{\textbf{95.58}}   &\textbf{93.95}               &\underline{\textbf{94.48}}   &\textbf{94.31}   
                                                &\textbf{98.75} 	          &\textbf{98.50}               &\textbf{99.00}               &\textbf{99.50}   \\
                                                
\rowcolor[HTML]{EFEFEF}\textbf{DHK-ConvNeXt-B}  &\textbf{95.00}               &\textbf{91.11}               &\textbf{88.75}               &\textbf{92.33}     
                                                &\textbf{94.74}               &\underline{\textbf{94.19}}   &\textbf{93.67}               &\underline{\textbf{94.58}}   
                                                &\underline{\textbf{99.75}}   &\textbf{99.75}               &\underline{\textbf{100.0}}   &\underline{\textbf{99.75}}   \\  
                                                
\midrule
MobileNetv2                                     &88.57 	                      &91.67                        &0.00                         &99.67        
                                                &88.07 	                      &88.41 	                    &87.25 	                      &88.58         
                                                &93.75 	                      &94.25 	                    &93.25 	                      &95.25   \\
                                                
ShuNetv2-1.5                                    &\underline{97.86} 	          &90.00                        &0.00                         &98.33                     
                                                &89.40 	                      &88.64 	                    &89.36 	                      &88.86         
                                                &88.50 	                      &87.75 	                    &94.25 	                      &93.25  \\

HKG-MobileNetv2                                 &93.57 	                      &88.33                        &23.75                        &96.00                     
                                                &90.36 	                      &89.64 	                    &89.56 	                      &89.34         
                                                &\underline{97.50} 	          &\underline{96.50} 	        &95.75 	                      &95.25  \\                                                
HKG-ShuNetv2-1.5                                &96.43 	                      &\underline{92.78}            &5.00                         &\underline{100.0}                
                                                &90.00 	                      &89.95 	                    &89.86 	                      &90.09         
                                                &93.75 	                      &92.75 	                    &92.75 	                      &93.25  \\                                                
                                                
\rowcolor[HTML]{EFEFEF}\textbf{DHK-MobileNetv2} &\textbf{87.14} 	          &\textbf{86.11}               &\textbf{88.75}               &\textbf{90.67}        
                                                &\textbf{90.64} 	          &\textbf{90.74}               &\textbf{89.86}               &\textbf{89.58}        
                                                &\textbf{96.75}               &\textbf{96.25}               &\underline{\textbf{98.50}}   &\underline{\textbf{98.00}}  \\
                                                
\rowcolor[HTML]{EFEFEF}\textbf{DHK-ShuNetv2-1.5}&\textbf{89.29} 	          &\textbf{86.11}               &\underline{\textbf{90.00}}   &\textbf{89.67}
                                                &\underline{\textbf{91.40}}   &\underline{\textbf{90.97}}   &\underline{\textbf{90.36}}   &\underline{\textbf{90.18}}    
                                                &\textbf{93.75} 	          &\textbf{95.50}               &\textbf{94.25}               &\textbf{94.75}  \\
\midrule
ViT-B                                           &\underline{100.0} 	          &94.44                        &6.25                         &\underline{100.0}        
                                                &91.77                        &92.86 	                    &92.17                        &93.04         
                                                &97.75                        &98.25 	                    &96.50 	                      &96.50 \\
                                                
Swin-B                                          &99.29 	                      &96.11                        &8.75                         &\underline{100.0}        
                                                &93.29                        &92.54 	                    &93.47                        &93.04         
                                                &99.50                        &98.75 	                    &98.75 	                      &99.75 \\
                                                
PerViT-B                                        &92.86 	                      &85.00                        &91.25                        &88.33                                
                                                &93.94                        &92.63 	                    &93.67                        &92.95          
                                                &98.50                        &98.00 	                    &98.00 	                      &98.50     \\
                                                
UniFormer-B                                     &88.57 	                      &90.56                        &88.75                        &88.00                                
                                                &93.01                        &93.47 	                    &94.48                        &93.86          
                                                &98.75                        &99.25 	                    &97.75 	                      &97.50     \\

HKG-ViT-B                                       &\underline{100.0} 	          &94.44                        &12.50                        &\underline{100.0}                     
                                                &93.37                        &93.23 	                    &92.77                        &92.68          
                                                &98.25                        &98.75 	                    &98.00	                      &98.75     \\

HKG-Swin-B                                      &\underline{100.0} 	          &\underline{97.78}            &13.75                        &\underline{100.0}                     
                                                &93.29                        &93.03 	                    &94.18                        &93.10          
                                                &\underline{100.0}            &\underline{100.0} 	        &99.50	                      &99.00     \\

\rowcolor[HTML]{EFEFEF}\textbf{DHK-ViT-B}       &\textbf{85.71} 	          &\textbf{93.33}               &\textbf{95.00}               &\textbf{92.00}
                                                &\textbf{93.78} 	          &\textbf{94.27}               &\textbf{93.47}               &\textbf{93.86}
                                                &\textbf{99.50}	              &\textbf{98.50}               &\textbf{99.25}               &\textbf{99.00}  \\
                                                
\rowcolor[HTML]{EFEFEF}\textbf{DHK-Swin-B}      &\textbf{96.43} 	          &\textbf{90.56}               &\textbf{91.25}               &\textbf{90.67}       
                                                &\textbf{93.98}	              &\textbf{94.39}               &\underline{\textbf{95.88}}   &\textbf{93.46}        
                                                &\underline{\textbf{100.0}}   &\textbf{99.50}               &\textbf{99.50}               &\underline{\textbf{100.0}}  \\
                                                
\rowcolor[HTML]{EFEFEF}\textbf{DHK-PerViT-B}    &\textbf{92.14} 	          &\textbf{92.22}               &\textbf{95.00}               &\textbf{92.00}       
                                                &\underline{\textbf{94.14}}   &\underline{\textbf{95.15}}   &\textbf{94.88}               &\textbf{94.64}        
                                                &\underline{\textbf{100.0}}   &\textbf{99.75}               &\underline{\textbf{100.0}}   &\textbf{99.75} \\
                                                
\rowcolor[HTML]{EFEFEF}\textbf{DHK-UniFormer-B} &\textbf{89.29} 	          &\textbf{93.33}               &\underline{\textbf{96.25}}   &\textbf{93.33}       
                                                &\textbf{93.86} 	          &\textbf{95.06}               &\textbf{94.78}               &\underline{\textbf{95.48}}        
                                                &\textbf{99.75} 	          &\underline{\textbf{100.0}}   &\underline{\textbf{100.0}}   &\underline{\textbf{100.0}}    \\
                                                
\bottomrule
\end{tabular}}
\end{table}

Furthermore, we show the results of various comparison methods on three cavitation datasets, see \hreftable{tab: Compared Methods Cavitation Results}. It can be clearly seen that DHK+ResNet34, DHK+ConvNeXt-B, DHK+Swin-B and DHK+UniFormer-B outperform all other fault intensity diagnostic methods across various metrics. This indicates that the DHK offers notable advantages for various backbones in cavitation intensity diagnosis, proving the importance of hierarchical structure constraints ($cf.$ \hrefequation{eq: FHT Loss}) and boundary structural knowledge ($cf.$ \hrefequation{eq: GTT Loss}). The cavitation fine state results are shown in \hrefappendix{app: Results}.
\begin{table}[htbp]
\caption{Results of different compared methods on PUB dataset averaged over three runs. }
\label{tab: PUB Results}
\scriptsize
\setlength{\tabcolsep}{0.15mm}{
\begin{tabular}{l|cc|cccccc|cccc}
\toprule
\multicolumn{1}{c}{\multirow{2}{*}{Methods}} & \multirow{2}{*}{Backbones} 
                                              & \multicolumn{1}{c}{\multirow{2}{*}{\begin{tabular}[c]{@{}c@{}}Image\\ size\end{tabular}}} 
                                              & \multicolumn{6}{c}{Accuracy of fine states}                                                                                                                        
                                              & \multicolumn{4}{c}{Overall}\\ 
\cmidrule{4-13} 
\multicolumn{1}{c}{}    &  & \multicolumn{1}{c}{} 
                        & \multicolumn{1}{c}{Health}
                        & \multicolumn{1}{c}{IR-1} 
                        & \multicolumn{1}{c}{IR-2} 
                        & \multicolumn{1}{c}{IR-3} 
                        & \multicolumn{1}{c}{OR-1} 
                        & \multicolumn{1}{c}{OR-2}
                        & Acc   & Pre   & Rec   & F1  \\ 
\midrule
DATMMD         & \multicolumn{1}{c|}{LeNet}          &${256^2}$        &90.63   &93.75  &93.75  &100.0  &93.75  &94.94  &93.52  &92.46  &94.47  &93.38      \\
LiftingNet     & \multicolumn{1}{c|}{AlexNet}        &${256^2}$        &95.83 	&90.18 	&93.75 	&93.75 	&94.64 	&92.41  &93.30  &91.18  &93.43  &92.09      \\
MIP-LCNet      & \multicolumn{1}{c|}{AlexNet}        &${256^2}$        &97.92 	&89.29 	&95.83 	&93.75 	&94.64 	&96.20  &93.48  &91.68  &94.61  &92.96      \\
ResNet-APReLU  & \multicolumn{1}{c|}{ResNet18}       &${256^2}$        &97.92 	&98.21 	&91.67 	&100.0 	&96.43 	&96.20  &96.76  &95.07  &96.74  &95.78      \\
ADMTL          & \multicolumn{1}{c|}{ResNet18}       &${256^2}$        &95.83 	&95.54 	&97.92 	&93.75 	&98.21 	&96.20  &96.54  &95.13  &96.24  &95.66      \\
RDRN           & \multicolumn{1}{c|}{ResNet18}       &${256^2}$        &97.92 	&96.43 	&95.83 	&100.0 	&98.21 	&94.94  &96.98  &94.32  &97.22  &95.57      \\
TDMSAE         & \multicolumn{1}{c|}{ResNet50}       &${256^2}$        &93.75 	&98.21 	&95.83 	&93.75 	&99.11 	&98.73  &97.19  &95.50  &96.56  &95.96      \\
LRSADTLM       & \multicolumn{1}{c|}{Transformer}    &${224^2}$        &97.92 	&97.32 	&95.83 	&100.0 	&99.11 	&96.20  &97.62  &96.40  &97.73  &97.00      \\
SACL           & \multicolumn{1}{c|}{Transformer}    &${224^2}$        &97.92 	&98.21 	&93.75 	&100.0 	&97.32 	&97.47  &97.41  &95.83  &97.45  &96.51      \\
TS-TCC         & \multicolumn{1}{c|}{Transformer}    &${224^2}$        &98.96 	&97.32 	&95.83 	&100.0  &97.32 	&98.73  &97.84  &96.56  &98.03  &97.23      \\
HKG+ViT-S      & \multicolumn{1}{c|}{Transformer}    &${224^2}$        &100.0   &98.21  &97.92  &100.0  &99.11  &98.73  &98.92  &98.30  &99.00  &98.63 \\
\midrule
\rowcolor[HTML]{EFEFEF}
\multicolumn{1}{c|}{}  
& \multicolumn{1}{c|}{\textbf{ResNet18}}            &${256^2}$  &\textbf{97.92} 
                                                                &\textbf{97.32} 
                                                                &\textbf{97.92} 
                                                                &\textbf{93.75} 
                                                                &\textbf{96.43} 
                                                                &\underline{\textbf{100.0}}        
                                                                &\textbf{97.62}       
                                                                &\textbf{96.50}       
                                                                &\textbf{97.22}       
                                                                &\textbf{96.84}      \\
\rowcolor[HTML]{EFEFEF}
\multicolumn{1}{c|}{\multirow{-1}{*}{\textbf{DHK}}} 
& \multicolumn{1}{c|}{\textbf{ViT-S}}            &${224^2}$    &\underline{\textbf{100.0}} 
                                                               &\underline{\textbf{99.11}}
                                                               &\underline{\textbf{100.0}} 
                                                               &\underline{\textbf{100.0}} 
                                                               &\underline{\textbf{99.11}}
                                                               &\textbf{98.73}
                                                               &\textbf{99.35}       
                                                               &\textbf{98.70}       
                                                               &\textbf{99.49}
                                                               &\textbf{99.08}      \\
\rowcolor[HTML]{EFEFEF}
\multicolumn{1}{c|}{}    
& \multicolumn{1}{c|}{\textbf{PerViT-B}}         &${224^2}$   &\underline{\textbf{100.0}} 
                                                              &\underline{\textbf{99.11}}
                                                              &\underline{\textbf{100.0}} 
                                                              &\underline{\textbf{100.0}} 
                                                              &\underline{\textbf{99.11}}
                                                              &\underline{\textbf{100.0}}   
                                                              &\underline{\textbf{99.57}}       
                                                              &\underline{\textbf{98.81}}       
                                                              &\underline{\textbf{99.70}}      
                                                              &\underline{\textbf{99.24}}      \\
\bottomrule
\end{tabular}}
\end{table}

\noindent\textbf{Results on PUB.} \hreftable{tab: PUB Results} presents the experimental outcomes across evaluation metrics. The DHK+PerViT-B delivers excellent accuracy with \textbf{99.57}$\%$, outperforming all SOAT methods and increasing the average accuracy by \textbf{3.25}$\%$. The DHK also achieves the best precision, recall and F1-score of \textbf{98.81}$\%$, \textbf{99.70}$\%$ and \textbf{99.24}$\%$, respectively.

Specifically, (1) Health: DHK+ViT-S and DHK+PerViT-B achieve an accuracy of \textbf{100}$\%$, with an average improvement of \textbf{3.22}$\%$ over all SOAT methods. (2) IR-1: While ResNet-APReLU, TDMSAE, SACL and HKG+ViT-S all reach \textbf{98.21}$\%$, the performance of DHK+ViT-S and DHK+PerViT-B is \textbf{99.11}$\%$. (3) IR-2: DHK+ViT-S and DHK+PerViT-B achieves \textbf{100}$\%$ accuracy, outperforming the comparison methods by an average of \textbf{4.74}$\%$. (4) IR-3: Both DHK+ViT-S and DHK+PerViT-B perform with \textbf{100}$\%$ accuracy, surpassing other SOAT methods. (5) OR-1: DHK+ViT-S and DHK+PerViT-B all attain a performance of \textbf{99.11}$\%$. (6) OR-2: DHK+ResNet18/PerViT-B and DHK+ViT-S achieve \textbf{100}$\%$ and \textbf{98.73}$\%$ accuracy, showing improvements of \textbf{3.57}$\%$ and \textbf{2.3}$\%$ over SOAT methods, respectively. 

\subsection{Ablation Analysis}
\noindent\textbf{Different Losses.} We report results for ResNet34 with different losses in \hreftable{tab: different loss}. It can be seen that ResNet34+DHK significantly outperforms ResNet34+CCE/Focal, which shows the DHK ($cf.$ \hrefequation{eq: Training Objective}) successfully embeds hierarchical knowledge constraints of same-property classes and boundary structure knowledge with different-property classes into the typical BCE loss ($cf.$ \hrefequation{eq: BCE Loss}).

\noindent\textbf{Weight Schemes.} We compare the performance of hierarchical tree loss ${\mathcal{L}^{HT}}$ ($cf.$ \hrefequation{eq: HT Loss}) and focal hierarchical tree loss ${\mathcal{L}^{FHT}}$ ($cf.$ \hrefequation{eq: FHT Loss}) using different weight schemes. As shown in \hreftable{tab: HT FHT Weight}, both weighting schemes improve the performance of ${\mathcal{L}^{HT}}$ and ${\mathcal{L}^{FHT}}$, with proportional height weight (PHW) showing a more significant improvement compared to normalized height weight (NHW). Compared to the NHW, the PHW eliminates absolute bias at specific hierarchies by balanced weight assignment and dynamic adjustment, leading to a more comprehensive understanding of class relationships.

\noindent\textbf{Key Component Analysis.} We analyse the essential components of DHK ($cf.$ \hrefequation{eq: Training Objective}) under the ResNet34 backbone, as shown in \hreftable{tab: HT FT GTT}. It can be observed that ${\mathcal{L}^{HT}}$ w/ PHW + ${\mathcal{L}^{GTT}}$ and ${\mathcal{L}^{FHT}}$ w/ PHW + ${\mathcal{L}^{GTT}}$ achieve better performance compared to other combinations. Furthermore, the ${\mathcal{L}^{FHT}}$ w/ PHW + ${\mathcal{L}^{GTT}}$ obtains the best results with \textbf{2.71}$\%$ improvement over the ${\mathcal{L}^{HT}}$ + ${\mathcal{L}^{GTT}}$. It further confirms the significance and necessity of hierarchical knowledge constraints and boundary structural knowledge.

\noindent\textbf{Variants Analysis.} We investigate the group tree triplet loss ${\mathcal{L}^{GTT}}$ ($cf.$ \hrefequation{eq: GTT Loss}) with different margin versions, see \hreftable{tab: GTT margin}. In \hreftable{tab: GTT margin}, "Vanilla" indicates the classical triplet loss with a constant margin. It is evident that our designed ${\mathcal{L}^{GTT}}$ with a constant margin outperforms "Vanilla". It indicates that the introduction of hierarchical tree group is essential. In addition, ${\mathcal{L}^{GTT}}$ with a hierarchical dynamic margin ($cf.$ \hrefequation{eq: Hierarchical Dynamic Margin}) has better performance than ${\mathcal{L}^{GTT}}$ with a constant margin, which illustrates the effectiveness of embedding the boundary hierarchy constraints. 

\noindent\textbf{Distance Measure.} We also explore the impact of the group tree triplet loss ${\mathcal{L}^{GTT}}$ ($cf.$ \hrefequation{eq: GTT Loss}) with different distance measurement $\left\langle { \cdot , \cdot } \right\rangle$. From \hreftable{tab: GTT distance}, we observe that cosine distance obtains better results than euclidean distance. This is primarily because the cosine distance is more suitable for a specific topology of hierarchical tree \cite{zhang2024representation,zhang2024automated}. 
\begin{table}[htbp]
\centering
\scriptsize
\caption{\textbf{Comparison of loss functions} on Cavitation-Short. All results are averaged over three runs.}
\vspace{-5pt}
\label{tab: different loss}
\setlength{\tabcolsep}{3.5mm}{
\begin{tabular}{c|c|cccc}
\toprule
Backbones                  & Loss  & Accuracy & Precision & Recall & F1-score \\ 
\midrule
\multirow{5}{*}{ResNet-34} & CCE   & 88.57    & 91.47     & 76.22  & 74.00    \\
                           & Focal & 90.43    & 88.81     & 89.91  & 89.32    \\
                           & SCE   & 89.86    & 87.57     & 89.95  & 88.63    \\
                           & CB    & 89.14    & 86.77     & 87.88  & 87.29    \\
                           & SL    & 91.14    & 88.33     & 91.03  & 89.46   \\
                           & HS    & 92.71    & 90.78     & 92.84  & 91.71   \\
                           &\cellcolor[HTML]{EFEFEF}\textbf{DHK} 
                           &\cellcolor[HTML]{EFEFEF}\textbf{93.57}    
                           &\cellcolor[HTML]{EFEFEF}\textbf{92.27}     
                           &\cellcolor[HTML]{EFEFEF}\textbf{93.79}  
                           &\cellcolor[HTML]{EFEFEF}\textbf{92.94}    \\
\midrule
\multirow{5}{*}{Swin-B}    & CCE   & 88.43    & 91.39     & 75.90  & 73.41    \\
                           & Focal & 90.57    & 88.12     & 91.25  & 89.44    \\
                           & SCE   & 89.71    & 87.47     & 90.19  & 88.64    \\
                           & CB    & 88.71    & 85.86     & 89.36  & 87.21    \\
                           & \cellcolor[HTML]{EFEFEF}\textbf{DHK}   
                           & \cellcolor[HTML]{EFEFEF}\textbf{91.86}    
                           & \cellcolor[HTML]{EFEFEF}\textbf{89.87}     
                           & \cellcolor[HTML]{EFEFEF}\textbf{92.23}  
                           & \cellcolor[HTML]{EFEFEF}\textbf{90.94}    \\ 
\bottomrule
\end{tabular}}
\end{table}

\begin{table}[htbp]
\centering
\scriptsize
\caption{\textbf{Analysis of weight schemes for ${{\mathcal L}^{{\rm{HT}}}}$ and ${{\mathcal L}^{{\rm{FHT}}}}$} on Cavitation-Short. All results are averaged over three runs.}
\label{tab: HT FHT Weight}
\setlength{\tabcolsep}{3.6mm}{
\begin{tabular}{c|c|cccc}
\toprule
\multicolumn{1}{c|}{Loss} & \multicolumn{1}{c|}{Weight} & Accuracy & Precision &Recall & F1-score \\ 
\midrule
\multirow{3}{*}{${{\mathcal L}^{{\rm{HT}}}}$}      
& -            
& 90.43
& 88.46
& 90.60  
& 89.38\\

& NHW          
& 90.86
& \textbf{89.61}
& \textbf{91.18}   
& \textbf{90.34}\\

& \cellcolor[HTML]{EFEFEF}\textbf{PHW} 
& \cellcolor[HTML]{EFEFEF}\textbf{91.29}
& \cellcolor[HTML]{EFEFEF}89.02
& \cellcolor[HTML]{EFEFEF}90.69
& \cellcolor[HTML]{EFEFEF}89.79\\ 

\midrule
\multirow{3}{*}{${{\mathcal L}^{{\rm{FHT}}}}$}     
& -            
& 91.57 
& 89.62
& 92.05 
& 90.69 \\

& NHW          
& 91.86 
& 89.87  
& 92.20  
& 90.93\\

& \cellcolor[HTML]{EFEFEF}\textbf{PHW} 
& \cellcolor[HTML]{EFEFEF}\textbf{92.29}
& \cellcolor[HTML]{EFEFEF}\textbf{90.09}
& \cellcolor[HTML]{EFEFEF}\textbf{92.38}
& \cellcolor[HTML]{EFEFEF}\textbf{91.12}\\ 
\bottomrule
\end{tabular}}
\end{table}

\begin{table}[htbp]
\centering
\scriptsize
\caption{\textbf{Analysis of essential components for DHK} on Cavitation-Short. All results are averaged over three runs.}
\label{tab: HT FT GTT}
\setlength{\tabcolsep}{1.8mm}{
\begin{tabular}{c|cccc|cccc}
\toprule
\multirow{3}{*}{Loss} & \multicolumn{3}{c}{Weight} & \multirow{3}{*}{${{\mathcal L}^{{\rm{GTT}}}}$} & \multirow{3}{*}{Accuracy} & \multirow{3}{*}{Precision} & \multirow{3}{*}{Recall} & \multirow{3}{*}{F1-score} \\ 
\cmidrule{2-4}
& -        & NHW        & PHW   & &   &  & &   \\ 
\midrule
\multirow{3}{*}{${{\mathcal L}^{{\rm{HT}}}}$}   
&\ding{51} &\ding{55} &\ding{55} &\ding{51} 
& 90.86
& 88.83  
& 91.16    
& 89.87\\
&\ding{55} &\ding{51} &\ding{55} &\ding{51} 
& 91.14
& 89.36 
& 91.55 
& 90.32\\
&\cellcolor[HTML]{EFEFEF}\ding{55} 
&\cellcolor[HTML]{EFEFEF}\ding{55} 
&\cellcolor[HTML]{EFEFEF}\ding{51} 
&\cellcolor[HTML]{EFEFEF}\ding{51} 
&\cellcolor[HTML]{EFEFEF}\textbf{91.71} 
&\cellcolor[HTML]{EFEFEF}\textbf{89.46}  
&\cellcolor[HTML]{EFEFEF}\textbf{92.43}  
&\cellcolor[HTML]{EFEFEF}\textbf{90.72}\\ 
\midrule
\multirow{3}{*}{${{\mathcal L}^{{\rm{FHT}}}}$}  
&\ding{51} &\ding{55} &\ding{55} &\ding{51} 
& 92.43
& 90.28
& 92.18
& 91.13\\
&\ding{55} &\ding{51} &\ding{55} &\ding{51} 
& 92.86 
& 90.97 
& 93.00  
& 91.88 \\
&\cellcolor[HTML]{EFEFEF}\ding{55} 
&\cellcolor[HTML]{EFEFEF}\ding{55} 
&\cellcolor[HTML]{EFEFEF}\ding{51} 
&\cellcolor[HTML]{EFEFEF}\ding{51} 
&\cellcolor[HTML]{EFEFEF}\textbf{93.57}
&\cellcolor[HTML]{EFEFEF}\textbf{92.27}
&\cellcolor[HTML]{EFEFEF}\textbf{93.79}
&\cellcolor[HTML]{EFEFEF}\textbf{92.94} \\ 
\bottomrule
\end{tabular}}
\end{table}

\begin{table}[htbp!]
\centering
\scriptsize
\caption{\textbf{Analysis of variants of ${{\mathcal L}^{{\rm{GTT}}}}$ under ${{\mathcal L}^{{\rm{FHT}}}}$ (w/ PHW)} on Cavitation-Short. All results are averaged over three runs.}
\label{tab: GTT margin}
\setlength{\tabcolsep}{2.9mm}{
\begin{tabular}{c|c|cccc}
\toprule
Triplet Loss         & Margin    & Accuracy & Precision & Recall & F1-score \\ 
\midrule
Vanilla   & Constant  
& 92.71 
& 90.78  
& 92.65  
& 91.62 \\ 
\midrule
\multirow{2}{*}{${{\mathcal L}^{{\rm{GTT}}}}$} 
& Constant  
& 93.14   
& 92.05   
& 93.73
& 92.83   \\

& \cellcolor[HTML]{EFEFEF}\textbf{Hierarchy} 
& \cellcolor[HTML]{EFEFEF}\textbf{93.57}
& \cellcolor[HTML]{EFEFEF}\textbf{92.27}
& \cellcolor[HTML]{EFEFEF}\textbf{93.79}
& \cellcolor[HTML]{EFEFEF}\textbf{92.94}\\ 
\bottomrule
\end{tabular}}
\end{table}

\begin{table}[htbp!]
\centering
\scriptsize
\caption{\textbf{Effects of distance measure for ${{\mathcal L}^{{\rm{GTT}}}}$} on Cavitation-Short. All results are averaged over three runs.}
\label{tab: GTT distance}
\setlength{\tabcolsep}{3.1mm}{
\begin{tabular}{c|c|cccc}
\toprule
Triplet Loss  & Distance  & Accuracy & Precision & Recall & F1-score \\ 
\midrule
\multirow{2}{*}{${{\mathcal L}^{{\rm{GTT}}}}$} 
& Euclidean 
& 93.29  
& 91.80     
& 93.67
& 92.79 \\

& \cellcolor[HTML]{EFEFEF}\textbf{Cosine}    
& \cellcolor[HTML]{EFEFEF}\textbf{93.57}
& \cellcolor[HTML]{EFEFEF}\textbf{92.27}
& \cellcolor[HTML]{EFEFEF}\textbf{93.79}
& \cellcolor[HTML]{EFEFEF}\textbf{92.94}\\ 
\bottomrule
\end{tabular}}
\end{table}

\begin{figure*}[htbp]
\centering
\subfigure[Parameter sensitivity]{
\begin{minipage}[t]{0.25\linewidth}
\centering
\includegraphics[width=\textwidth,height=32mm]{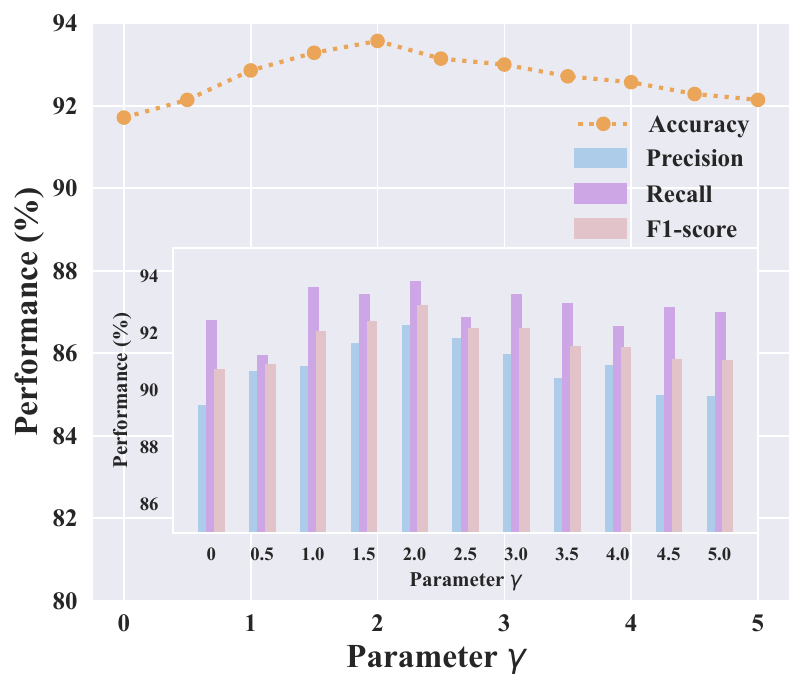}
\end{minipage}%
}%
\subfigure[STFT parameter analysis]{
\begin{minipage}[t]{0.25\linewidth}
\centering
\includegraphics[width=\textwidth,height=32mm]{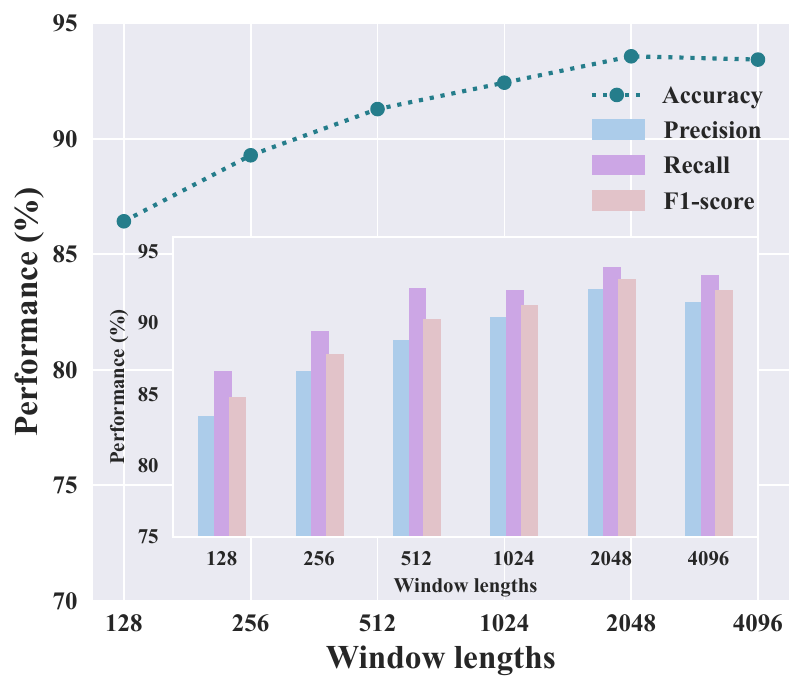}
\end{minipage}%
}%
\subfigure[Window size analysis]{
\begin{minipage}[t]{0.25\linewidth}
\centering
\includegraphics[width=\textwidth,height=32mm]{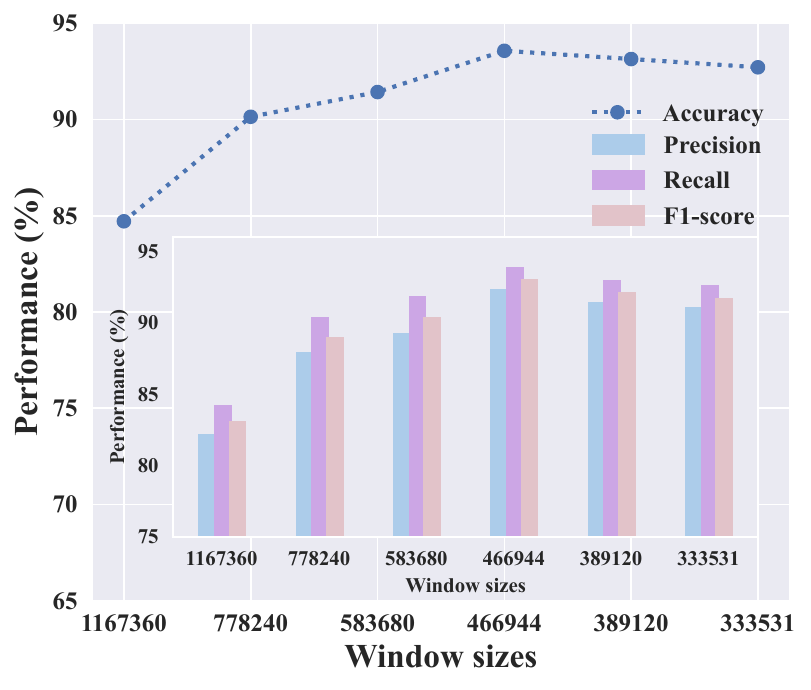}
\end{minipage}%
}%
\subfigure[Downsampling effects]{
\begin{minipage}[t]{0.25\linewidth}
\centering
\includegraphics[width=\textwidth,height=32mm]{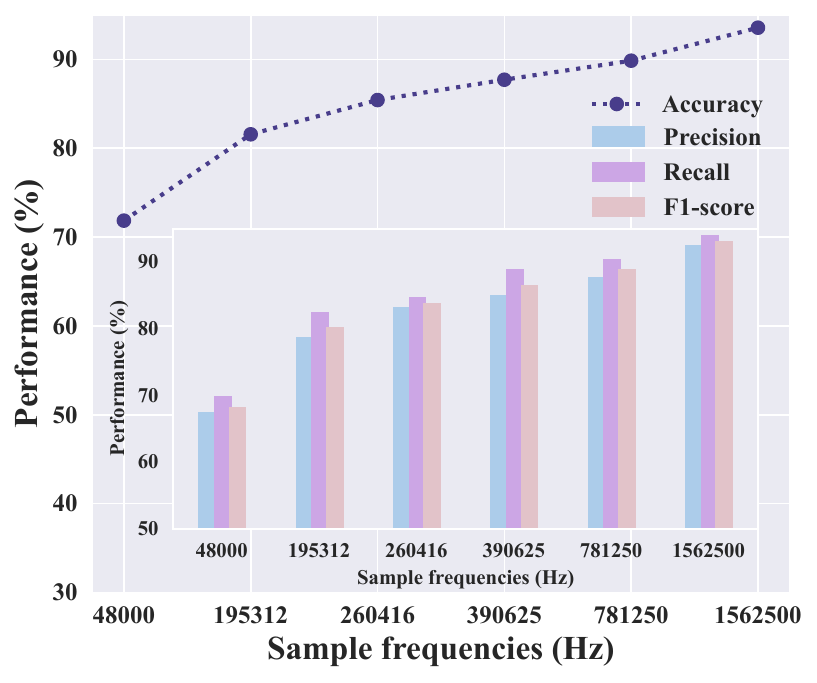}
\end{minipage}%
}%
\centering
\vspace{-12pt}
\caption{Results of different evaluation metrics from various ablation experiments on Cavitation-Short. (a)-(b) and (d) are all performed with a window size of 466944 and a step size of 466944. All results are the average of three runs.} 
\label{fig: ablation analysis}
\vspace{-7pt}
\end{figure*}

\noindent\textbf{Parameter Sensitivity.} We analyse the effect of different values of the tunable focusing factor $\gamma$ in \hrefequation{eq: FHT Loss} on Cavitation-Short, see \hreffigurea{fig: ablation analysis}. The DHK achieves the best accuracy with $\gamma  = 2$. When $\gamma$ is set too small, the model focuses excessively on easy-to-classify samples, leading to insufficient learning of hard samples. Conversely, when $\gamma$ is set too large, although the model places more emphasis on hard samples, it may introduce too much noise, resulting in an unstable training process and potentially hindering the learning of easy-to-classify samples.

\noindent\textbf{STFT Parameter Analysis.} We examine the impact of the window length of the STFT on the performance of DHK. The step length is one-quarter of the window length in all experiments. As shown in \hreffigureb{fig: ablation analysis}, the window length plays a crucial role in influencing the results and the most optimal window length is 2048. In general, a longer window length results in higher frequency resolution but lower time resolution, while a shorter window length leads to higher time resolution but lower frequency resolution.

\noindent\textbf{Window Size Analysis.} We evaluate the performance of our method with various window sizes on Cavitation-Short. \hreffigurec{fig: ablation analysis} clearly shows that the window size has a significant effect on the performance of DHK and the optimal window size is 466944. It indicates that this window size preserves the important physical states of the cavitation flow.

\noindent\textbf{Downsampling Effects.} In practical applications, the ability to recognize signals captured from low-level sensors is very significant and challenging. We study our method under the original sampling frequency ($Fs$ = \SI{1562500}{Hz}), one-half, one-quarter, one-sixth, one-eighth of the original sampling frequency (\SI{781250}{Hz}, \SI{390625}{Hz}, \SI{260416}{Hz}, \SI{195312}{Hz}) and the maximum frequency tolerable by a mobile phone (\SI{48000}{Hz} $\approx Fs/32$). \hreffigured{fig: ablation analysis} illustrates that although the performance of DHK gradually declines as the sampling frequency decreases, the accuracy consistently remains above \textbf{80}$\%$. Moreover, our method also achieves \textbf{71.86}$\%$ accuracy at the maximum sampling frequency of mobile phones.

\noindent\textbf{Computational Complexity.} The training and inference time of DHK are given in \hreftable{apptab: training time} and \hreftable{apptab: inference time}, respectively. From \hreftable{apptab: training time}, it can be found that DHK loss only introduces a minimal training time cost compared to CCE loss. From \hreftable{apptab: inference time}, it can clearly be seen that there is almost no difference in the inference time between DHK and CCE. The computational complexity of DHK from theoretical angles are given in \hrefappendix{subsubsection: Computational Complexity}-\ref{subsubsection: Inference Time}.
\begin{table}[htbp]
\centering
\footnotesize
\centering
\caption{The training time on Cavitation-Short. All results are averaged over three runs.}
\vspace{-5pt}
\label{apptab: training time}
\setlength{\tabcolsep}{5.5mm}{
\begin{tabular}{c|c|cc}
\toprule
Backbone & Loss & ave\_epoch (min) & total (h) \\ 
\midrule
\multirow{2}{*}{ResNet18} 
& CCE  & 2.79 & 4.64 \\
& DHK  & 2.87 & 4.87 \\
\bottomrule
\end{tabular}}
\end{table}

\begin{table}[htbp]
\centering
\footnotesize
\centering
\caption{The inference time on Cavitation-Short. All results are averaged over three runs.}
\vspace{-5pt}
\label{apptab: inference time}
\footnotesize
\setlength{\tabcolsep}{5.9mm}{
\begin{tabular}{c|c|cc}
\toprule
Backbone & Loss & ave\_batch (s) & total (s) \\ 
\midrule
\multirow{2}{*}{ResNet18} 
& CCE  & 0.0076 & 3.05 \\
& DHK  & 0.0078 & 3.13 \\
\bottomrule
\end{tabular}}
\end{table}

\noindent\textbf{Label Noise.} To evaluate the robustness of DHK under label noise, we randomly flip a certain percentage of labels in the training samples to simulate noisy labels and simultaneously keep the test sets clean, see \hreftable{apptab: label noise}. From \hreftable{apptab: label noise}, the DHK consistently outperforms baseline across different noise levels, it demonstrate the potential of DHK in real-world scenarios with imperfect data annotation.
\begin{table}[htbp]
\centering
\scriptsize
\caption{Results of ResNet18 and ResNet18+DHK with different label noise ratios on Cavitation-Short.}
\label{apptab: label noise}
\vspace{-5pt}
\setlength{\tabcolsep}{0.38mm}{
\begin{tabular}{c|llll|llll|llll}
\toprule 
Noise Ratios & \multicolumn{4}{c|}{0\%}  & \multicolumn{4}{c|}{5\%}  & \multicolumn{4}{c}{10\%}    \\ 
\midrule
Method       & \multicolumn{1}{c}{Acc} & \multicolumn{1}{c}{Pre} & \multicolumn{1}{c}{Rec} & \multicolumn{1}{c|}{F1} 
             & \multicolumn{1}{c}{Acc} & \multicolumn{1}{c}{Pre} & \multicolumn{1}{c}{Rec} & \multicolumn{1}{c|}{F1} 
             & \multicolumn{1}{c}{Acc} & \multicolumn{1}{c}{Pre} & \multicolumn{1}{c}{Rec} & \multicolumn{1}{c}{F1}  \\ 
\midrule
ResNet18     
&87.57        &90.25        &77.16          &77.95       
&85.43        &82.67        &84.87          &83.62                         
&81.29        &78.17        &82.90          &79.66                      \\
\rowcolor[HTML]{EFEFEF}
\textbf{ResNet18+DHK} 
&\textbf{92.57}        &\textbf{90.67}        &\textbf{91.95}          &\textbf{91.27}       
&\textbf{88.85}        &\textbf{86.84}        &\textbf{88.48}          &\textbf{87.55}                          
&\textbf{83.86}        &\textbf{80.57}        &\textbf{83.44}          &\textbf{81.59}      \\
\bottomrule
\end{tabular}}
\end{table}

\noindent\textbf{Other Analysis.} The statistical p-value analysis of DHK results are given in \hrefappendix{subsubsection: Statistical p-value}. The discussion of DHK refers to \hrefappendix{app: Discussion} (include assumption, limitations, extensibility, broader impact and future improvements).
\section{Related Work}
\label{sec: Related Work}
\noindent\textbf{Fault Intensity Diagnosis.} The FID is treated as a special recognition task based on signal fine-grained fault classification, which can be broadly organized into two categories: data-driven and knowledge-data hybrid-driven. For data-driven methods, numerous studies explore various deep learning architectures by exploiting the ability of convolution to capture local features \cite{pan2017liftingnet,pan2019novel,li2020intelligent,zhao2020deep,mohammad2023one,yu2023tdmsae} or using the advantage of transformer in modeling global features and long-term dependencies \cite{yu2023adaptive,eldele2023self,cui2024self}. For knowledge-data hybrid-driven methods, these studies focus on the combination of domain knowledge and data features achieve efficient modeling of complex signal features by introducing physical models, a priori knowledge or expert experience \cite{li2024physics,li2024predicting,sun2024target,guo2024causal,sha2024hierarchical}. 

\noindent\textbf{Hierarchical Classification.} The hierarchical information of target class embedded into deep learning models is a widely discussed topic \cite{gu2025adaptive,zhang2025understanding,he2025lorentzian,springstein2024visual}. 
Deep learning-based hierarchical classification methods can be broadly grouped into three types: hierarchical structures, label embedding and hierarchical loss. The hierarchical structure methods \cite{he2024language,wang2025home,yehezkel2019modeling} are deep neural networks designed to align with a task's class hierarchy. The label embedding methods \cite{yang2025mixed,kharbanda2024gandalf,bei2025correlation} encode hierarchical label information as semantic vectors with positional relationships. The hierarchical loss methods \cite{kamarthi2024large,sun2024self,Li_2022_CVPR} ensure prediction consistency with the class hierarchy by soft knowledge gradient propagation. In this work, our proposed DHK is one of the hierarchical loss methods.

\section{Conclusion}
\label{sec: Conclusion}
In this paper, we present a novel framework with deep hierarchical knowledge loss for fault intensity diagnosis, which incorporates the focal hierarchical tree loss with two adaptive weighting schemes and the group tree triplet loss with a hierarchical dynamic margin to model same-class knowledge constraints and cross-class boundary structural knowledge, respectively. Extensive experiments across various setting and practical datasets demonstrate the superiority of the proposed method. 
\section*{Acknowledgements}
\label{sec:acknowledgements}
This research is supported by the China Postdoctoral Science Foundation under No.\ 2025M781519 (Y. S.), by the National Natural Science Foundation of China under No.\ 92570117, by the CUHK-Shenzhen University Development Fund under No.\ UDF01003041 and UDF03003041, by Shenzhen Peacock Fund under No.\ 2023TC0007 (Y. S, K. Z.), by AI grant at FIAS through SAMSON AG (K. Z.), by the BMBF funded KISS consortium (05D23RI1) in the ErUM-Data action plan (K. Z.), by SAMSON AG (D. V., A. W.), by the Walter GreinerGesellschaft zur F\"orderung der physikalischen Grundla - genforschung e.V. through the Judah M. Eisenberg Lau-reatus Chair at Goethe Universit\"at Frankfurt am Main (H. S.) and by the NVIDIA GPU grant through NVIDIA Corporation (K. Z.).

\bibliographystyle{ACM-Reference-Format}
\bibliography{refs}

\clearpage
\appendix
\titlecontents{section}[1.5em]{\small\bfseries}{\contentslabel{1.5em}}{\hspace*{-1.5em}}{\titlerule*[0.5pc]{.}\contentspage}
\titlecontents{subsection}[3.8em]{\small}{\contentslabel{2.3em}}{\hspace*{-2.3em}}{\titlerule*[0.5pc]{.}\contentspage}
\section*{Contents of Appendix}
\addcontentsline{toc}{section}{Contents of Appendix}  
\startcontents[sections]  
\printcontents[sections]{}{1}{}  %
\setcounter{table}{0}
\setcounter{figure}{0}
\setcounter{equation}{0}
\renewcommand{\thetable}{A\arabic{table}}
\renewcommand{\thefigure}{A\arabic{figure}}
\renewcommand{\theequation}{A.\arabic{equation}}
\newpage
\section{Preliminaries}
\label{app: Preliminaries}
\subsection{Definition of Cavitation Intensity}
\label{app: Cavitation Intensity Definition}
\begin{figure}[htbp]
    \centering
    \includegraphics[width=0.45\textwidth,height=40mm]{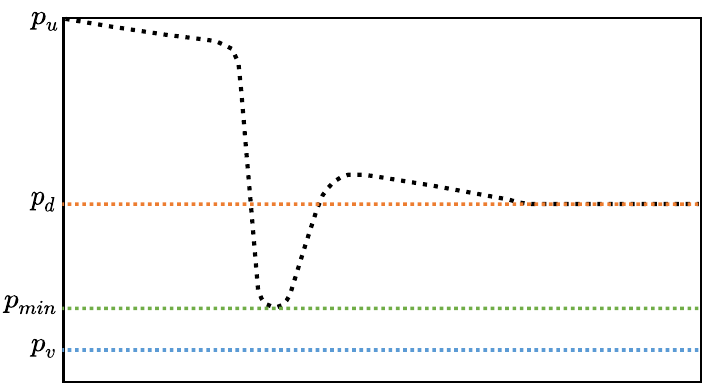}
    \caption{Pressure flow model in a valve. The solid black, dotted orange, dotted green and dotted blue lines represent the upstream pressure ${p}_{u}$, the downstream pressure ${p}_{d}$, the minimum pressure ${p}_{min}$ and the vapor pressure ${p}_{v}$, respectively.}
    \label{fig: civitation knowledge}
\end{figure}

\hreffigure{fig: civitation knowledge} demonstrates the variation in local pressure in a one-dimensional flow. Cavitation does not occur and the valve functions normally when the minimum pressure ${p}_{min}$ exceeds the vapor pressure ${p}_{v}$. However, when ${p}_{v}>{p}_{min}$, cavitation begins. In practice, direct measurement is difficult because the minimum pressure is downstream of the restriction. Therefore, the cavitation coefficient ${X}_{FZ}$ is proposed, which represents the ratio of the external pressure difference to the internal pressure difference. This coefficient can be determined empirically by assuming that cavitation noise begins when the minimum pressure ${p}_{min}$ matches the vapor pressure ${p}_{v}$. Consequently, the cavitation coefficient ${X}_{FZ}$ can be measured through the noise, which varies depending on the valve load. The formulas for the cavitation coefficient ${X}_{FZ}$ and the operating pressure ratio ${X}_{F}$ are given below:
\begin{equation}
{X}_{FZ}=\frac{{p}_{u}-{p}_{d}}{{p}_{u}-{p}_{min}}, \,\,\,\,\,\,\,\,\,\,\,\,\,\,\,\,\,\,\,\,\,\,\,\,\,\,\,\, {X}_{F}=\frac{{p}_{u}-{p}_{d}}{{p}_{u}-{p}_{v}},
\end{equation}
where ${p}_{u}$ represents the upstream pressure, ${p}_{d}$ denotes the downstream pressure, ${p}_{min}$ refers to the minimum pressure within the valve and ${p}_{v}$ stands for the vapor pressure. 

When all coefficients are determined across the entire range of valve opening, the following conclusions can be drawn:
\begin{itemize}
\item ${X}_{F}<{X}_{FZ}$: The valve operates without cavitation and the flow may be either turbulent or laminar.
\item ${X}_{F}\geq {X}_{FZ}$: When ${X}_{F}={X}_{FZ}$, the valve operates exhibits incipient cavitation. As the difference between ${X}_{FZ}$ and ${X}_{F}$ increases, the cavitation region expands due to pressure drop from higher flow velocities.
\item ${X}_{F}>1$: Here the bubbles do not implode in the valve but rather continue to flow into the pipe because the downstream pressure ${p}_{d}$ is lower than the vapor pressure ${p}_{v}$. This phenomenon is called flashing.
\end{itemize}
The cavitation coefficient ${X}_{FZ}$ is applied only to the fluid, where it is measured empirically. Its valve varies for different liquid mediums due to changes in viscosity, dissolved gas content and other factors.

\subsection{Cavitation Event Intensity Diagnosis}
\label{app: Cavitation Event Intensity Diagnosis}
Cavitation is defined as the phenomenon of involving the formation, growth and collapse of local bubbles or vapor cavities in a liquid \cite{plesset1977bubble}. In piping systems, the acoustic signals of different flow conditions (different levels of cavitation or non-cavitation) are recorded as continuous waveforms using acoustic sensors. Each observed acoustic signal records the entire physical process from the beginning to the end of the event of the corresponding flow state. In our experiments, cavitation intensity recognition mainly distinguishes incipient cavitation, constant cavitation, choked flow cavitation and non-cavitation. Whether severe or subtle, any form of cavitation would indicate a potential issue or failure in system operation. Therefore, it is crucial for operators of industrial systems to effectively and precisely recognize different intensities of cavitation in order to implement appropriate countermeasures. 

\subsection{Acoustic Signals Augmentation}
\label{app: Acoustic Signals Augmentation}
Formally, there is $x\subseteq {\mathbb{R}}^{M\times N}$ with $M$ measurements for each acoustic signal. Considering the purposely maintained steady flow status (i.e., it’s always the same fluid status class within the individual measurement duration with \SI{3}{s} or \SI{25}{s}) in each recorded stream and fine resolution for the sensor. Each signal can split each stream into several pieces, each containing sufficient information for detection. The segment length is not too short to account for the inherent randomness in the noise emission and the features of each segment are independent. Therefore, we apply a sliding window (SW) with window size ${w}$ and step size ${s}$ to divide the acoustic signal sequence into a set of sub-sequences ${\mathbf{X}}_{sw}=\left\{{x}_{i,j},i=1,2,\ldots,N;j=1,2,\ldots,k\right \}\subseteq {\mathbb{R}}^{{w}\times kN}$, where $k=\frac{\left(M-{w}\right)}{{s}}$ is the number of sub-sequences and $N$ is stream. The SW technique is a crucial part of the acoustic signal pre-processing.

\subsection{Time-Frequency Transform}
\label{app: Time-Frequency Transform}
Time-Frequency (T-F) transform provides more detailed and comprehensive information across both time and frequency dimensions. The most widely used T-F transform is computed by the short time Fourier transform (STFT), which can be converted back to time-domain signals by the inverse STFT (iSTFT). Given a sequence of signal $ {x[n]}_{n=0}^{N-1}$, the STFT converts the signal sequence into the T-F domain and is defined by the formula:
\begin{equation}
\label{eq: STFT}
{\tilde X}_{w}[k] = \sum\limits_{n = 0}^{N - 1} {{{x}_{w}\left [n\right]}{e^{ - \frac{{2\pi j}}{N}nk}}}  := \sum\limits_{n = 0}^{N - 1} {{{x}_{w}\left [n\right]}W_N^{kn}},
\end{equation}
where ${x}_{w}\left [n\right]=x\left [n\right]\cdot w\left [n-m\right]$ denotes the weighted signals to the window function $w\left [n-m\right]$, ${\tilde X}_{w}[k]$ is the result in the frequency domain, $j$ is the imaginary unit and ${W_N} = {e^{ - \frac{{2\pi j}}{N}}}$. The essence of the STFT is to apply a Discrete Fourier Transform (DFT) on the resulting windowed signal ${x}_{w}\left [n\right]$. The DFT formulation in \hrefequation{eq: STFT} can be derived from the Fourier transform for continuous signals. For our method, the STFT plays an essential role in the pre-processing of the acoustic signal.

\section{Proofs}
\label{app: Proofs}
\subsection{Proof of Assumption 3.1}
\label{app: Hierarchical Inference}
$Proof.$ Let $\mathcal{T} = (\mathcal{V},\mathcal{E})$ be a hierarchical tree, i.e. a directed acyclic graph, where $\mathcal{V}$ is the set of nodes representing possible class labels and $\mathcal{E}$ is the set of edges representing hierarchical relationships among the classes. Each path $\mathcal{P} \subseteq \mathcal{T}$ is a valid path from the root node ${v^r}={v_1}$ to a leaf node ${v_c}$, with each node ${v_i}$ being a node in the path.

Since each node ${v_i}$ depends only on the input $\tilde {x}$ and its preceding node ${v_{i-1}}$, the probability of the path from the root node ${v^r}$ to the leaf node ${v_c}$ can be written as:
\begin{equation}
\label{eq: root to leaf}
p({v_1} \to {v_2} \to  \cdots  \to {v_c}|\tilde x) = p({v_1}|\tilde x) \cdot p({v_2}|{v_1},\tilde x) \cdots p({v_c}|{v_{c - 1}},\tilde x).
\end{equation}
Based on conditional independence, we conclude that:
\begin{equation}
\label{eq: conditional independence}
p({v_i}|{v_{i - 1}},\tilde x) = p({v_i}|\tilde x).
\end{equation}
By applying \hrefequation{eq: root to leaf} and \hrefequation{eq: conditional independence}, the path probability formula can be simplified to:
\begin{equation}
\label{eq: path probability simplified}
\begin{split}
p({v_1} \to {v_2} \to  \cdots  \to {v_c}|\tilde x) &= p({v_1}|\tilde x) \cdot p({v_2}|\tilde x) \cdots p({v_c}|\tilde x) \\
&= \prod\nolimits_{i = 1}^{|\mathcal{P}|} {p({v_i}|\tilde x)}.
\end{split}
\end{equation}
When the hierarchical tree $\mathcal{T}$ consists of only one hierarchy, all leaf nodes are directly connected to the root node, i.e. the path length is 1. In this case, hierarchical inference reduces to a typical classification inference:
\begin{equation}
\label{eq: typical classification inference}
p({v_1} \to {v_c}|\tilde x) = p({v_1}|\tilde x)p({v_c}|{v_1},\tilde x) = p({v_c}|\tilde x).
\end{equation}
It follows that traditional classification inference ($cf.$ \hrefequation{eq: typical classification inference}) is a special case of hierarchical inference ($cf.$ \hrefequation{eq: path probability simplified}).

\subsection{Derivation of \hrefequation{eq: HT Loss}}
\label{app: HT Loss}
$Proof.$ First, the traditional binary cross-entropy loss function can be written as:
\begin{equation}
\label{eqapp: BCE Loss}
{\mathcal{L}^{BCE}} = \sum\limits_{v \in \mathcal{V}} { - {{\tilde y}_v}\log ({s_v}) - (1 - {{\tilde y}_v})\log (1 - {s_v})},
\end{equation}
where ${s_v}$ represents the predicted probability for node $v$ and ${y_v} \in \{ 0,1\}$ denotes the true label for node $v$. 

In addition, we impose the following hierarchical constraints:
\begin{equation}
\label{eqapp: updated prediction score}
\left\{\begin{aligned}
{{\hat s}_v} &= \mathop{\min}\limits_{u \in {\mathcal{V}_A}} ({s_u}) \;\;\;\;\;\;\;\;\;\;\;\;\;\;\;\;\;{{\tilde y}_v} = 1, \\
1 - {{\hat s}_v} &= \mathop{\min}\limits_{u \in {\mathcal{V}_R}} (1 - {s_u}) \;\;\;\;\;\;\;\;\;\;\;\;{{\tilde y}_v} = 0.
\end{aligned}\right.
\end{equation}
In order to facilitate the incorporation of \hrefequation{eqapp: updated prediction score} into \hrefequation{eqapp: BCE Loss}, \hrefequation{eqapp: updated prediction score} is converted as follows:
\begin{equation}
\label{eqapp: converted updated prediction score}
\left\{\begin{aligned}
{{\hat s}_v} &= \mathop{\min}\limits_{u \in {\mathcal{V}_A}} ({s_u}) \;\;\;\;\;\;\;\;\;\;\;\;\;\;\;\;\;\;\;\;\;\;\;\;\;\;\;\;\;\;\;\;\;\;\;\;\;\;\;\;\;{{\tilde y}_v} = 1, \\
{{\hat s}_v} &= 1-\mathop{\min}\limits_{u \in {\mathcal{V}_R}} (1 - {s_u})= \mathop{\max}\limits_{u \in {\mathcal{V}_R}} ({s_u})\;\;\;\;\;\;\;\;\;\;\;\;{{\tilde y}_v} = 0.
\end{aligned}\right.
\end{equation}
By incorporating the updated prediction scores ($cf.$ \hrefequation{eqapp: converted updated prediction score}) into the binary cross-entropy loss ${\mathcal{L}^{BCE}}$ ($cf.$ \hrefequation{eqapp: BCE Loss}), we obtain the hierarchical tree loss ${\mathcal{L}^{HT}}$ ($cf.$ \hrefequation{eq: HT Loss}):
\begin{equation}
\label{eqapp: Derivation HT Loss}
\begin{aligned}
\mathcal{L}^{HT} &= \sum\limits_{v \in \mathcal{V}} { - {{\tilde y}_v}\log ({s_v}) - (1 - {{\tilde y}_v})\log (1 - {s_v})}, \\
&= \sum\limits_{v \in \mathcal{V}} { - {{\tilde y}_v}\log (\mathop {\min }\limits_{u \in {\mathcal{V}_A}} ({s_u}))}  - \;(1 - {{\tilde y}_v})\log (1 - \mathop {\max }\limits_{u \in {\mathcal{V}_R}} ({s_u})).
\end{aligned}
\end{equation}

\subsection{Convergence of \hrefequation{eq: HT Loss}}
\label{app: Convergence HT Loss}
$Proof.$ To analyse the convergence of the hierarchical tree loss $\mathcal{L}^{HT}$, we need to prove that it satisfies the following five conditions: non-negativity, minimum value of 0, Lipschitz continuity, bounded gradients and deterministic convergence. Next, we will conduct a detailed analysis based on these five aspects.

\noindent\textbf{Analysis of Non-Negativity.} Each term in the hierarchical tree loss $\mathcal{L}^{HT}$ is non-negative. Specifically:
\begin{itemize}
    \item When ${{\tilde y}_v} = 1$, the following holds:
    \begin{equation}
    \label{eqapp: first trem}
    { - {{\tilde y}_v}\log (\mathop {\min }\limits_{u \in {\mathcal{V}_A}} ({s_u}))}={-\log(\mathop {\min }\limits_{u \in {\mathcal{V}_A}} ({s_u}))}.
    \end{equation}
    Since ${s_u} \in [0,1]$, we have $\mathop {\min }\limits_{u \in {\mathcal{V}_A}} ({s_u}) \in [0,1]$. Moreover, $- \log ( \cdot )$ is defined over the interval $\left( {0,1} \right]$, it follows that ${ - {{\tilde y}_v}\log (\mathop {\min }\limits_{u \in {\mathcal{V}_A}} ({s_u}))} \ge 0$. Therefore, this term is non-negative.
    \item When ${{\tilde y}_v} = 0$, the following holds:
    \begin{equation}
    \label{eqapp: second term}
    - \;(1 - {{\tilde y}_v})\log (1 - \mathop {\max }\limits_{u \in {\mathcal{V}_R}} ({s_u}))=-\log (1 - \mathop {\max }\limits_{u \in {\mathcal{V}_R}} ({s_u})).
    \end{equation}
    Since ${s_u} \in [0,1]$, we have $\mathop {\max }\limits_{u \in {\mathcal{V}_R}} ({s_u}) \in [0,1]$, which implies that $1 - \mathop {\max }\limits_{u \in {\mathcal{V}_R}} ({s_u}) \in [0,1]$. Over this interval, $-\log (1 - \mathop {\max }\limits_{u \in {\mathcal{V}_R}} ({s_u}))\ge 0$. Therefore, this term is also non-negative. 
\end{itemize}
Consequently, the hierarchical tree loss $\mathcal{L}^{HT}$ is non-negative.

\noindent\textbf{Conditions for Minimum Valve of 0.} For the hierarchical tree loss $\mathcal{L}^{HT}$ to achieve its minimum value of 0, each term must be 0. This requires the following conditions:
\begin{itemize}
    \item When ${{\tilde y}_v} = 1$, to satisfy $-\log(\mathop {\min }\limits_{u \in {\mathcal{V}_A}} ({s_u}))=0$, we need ${\mathop {\min }\limits_{u \in {\mathcal{V}_A}} ({s_u})}=1$, which implies that the predicted probability ${s_u}=1$ for all ancestor nodes.
    \item When ${{\tilde y}_v} = 0$, to satisfy $-\log (1 - \mathop {\max }\limits_{u \in {\mathcal{V}_R}} ({s_u}))=0$, we need $1 - \mathop {\max }\limits_{u \in {\mathcal{V}_R}} ({s_u})=1$, i.e. $\mathop {\max }\limits_{u \in {\mathcal{V}_R}} ({s_u})=0$. This implies that the predicted probability ${s_u}=0$ for all child nodes.
\end{itemize}
Based on the above, when these conditions are met, the hierarchical tree loss $\mathcal{L}^{HT}=0$.

\noindent\textbf{Analysis of Lipschitz continuity.} First, we decompose the function $\mathcal{L}^{HT}$ into two parts:
\begin{equation}
\label{eqapp: decompose function}
{\mathcal{L}^{HT}} = \underbrace { - {{\tilde y}_v}\log (\mathop {\min }\limits_{u \in {\mathcal{V}_A}} {s_u})}_{: = {f_1}(\theta )} + \underbrace { - (1 - {{\tilde y}_v})\log (1 - \mathop {\max }\limits_{u \in {\mathcal{V}_R}} {s_u})}_{: = {f_2}(\theta )}.
\end{equation}
Then, it suffices to prove that the gradients of ${f_1}$ and ${f_2}$ are Lipschitz continuous, respectively.
\begin{itemize}
    \item Analysis of the gradient of ${f_1}(\theta ) =  - {{\tilde y}_v}\log (\mathop {\min }\limits_{u \in {\mathcal{V}_A}} {s_u})$. Assuming ${s_u} = \sigma ({z_u})$ ($\sigma$ is the sigmoid function) and ${z_u} = w_u^Th + {b_u}$ ($h$ is the hidden layer output). According to the chain rule, we have:
    \begin{equation}
    \label{eqapp: part 1 gradient}
    {\nabla _{{w_u}}}{f_1} =  - \frac{1}{{\min {s_u}}} \cdot \mathbbm{1}({s_u} = \min {s_u}) \cdot \sigma '({z_u}) \cdot h,
    \end{equation}
    where $\mathbbm{1}(\cdot)$ is an indicator function. Since $\sigma '({z_u}) = {s_u}(1 - {s_u}) \le \frac{1}{4}$ and after numerical stabilization $\min {s_u} \ge \varepsilon$, we obtain:
    \begin{equation}
     \label{eqapp: low boundary} 
     \left\| {{\nabla _{{w_u}}}{f_1}} \right\| \le \frac{1}{\varepsilon } \cdot \frac{1}{4} \cdot \left\| h \right\|.
    \end{equation}
    Since the hidden layer output $h$ is bounded, then $\nabla {f_1}$ is bounded. According to the boundedness of the gradient and the mean value theorem, for any ${\theta _1}$ and ${\theta _2}$, we have:
    \begin{equation}
    \label{eqapp: mean value theory}    
    \left\| {\nabla {f_1}({\theta _1}) - \nabla {f_1}({\theta _2})} \right\| \le \alpha_1 \left\| {{\theta _1} - {\theta _2}} \right\|,
    \end{equation}
    where $\alpha$ depends on $\varepsilon$ and the Lipschitz constant of the network parameters.
    \item Analysis of the gradient of ${f_2}(\theta ) =  - (1 - {{\tilde y}_v})\log (1 - \mathop {\max }\limits_{u \in {V_R}} {s_u})$. Similarly, following the same reasoning as above, the Lipschitz constant $\alpha_2$ of $\nabla {f_2}$ is determined by the max operator and the boundedness of the gradient of $1 - {s_u}$.
\end{itemize}
By setting $\alpha  = {\alpha _1} + {\alpha _2}$, we conclude that the gradient of $\mathcal{L}^{HT}$ is Lipschitz continuous.

\noindent\textbf{Analysis of Gradient Bound.} The gradient boundary of $\mathcal{L}^{HT}$ is composed of the bounding of $\nabla {f_1}$ and the bounding of $\nabla {f_2}$.
\begin{itemize}
    \item Analysis of bounding $\nabla {f_1}$. Assuming $\left\| h \right\| \le c$, based on \hrefequation{eqapp: low boundary}, we have:
    \begin{equation}
    \label{eqapp: f1 bounding} 
    \left\| {\nabla {f_1}} \right\| \le \frac{c}{{4\varepsilon }}.
    \end{equation}
    \item Analysis of bounding $\nabla {f_2}$. Similarly, we have:
    \begin{equation}
    \label{eqapp: f2 bounding} 
    \left\| {\nabla {f_2}} \right\| \le \frac{c}{{4\varepsilon }}.
    \end{equation}
\end{itemize}
Since $\mathcal{L}^{HT}=f_1 + f_2$ and $C = \frac{c}{{2\varepsilon }}$, we have $\left\| {\nabla {\mathcal{L}^{HT}}} \right\| \le C$. Therefore, the gradient of $\mathcal{L}^{HT}$ is bounded.

\noindent\textbf{Analysis of Deterministic Convergence.} Since ${\nabla {\mathcal{L}^{HT}}}$ is Lipschitz continuous, for any $\theta$ and ${\theta '}$, we have:
\begin{equation}
\label{eqapp: Descent Lemma1}
{\mathcal{L}^{HT}}(\theta ') \le {\mathcal{L}^{HT}}(\theta ) + \nabla {\mathcal{L}^{HT}}{(\theta )^T}(\theta ' - \theta ) + \frac{\alpha }{2}{\left\| {\theta ' - \theta } \right\|^2}.
\end{equation}
Based on gradient descent update $\theta ' = \theta  - \eta \nabla {L^{HT}}(\theta )$, we have:
\begin{equation}
\label{eqapp: Descent Lemma2}
{\mathcal{L}^{HT}}({\theta _{k + 1}}) \le {\mathcal{L}^{HT}}({\theta _k}) - \eta (1 - \frac{{\eta \alpha }}{2}){\left\| {\nabla {\mathcal{L}^{HT}}({\theta _k})} \right\|^2}.
\end{equation}
When $\eta  < \frac{2}{\alpha }$, then ${\mathcal{L}^{HT}}({\theta _{k + 1}}) < {\mathcal{L}^{HT}}({\theta _k})$, i.e. strictly decreasing. Since $\mathcal{L}^{HT}$ is bounded, then the monotonically decreasing sequence $\{ {L^{HT}}({\theta _k})\}$ converges to a certain value ${\alpha ^*}$. Next, the sum of \hrefequation{eqapp: Descent Lemma2}, we have:
\begin{equation}
\label{eqapp: Descent Lemma3} 
\sum\limits_{k = 0}^\infty  {\eta (1 - \frac{{\eta \alpha }}{2})} {\left\| {\nabla {\mathcal{L}^{HT}}({\theta _k})} \right\|^2} \le {\mathcal{L}^{HT}}({\theta _0}) - {\alpha ^*} < \infty. 
\end{equation}
Therefore, we have $\mathop {\lim }\limits_{k \to \infty } \left\| {\nabla {L^{HT}}({\theta _k})} \right\| = 0$, i.e. $\mathcal{L}^{HT}$ satisfies deterministic convergence.

In summary, the hierarchical tree loss $\mathcal{L}^{HT}$ is convergent. Similarly, the focal hierarchical tree loss $\mathcal{L}^{FHT}$ is also convergent.

\subsection{Differentiability Analysis of \hrefequation{eq: HT Loss}}
\label{app: Differentiability HT Loss}
$Proof.$ The min and max operations in \hrefequation{eq: HT Loss} can introduce non-differentiable points due to sudden changes in output as internal node scores change. Specifically, the results of the min and max operations may shift abruptly as the values of the internal nodes vary, leading to non-differentiable points. However, in deep learning optimization, we can handle such non-differentiable cases using subgradients or alternative methods.
\begin{itemize}
    \item \textbf{Smooth Approximation:} The min and max operators can be approximated with differentiable smooth functions, such as the Softplus function or the LogSumExp function, as follows:
    \begin{equation}
    \label{eqapp: Softplus function}
    \begin{aligned}
    \min ({x_1},{x_2}, \cdots ,{x_n}) & \approx  - \frac{1}{\alpha }\log (\sum\limits_{i = 1}^n {{e^{ - \alpha {x_i}}}} ),\\
    \max ({x_1},{x_2}, \cdots ,{x_n}) & \approx \frac{1}{\alpha }\log (\sum\limits_{i = 1}^n {{e^{\alpha {x_i}}}} ),
    \end{aligned}
    \end{equation}
    where $\alpha$ is a smoothing parameter and as $\alpha  \to \infty $, the approximation becomes increasingly accurate.
    \item \textbf{Gradient Computation:} In practical optimization, even if non-differentiable points exist, automatic differentiation frameworks (e.g. PyTorch and TensorFlow) can generally handle the gradients of piecewise smooth function, enabling parameter updates to proceed without obstruction.
\end{itemize}
Although the hierarchical tree loss $\mathcal{L}^{HT}$ contains min and max operations that lead to non-differentiable points, it can still be optimized using smooth approximations or subgradient methods. 

Specifically, considering the gradient derivation of the min and max parts in the hierarchical tree loss $\mathcal{L}^{HT}$ ($cf.$ \hrefequation{eq: HT Loss}):
\begin{itemize}
    \item When ${{\tilde y}_v} = 1$, the loss term is:
    \begin{equation}
    \label{eqapp: first trem diff}
    { - {{\tilde y}_v}\log (\mathop {\min }\limits_{u \in {\mathcal{V}_A}} ({s_u}))}={-\log(\mathop {\min }\limits_{u \in {\mathcal{V}_A}} ({s_u}))}.
    \end{equation}
    Using the LogSumExp approximation as follows:
    \begin{equation}
    \label{eqapp: appro first trem}
    {-\log(\mathop {\min }\limits_{u \in {\mathcal{V}_A}} ({s_u}))}\approx  \frac{1}{\alpha }\log (\sum\limits_{u \in {{\mathcal{V}_A}}} {{e^{ - \alpha {s_u}}}} ).
    \end{equation}
    The gradient of the loss function is:
    \begin{equation}
    \label{eqapp: gradient first term}
    \frac{{\partial {\mathcal{L}^{HT}}}}{{\partial {s_u}}} =  - \frac{{{e^{ - \alpha {s_u}}}}}{{\sum\nolimits_{u \in {{\mathcal{V}_A}}} {{e^{ - \alpha {s_u}}}} }}.
    \end{equation}
    This is a smooth and differentiable gradient.
    \item When ${{\tilde y}_v} = 0$, the loss term is:
    \begin{equation}
    \label{eqapp: second term diff}
    - \;(1 - {{\tilde y}_v})\log (1 - \mathop {\max }\limits_{u \in {\mathcal{V}_R}} ({s_u}))=-\log (1 - \mathop {\max }\limits_{u \in {\mathcal{V}_R}} ({s_u})).
    \end{equation}
    Applying the LogSumExp approximation as follows:
    \begin{equation}
    \label{eqapp: appro second trem}
    -\log (1 - \mathop {\max }\limits_{u \in {\mathcal{V}_R}} ({s_u}))\approx  \frac{1}{\alpha }\log (\sum\limits_{u \in {{\mathcal{V}_R}}} {{e^{\alpha {s_u}}}} ).
    \end{equation}
    The gradient of the loss function is:
    \begin{equation}
    \label{eqapp: gradient second term}
    \frac{{\partial {\mathcal{L}^{HT}}}}{{\partial {s_u}}} =  - \frac{{{e^{\alpha {s_u}}}}}{{\sum\nolimits_{u \in {{\mathcal{V}_R}}} {{e^{\alpha {s_u}}}} }}.
    \end{equation}
    This is a smooth and differentiable gradient.
\end{itemize}
In summary, the min and max operations in \hrefequation{eq: HT Loss} are generally non-differentiable. However, we can replace them with differentiable expressions, ensuring the loss function is differentiable over the entire domain. This smooth approximation stabilizes gradient computation, allowing effective gradient calculation in deep learning through automatic differentiation tools.

In many cases, loss functions contain min or max operations, which are often introduced to incorporate some form of non-linearity or heuristic constraints. Although these operations can lead to points of non-differentiability in the loss function, they remain important in deep learning and optimization, as follows:
\begin{itemize}
    \item \textbf{Introducing Discontinuities:} The max or min operations can cause the slope of the function to change abruptly at certain points, leading to non-differentiable points. In other words, the loss function can change its shape under specific conditions when the predicted value equals a threshold.
    \item \textbf{Enhancing Model Capability:} By maximizing or minimizing certain quantities, loss functions can better capture complex patterns in the data, which can guide the model to learn more effectively.
    \item \textbf{Improving Model Robustness:} The max or min operations make the model more robust to noise and uncertainty in the input data. 
    \item \textbf{Incorporating Prior Knowledge:} The max or min operations can enforce prior knowledge or constraints. 
\end{itemize}
In summary, although max or min operations may introduce non-differentiable points in the loss function, their advantages and effectiveness in the optimization process often outweigh this drawback. In practice, many optimization algorithm (e,g, Adam and RMSProp, etc.) can handle these discontinuities effectively, resulting in good training results.

\subsection{Proof of Triangle Inequality for $\psi ( \cdot , \cdot )$}
\label{app: triangle inequality}
$Proof.$ For any three nodes $u$, $v$, $w$ in a hierarchical tree $\mathcal{T}$, the distance between any two nodes $u$ and $v$ is defined as:
\begin{equation}
\label{eqapp: tree distance}
\psi (u,v) = \psi (u,lca(u,v)) + \psi (v,lca(u,v)),
\end{equation}
where $\psi (u , v )$ represents the path length between nodes $u$, $v$ and $lca(u,v)$ denotes the lowest common ancestor of $u$ and $v$. Now, consider any three nodes $u,v,w \in \mathcal{T}$ and define their pairwise lowest common ancestors: 
\begin{equation}
\left\{ \begin{array}{l}
{L_1} = lca(u,v)\\
{L_2} = lca(v,w)\\
{L_3} = lca(u,w)
\end{array} \right..
\end{equation}
We can express the path from node $u$ to node $w$ as:
\begin{equation}
\psi (u,w) = \psi (u,{L_3}) + \psi ({L_3},w),
\end{equation}
where ${L_3} = lca(u,w)$ is the starting point of the shared path between $u$ and $w$. Based on the properties of lowest common ancestors, the path from $u$ to ${L_3}$ can be decomposed as:
\begin{equation}
\psi(u,{L_1}) + \psi({L_1},{L_2}) + \psi({L_2},w).
\end{equation}
Thus, we have:
\begin{equation}
\label{eqapp: A.24}
\begin{aligned}
\psi (u,w) &= \psi (u,{L_3}) + \psi (w,{L_3})  \\
           &\le (\psi (u,{L_1}) + \psi ({L_1},{L_2}) + \psi ({L_2},w)) + \psi (w,{L_3}).
\end{aligned}  
\end{equation}
\hrefequation{eqapp: A.24} can be further expressed as: 
\begin{equation}
\label{eqapp: A.25}
\begin{aligned}
\psi (u,w) &\le \psi (u,{L_1}) + \psi (v,{L_1}) + \psi (v,{L_2}) + \psi (w,{L_2}) \\
           &= \psi (u,v) + \psi (v,w).
\end{aligned} 
\end{equation}
Therefore, the tree distance $\psi ( \cdot , \cdot )$ satisfies the triangle inequality for any three nodes $u,v,w \in \mathcal{T}$, which proves the validity of this distance measure.

\subsection{Proof of ${m_\sigma }$ Boundary}
\label{app: Boundary}
$Proof.$ To determine the boundary of $m_\sigma  = \frac{{\psi ({{\hat v}_c},\hat v_c^ - ) - \psi ({{\hat v}_c},\hat v_c^ + )}}{2H}$, we need to calculate the maximum distance $\max \psi ({{\hat v}_c},\hat v_c^ + )$ and minimum distance $\min \psi ({{\hat v}_c},\hat v_c^ + )$ between the anchor sample $i$ and the positive sample $i^ +$, as well as the maximum distance $\max \psi ({{\hat v}_c},\hat v_c^ - )$ and minimum distance $\min \psi ({{\hat v}_c},\hat v_c^ - )$ between the anchor sample $i$ and the negative sample $i^ -$, respectively.

\noindent\textbf{Distance between anchor and positive samples.} Given the anchor sample $i$ and the positive sample $i^ +$ are located at their respective leaf nodes ${{\hat v}_c}$ and $\hat v_c^ + $ and that they satisfy the sample selection strategy ${g_c}(i) = {g_s}({i^ + })$ ($cf.$ \hrefdefinition{def: Sample Selection Strategy}), the minimum distance $\min \psi ({{\hat v}_c},\hat v_c^ + )$ between the anchor and positive samples occurs only when both samples are located at the same leaf node. In other words, if the anchor sample $i$ and positive sample $i^ +$ both reside at node ${{\hat v}_c}$, then we have:
\begin{equation}
\label{eqapp: min distance anchor positive}
\min \psi (\hat{v}_c, \hat{v}_c^+) = \psi (\hat{v}_c, \hat{v}_c) = \psi (\hat{v}_c^+, \hat{v}_c^+) = 0.
\end{equation}

\begin{figure}[htbp]
    \centering
    \includegraphics[width=0.45\textwidth,height=25mm]{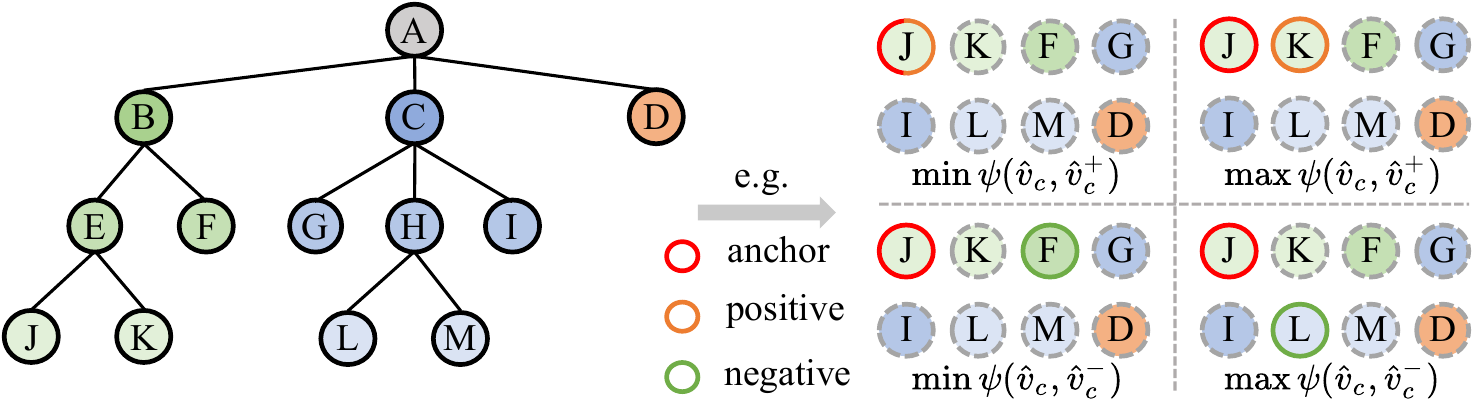}
    \caption{Schematic diagram of the maximum and minimum boundaries for anchor sample nodes, positive sample nodes and negative sample nodes. The left part shows a given hierarchical tree and the right part provides examples of maximum and minimum distances.}
    \label{app: example boundary}
\end{figure}

For the maximum distance $\max \psi ({{\hat v}_c},\hat v_c^ + )$, we consider the scenario where the anchor sample $i$ and the positive sample $i^ +$ share their lowest common ancestor node $u$. The path between ${{\hat v}_c}$ and $\hat v_c^ +$ in the tree can be represented as ${{\hat v}_c} \to u \to \hat v_c^ +$. Consequently, the maximum distance is given by:
\begin{equation}
\label{eqapp: max distance anchor positive}
\max \psi (\hat{v}_c, \hat{v}_c^+) = \psi (\hat{v}_c, u) + \psi (u, \hat{v}_c^+) = 1 + 1 = 2.
\end{equation}
Therefore, the maximum value of $\psi ({{\hat v}_c},\hat v_c^ + )$ is 2 and the minimum value of $\psi ({{\hat v}_c},\hat v_c^ + )$ is 0, i.e. $\max \psi ({{\hat v}_c},\hat v_c^ + )=2$ and $\min \psi ({{\hat v}_c},\hat v_c^ + )=0$. An example is shown in \hreffigure{app: example boundary}.

\noindent\textbf{Distance between anchor and negative samples.} For the minimum distance $\min \psi ({{\hat v}_c},\hat v_c^ - )$, given $u$ is the parent node of node ${{\hat v}_c}$ corresponding to the anchor sample $i$, ${\dot u}$ is the sibling node of $u$ and ${\ddot u}$ is the ancestor nodes of both $u$ and ${\dot u}$. The anchor sample $i$ and negative sample $i^ -$ can only come from their leaf nodes ${{\hat v}_c}$ and $\hat v_c^ -$. Moreover, they satisfy the sampling strategy ${g_c}(i) \ne {g_s}({i^ - })$ ($cf.$ \hrefdefinition{def: Sample Selection Strategy}). The path from ${{\hat v}_c}$ to $\hat v_c^ -$ can be described as ${{\hat v}_c} \to u \to \dot u \to \ddot u$, i.e. ${{\hat v}_c} \to u \to \dot u \to \hat v_c^ -$. Therefore, we have:
\begin{equation}
\label{eqapp: min distance anchor negative}
\min \psi ({{\hat v}_c},\hat v_c^ - ) = \psi ({{\hat v}_c},u) + \psi (u,\dot u) + \psi (\dot u,\hat v_c^ - ) = 1 + 1 + 1 = 3.
\end{equation}
In other words, a complete binary tree is a necessary condition for achieving the minimum distance between the anchor sample node and the negative sample node.

For the maximum distance $\max \psi ({{\hat v}_c},\hat v_c^ -)$, since the anchor sample $i$ and the negative sample $i^ -$ can only come from the corresponding leaf nodes ${{\hat v}_c}$ and $\hat v_c^ -$. In addition, they must also satisfy ${g_c}(i) \ne {g_s}({i^ - })$ ($cf.$ \hrefdefinition{def: Sample Selection Strategy}). The distance $\psi ({{\hat v}_c},\hat v_c^ -)$ is maximized only when the anchor sample node ${{\hat v}_c}$ and the negative sample node $\hat v_c^ -$ are located in different subtrees of depth $H$, as follows:
\begin{equation}
\label{eqapp: max distance anchor negative}
\max \psi ({{\hat v}_c},\hat v_c^ - ) = H + H = 2H.
\end{equation}
Therefore, the maximum value of $\psi ({{\hat v}_c},\hat v_c^ - )$ is $2H$ and the minimum value of $\psi ({{\hat v}_c},\hat v_c^ - )$ is 3, i.e. $\max \psi ({{\hat v}_c},\hat v_c^ - ) = 2H$ and $\min \psi ({{\hat v}_c},\hat v_c^ - ) = 3$. An example is shown in \hreffigure{app: example boundary}.

Based on the above, the minimum value of ${m_\sigma }$ is as follows:
\begin{equation}
\label{eqapp: min m_sigma}
\begin{aligned}
\min {m_\sigma } &= \min \frac{{\psi ({{\hat v}_c},\hat v_c^ - ) - \psi ({{\hat v}_c},\hat v_c^ + )}}{{2H}}\\
                 &= \frac{{\min \psi ({{\hat v}_c},\hat v_c^ - ) - \max \psi ({{\hat v}_c},\hat v_c^ + )}}{{2H}}\\
                 &= \frac{1}{{2H}}.
\end{aligned}
\end{equation}
Similarly, the maximum value of ${m_\sigma }$ can be expressed as follows:
\begin{equation}
\label{eqapp: max m_sigma}
\begin{aligned}
\max {m_\sigma } &= \max \frac{{\psi ({{\hat v}_c},\hat v_c^ - ) - \psi ({{\hat v}_c},\hat v_c^ + )}}{{2H}}\\
                 &= \frac{{\max \psi ({{\hat v}_c},\hat v_c^ - ) - \min \psi ({{\hat v}_c},\hat v_c^ + )}}{{2H}}\\
                 &= \frac{{2H}}{{2H}} = 1.
\end{aligned}
\end{equation}
When $H \to \infty$, we have $\lim (\min {m_\sigma }) = \lim \frac{1}{{2H}} \to 0$ and $\lim (\max {m_\sigma }) = 1$. Therefore, we have ${m_\sigma } \in (0,1]$.

\section{Method}
\label{app: Method}
\subsection{Framework of Model}
\label{app: Framework of Model}
\begin{figure*}[htbp]
    \centering
    \vspace{10pt}
    \includegraphics[width=\textwidth,height=30mm]{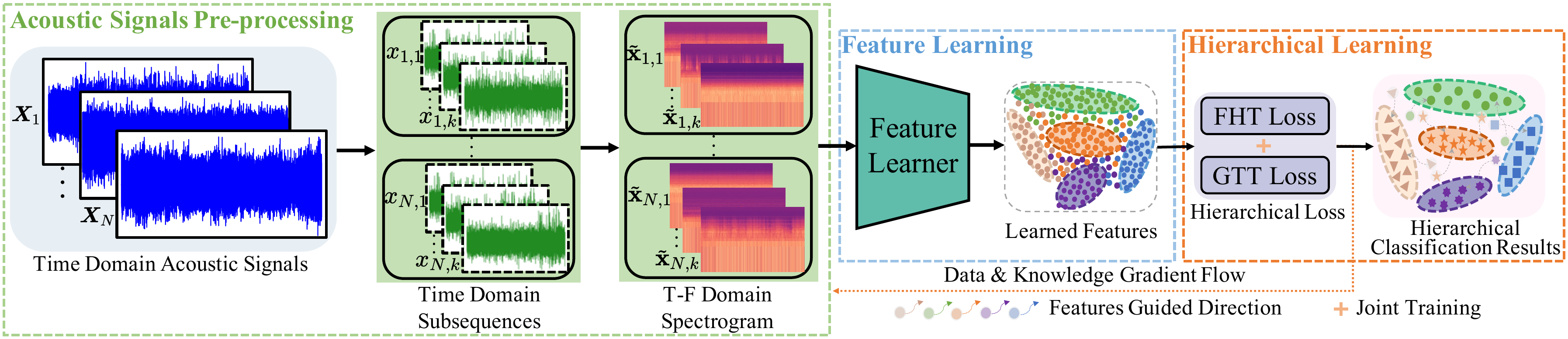}
    \caption{Overall framework of the DHK. The T-F domain spectrograms are fed into feature representation learning module to extract deep learned features. Meanwhile, the features are jointly trained and optimized by FHT and GTT loss functions for hierarchical fault intensity recognition.}
    \label{app: framework}
\end{figure*}
The overall architecture of our model is depicted in \hreffigure{app: framework}. Given a signal dataset ${\mathcal{X}} = \{ {X_i},i = 1,2, \ldots ,N\}  \subseteq {\mathbb{R}^{M \times N}}$ and the corresponding label ${\mathcal{Y}} \subseteq {\mathbb{R}^{C \times N}}$ with $N$ streams, $M$ measurements for each stream and $C$ fault classes. For feature representation learning, the signal dataset $\mathcal{X}$ is input into the acoustic signals pre-processing module and ${\mathcal{\tilde X}} \subseteq {\mathbb{R}^{T \times F \times 3}}$ is the output after the sliding window (SW) and STFT operations, as follows:
\begin{equation}
\label{eqapp: pre-processing}
\begin{array}{l}
{\mathcal{\tilde X}} = 10 \times {\log _{10}}(\frac{{\left| {\mathrm{STFT}(\mathrm{SW}(\mathcal{X}))} \right|}}{{\max (\left| {\mathrm{STFT}(\mathrm{SW}(\mathcal{X}))} \right|)}})\\
\;\;\;: = 10 \times {\log _{10}}(\frac{{\left| {{\mathcal{X}_{sw}}[n,m]} \right|}}{{\max (\left| {{\mathcal{X}_{sw}}[n,m]} \right|)}})
\end{array},
\end{equation}
where ${{\mathcal{X}}_{sw}}[n,m]$ indicates the $n$-th row and $m$-th column elements of the STFT result matrix and $\left| {{\mathcal{X}_{sw}}[n,m]} \right|$ denotes the elements of the amplitude spectrum matrix. Then, ${\mathcal{\tilde X}}$ is fed into the feature learning module (FL) to produce learned features ${\boldsymbol{F}} \in {{\mathbb{R}}^D}$ with $D$ denoting the dimension of the T-F domain spectrogram. Finally, the hierarchical classification score $\bm{s} \in \mathcal{V} \in {[0,1]^{|\mathcal{V}|}}$ cen be computed by $\bm{s}=\mathrm{FL}({\mathcal{\tilde X}})$. The whole process is trained and optimized through a training objective, i.e. $\mathcal{L} = \frac{{{h_i}}}{{\sum\nolimits_i {{h_i}} }} \times {\mathcal{L}^{FHT}} + \alpha {\mathcal{L}^{GTT}}$, consisting of the focal hierarchical tree loss ${\mathcal{L}^{FHT}}$ ($cf.$ \hrefequation{eq: FHT Loss}) and the group tree triplet loss ${\mathcal{L}^{GTT}}$ ($cf.$ \hrefequation{eq: GTT Loss}).


\section{Experiments}
\label{app: Experiments}
\subsection{Evaluation Metrics}
\label{app: Evaluation Metrics}
As mentioned in the evaluation metrics, we apply dynamic thresholds to assess the performance of fault intensity diagnosis for the proposed HKG model. For any given threshold, we can determine the counts of True Positives (TP), False Positives (FP), True Negatives (TN), and False Negatives (FN). In addition, the following metrics are calculated:
\begin{equation}
\begin{aligned}
\mathrm{Accuracy} &= \frac{\mathrm{TP} + \mathrm{TN}}{\mathrm{TP} + \mathrm{TN} + \mathrm{FP} + \mathrm{FN}}, \\
\mathrm{Precision} &= \frac{\mathrm{TP}}{\mathrm{TP} + \mathrm{FP}}, \\
\mathrm{Recall} &= \frac{\mathrm{TP}}{\mathrm{TP} + \mathrm{FN}}, \\
\mathrm{F1\text{-}score} &= \frac{2 \times \mathrm{Precision} \times \mathrm{Recall}}{\mathrm{Precision} + \mathrm{Recall}}.
\end{aligned}
\end{equation}
By evaluating across all possible thresholds, we can generate a precision-recall curve, with precision plotted on the $y$-axis and recall on the $x$-axis. The Average Precision (AP) is calculated as $AP = \sum\nolimits_n \left( \frac{{R_n - R_{n-1}}}{{P_n}} \right)$, where $P_n$ and $R_n$ represent the precision and recall at the $n$-th threshold, respectively.

\subsection{Datasets}
\label{app: Datasets}
\subsubsection{Cavitation Datasets} 
\begin{figure}[htbp]
\centering
\includegraphics[width=0.5\textwidth,height=30mm]{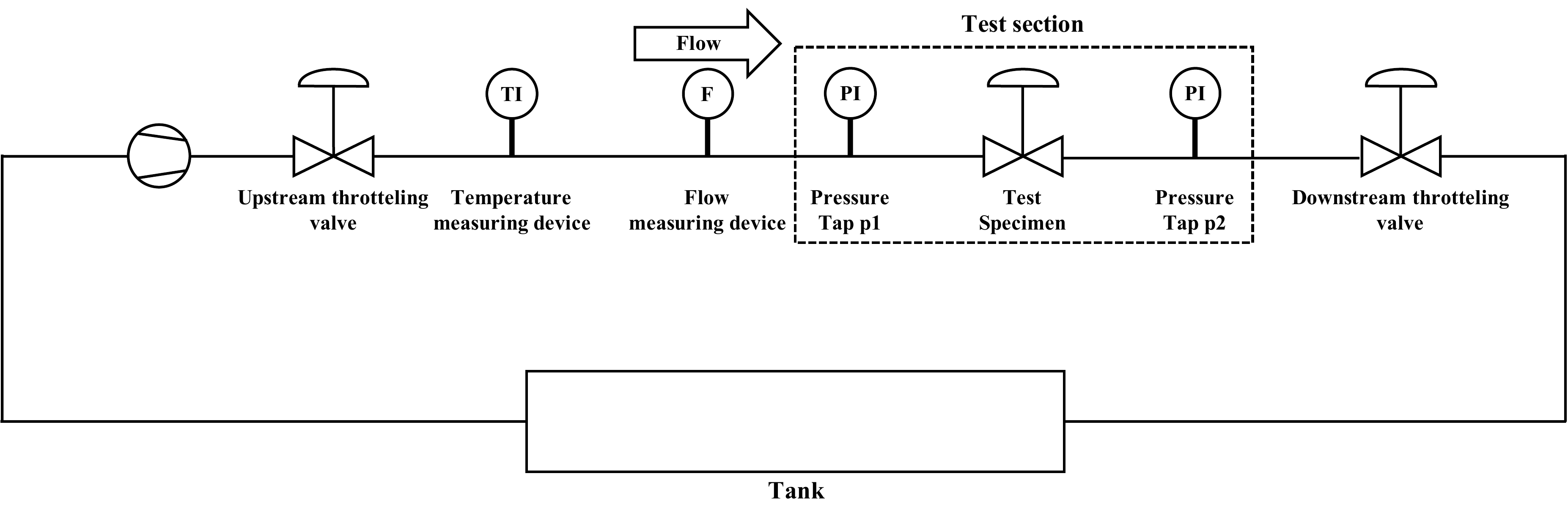}
\caption{Schematic view of the test rack at SAMSON AG.}
\label{fig: system}
\end{figure}

The cavitation datasets are provided by SAMSON AG in Frankfurt. The schematic of the experimental setup is illustrated in \hreffigure{fig: system}. The acoustic signals are recorded under five distinct flow conditions, generated by adjusting the differential pressure at various constant upstream pressures of the control valve. These conditions include choked flow cavitation, constant cavitation, incipient cavitation, turbulent flow, and no flow (see \hreftable{tab: CavitationDatasets-FlowStatus} and Table \ref{tab: CavitationDatasets-operation}). The detailed dataset statistics and label distributions of three real-world cavitation datasets without data augmentation are provided in \hreftable{tab: training and test sets}. In addition, the hierarchical tree of cavitation datasets is shown in \hreffigure{fig: HierarchicalCavitationTree}. 
\begin{table}[htbp]
\centering
\footnotesize
\caption{The content details of the three real-world cavitation datasets for each flow state.}
\label{tab: CavitationDatasets-FlowStatus}
\setlength{\tabcolsep}{1.8mm}{
\begin{tabular}{lccccc}
\toprule 
\multirow{2}{*}{Dataset} & \multicolumn{3}{c}{Cavitation} & \multicolumn{2}{c}{Non-cavitation}  \\ 
\cmidrule{2-6} 
& \multicolumn{1}{l}{choked flow} & \multicolumn{1}{l}{constant} & \multicolumn{1}{l}{incipient} & \multicolumn{1}{l}{turbulent} & \multicolumn{1}{l}{no flow} \\ 
\midrule
Cavitation-Short           & 72    & 93   & 40   & 118   & 33  \\ 
Cavitation-Long            & 148   & 396  & 64   & 183   & 15  \\ 
Cavitation-Noise           & 40    & 40   & 40   & 40    & 0  \\ 
\bottomrule
\end{tabular}}
\end{table}

\begin{table}[htbp]
\centering
\footnotesize
\caption{Details of three real-world cavitation datasets for valve operation with various upstream pressures.}
\label{tab: CavitationDatasets-operation}
\setlength{\tabcolsep}{0.8mm}{
\begin{tabular}{lccc}
\toprule 
\multicolumn{1}{c}{\multirow{3}{*}{Dataset}} & \multicolumn{3}{c}{Operation parameters} \\ 
\cmidrule{2-4} 
\multicolumn{1}{c}{}    & \multicolumn{1}{c}{\begin{tabular}[c]{@{}c@{}}Valve stroke\\ (\SI{}{\mm})\end{tabular}} 
                        & \multicolumn{1}{c}{\begin{tabular}[c]{@{}c@{}}Upstream pressure\\ (\SI{}{\bar})\end{tabular}} 
                        & \multicolumn{1}{c}{\begin{tabular}[c]{@{}c@{}}Temperature\\ (\SI{}{\degreeCelsius})\end{tabular}} \\ 
\midrule
Cavitation-Short        & {[}15,13.5,11.25,7.5,3.75,1.5,0.75{]}      & {[}10,9,6,4{]}        & 25-50       \\
Cavitation-Long         & {[}60,55,45,30,25,15,6{]}                  & {[}10,6,4{]}          & 23-52        \\
Cavitation-Noise        & 15                                         & 10                    & 32-39         \\ 
\bottomrule
\end{tabular}}
\end{table}
\begin{table}[htbp]
\caption{Details of the training and test sets. $(\cdot )$ denotes the number after the sliding window (window size is 466944).}
\label{tab: training and test sets}
\scriptsize
\centering
\setlength{\tabcolsep}{0.1mm}{
\begin{tabular}{lcccccccc}
\toprule
                 \multicolumn{1}{c}{\multirow{3}{*}{Dataset}} & \multicolumn{4}{c}{Training set}      & \multicolumn{4}{c}{Testing set} \\ 
\cmidrule{2-9} 
                 & \multicolumn{3}{c}{Cavitation}       
                 & \multicolumn{1}{c|}{\multirow{2}{*}{Non}}
                 & \multicolumn{3}{c}{Cavitation}     
                 & \multirow{2}{*}{Non} \\ 
\cmidrule{2-4}
\cmidrule{6-8}
                 & Choked flow & Constant   & Incipient & \multicolumn{1}{c|}{}     & Choked flow & Constant & Incipient &    \\ 
\midrule
Cavitation-Short & 58($\times$10)    & 75($\times$10)   & 32($\times$10)  & \multicolumn{1}{c|}{121($\times$10)}  & 14($\times$10)    & 18($\times$10) & 8($\times$10)    & 30($\times$10)    \\
Cavitation-Long  & 118($\times$83)   & 317($\times$83) & 52($\times$83)  & \multicolumn{1}{c|}{158($\times$83)}  & 30($\times$83)    & 79($\times$83) & 12($\times$83)  & 40($\times$83)    \\
Cavitation-Noise & 32($\times$83)    & 32($\times$83)   & 32($\times$83)  & \multicolumn{1}{c|}{32($\times$83)}    & 8($\times$83)      & 8($\times$83)   & 8($\times$83)    & 8($\times$83)      \\ 
\bottomrule
\end{tabular}
}
\end{table}

\subsubsection{PUB Dataset} 
\begin{figure}[htbp]
    \centering
    \begin{minipage}{0.5\textwidth}
        \centering
        \includegraphics[width=\textwidth,height=40mm]{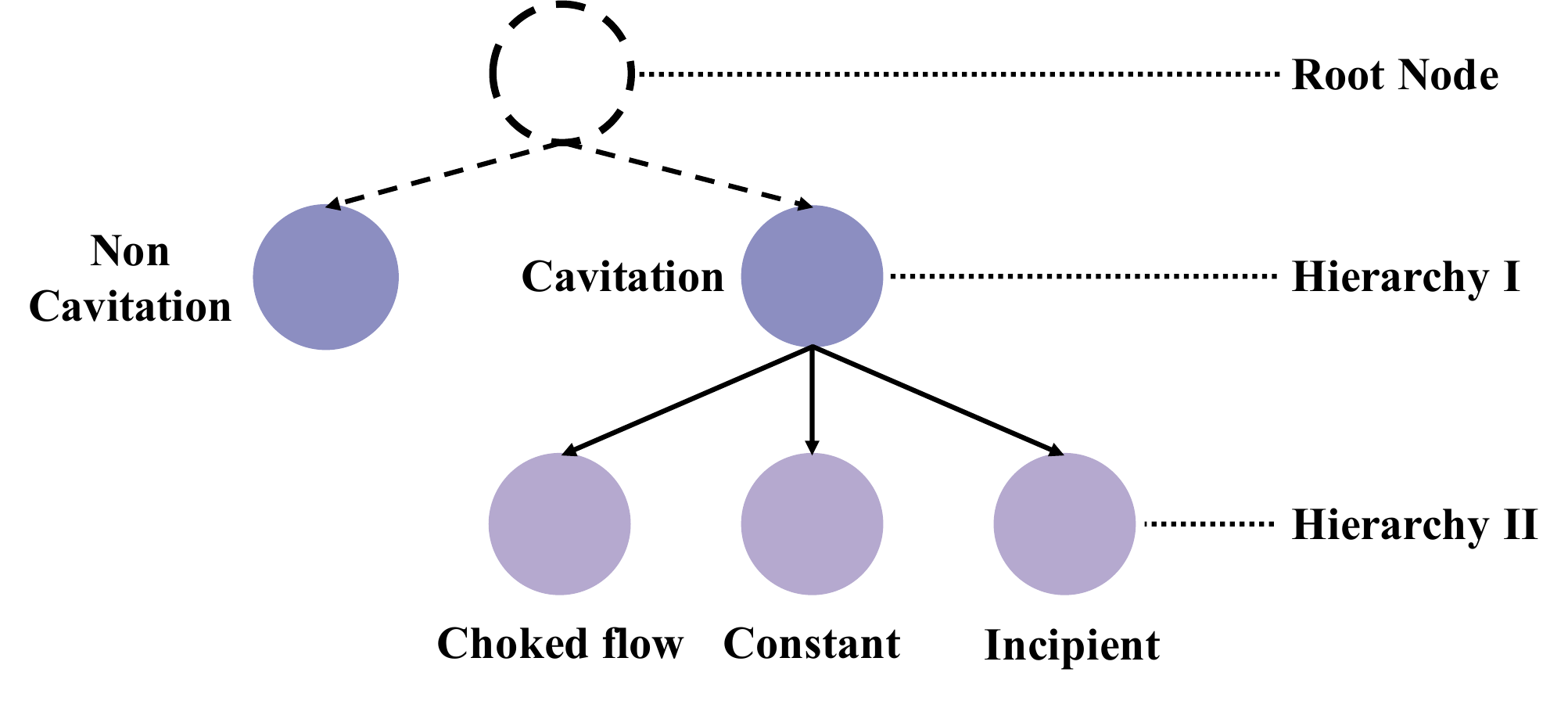}
        \caption{A hierarchical cavitation label tree with classes are drawn from different cavitation datasets.}
        \label{fig: HierarchicalCavitationTree}
    \end{minipage}%
    \hfill 
    \begin{minipage}{0.5\textwidth}
        \centering
        \includegraphics[width=\textwidth,height=45mm]{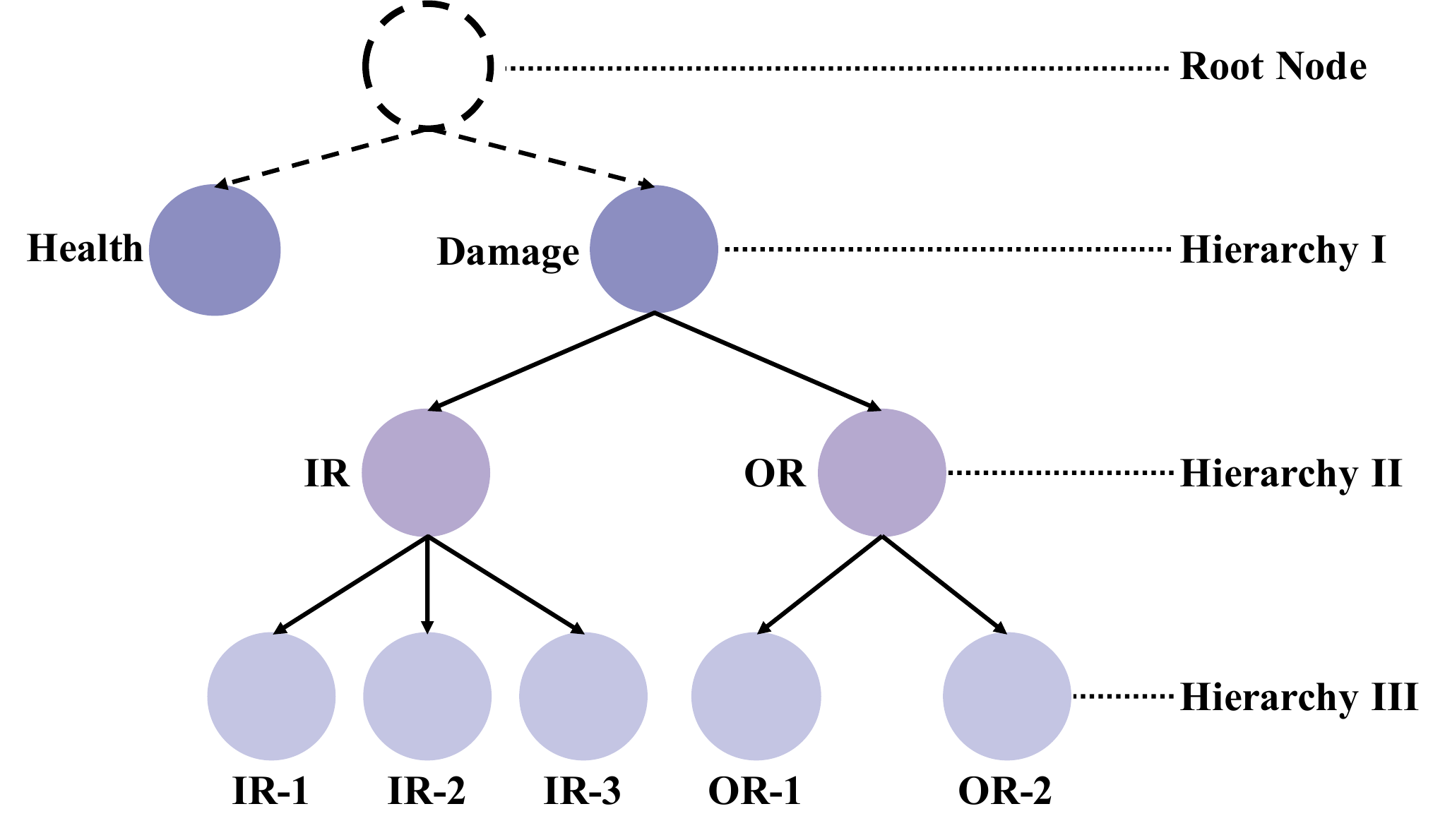}
        \caption{A hierarchical bearing fault tree from the PUB.}
        \label{fig: PUBD Hierarchical Tree}
    \end{minipage}
\vspace{10pt}
\end{figure}

This dataset is used to evaluate the scalability of our method. The levels of bearing damage are detailed in \hreftable{tab: damage levels PUB}. The file codes and corresponding fault types used in our experiment are provided in \hreftable{tab: fault type and file code}. The PUB is organized into three hierarchies: bearing diagnosis (Hierarchy I), bearing damage type diagnosis (Hierarchy II) and IR/OR intensity diagnosis (Hierarchy III-IR/III-OR), as depicted in \hreffigure{fig: PUBD Hierarchical Tree}.
\begin{table}[htbp]
    \centering
        \centering
        \caption{The bearing fault damage levels in the PUB.}
        \label{tab: damage levels PUB}
        \footnotesize
        \setlength{\tabcolsep}{4mm}{
        \begin{tabular}{ccc}
        \toprule
        Damage level & Percentage values  & Bearing limitations   \\ 
        \midrule
        1            & 0-2$\%$              & $\le$\SI{2}{mm}      \\
        2            & 2-5$\%$              & $>$\SI{2}{mm}    \\
        3            & 5-15$\%$             & $>$\SI{4.5}{mm}\\
        \bottomrule
        \end{tabular}}
\end{table}

\begin{table}[htbp]
    \centering
        \centering
        \caption{The bearing fault types and file codes in the PUB.}
        \label{tab: fault type and file code}
         \footnotesize
        \setlength{\tabcolsep}{3mm}{
        \begin{tabular}{ccccccc}
        \toprule
        \multirow{2}{*}{Fault type} & \multirow{2}{*}{Healthy} & \multicolumn{2}{c}{OR damage} & \multicolumn{3}{c}{IR damage} \\
        \cmidrule{3-7} 
        \multicolumn{1}{c}{}                            &                          & OR-1          & OR-2          & IR-1     & IR-2     & IR-3    \\
        \midrule
        \multicolumn{1}{l}{\multirow{7}{*}{File code}} & K001                     & KA01          & KA03          & KI01     & KI07     & KI16    \\
        \multicolumn{1}{l}{}                           & K002                     & KA05          & KA06          & KI03     & KI08     & -       \\
        \multicolumn{1}{l}{}                           & K003                     & KA04          & KA08          & KI04     & KI18     & -       \\
        \multicolumn{1}{l}{}                           & K004                     & KA07          & KA09          & KI05     & -        & -       \\
        \multicolumn{1}{l}{}                           & K005                     & KA15          & KA16          & KI14     & -        & -       \\
        \multicolumn{1}{l}{}                           & K006                     & KA22          & -             & KI17     & -        & -       \\
        \multicolumn{1}{l}{}                           & -                        & KA30          & -             & KI21     & -        & -       \\
        \bottomrule
        \end{tabular}}
        \end{table}

\subsection{Results}
\label{app: Results}

\subsubsection{Statistical p-value} 
\label{subsubsection: Statistical p-value}
\begin{figure*}[htbp]
\centering
\includegraphics[width=\textwidth,height=70mm]{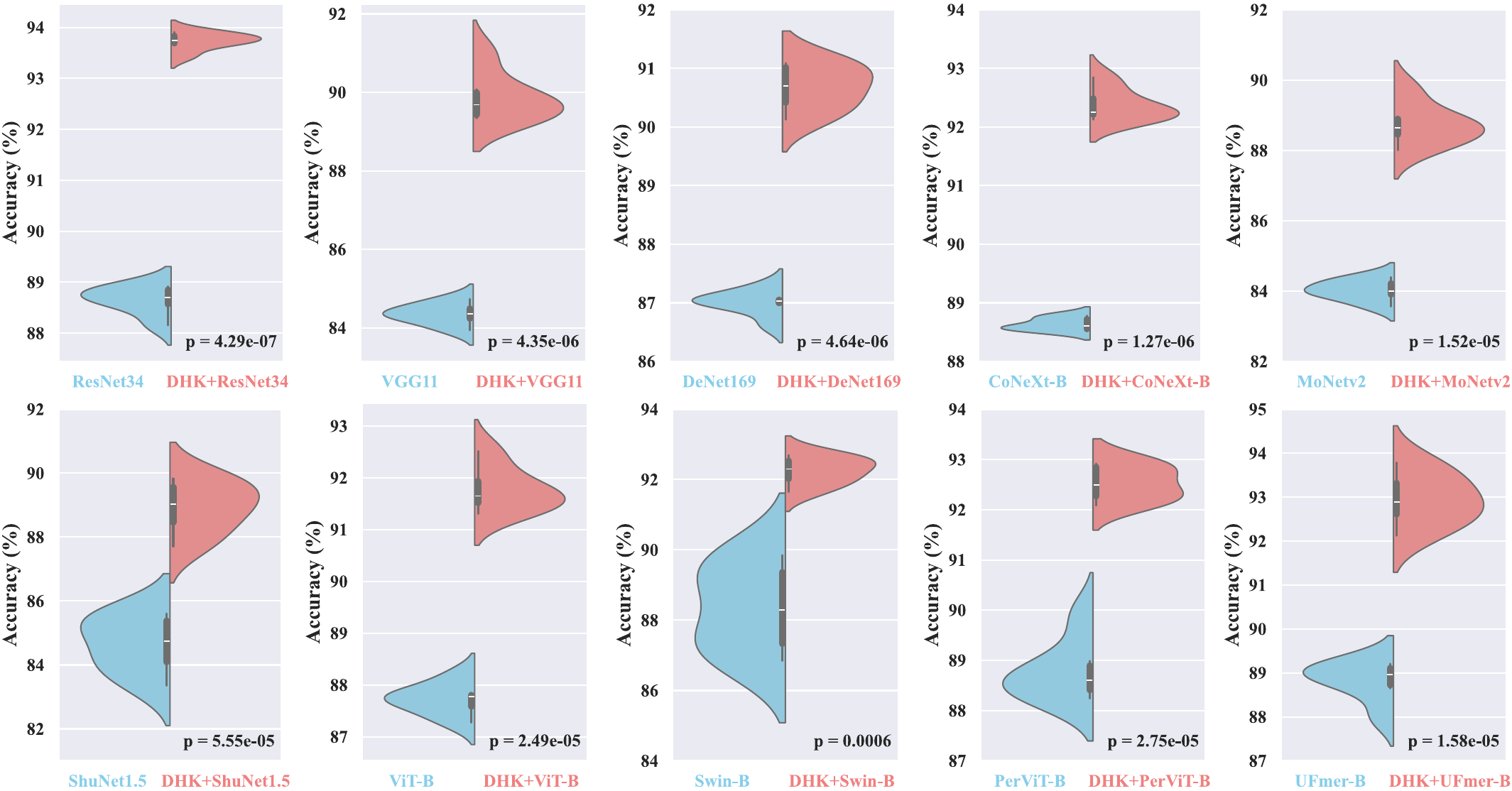}
\caption{Statistical p-value comparison of different methods under six runs. The baselines include convolutional neural network (ResNet, VGG, DenseNet and ConvNeXt), lightweight neural network (MobileNet-v2 and ShuffleNet-v2) and Transformer (ViT, Swin-Transformer, PerViT and UniFormer).}
\label{app: PValue}
\end{figure*}

\hreffigure{app: PValue} shows statistical p-value of different baselines over six runs by t-test on Cavitation-Short. It can be clearly found that statistical p-value of the performance metrics calculated through DHK with various backbones and corresponding baselines are less than 0.001. It indicates that the improvement effect of DHK with different backbones is statistically significant. The results confirm the robustness and generalizability of DHK in enhancing different backbone networks, further validating its effectiveness in improving model performance in a reliable and repeatable manner.


\subsubsection{Confusion Matrix}  
\label{subsubsection: confusion matrix}
\begin{figure*}[htbp]
\subfigure[\tiny UniFormer-B (CS)]{
\begin{minipage}[t]{0.25\linewidth}
\centering
\includegraphics[width=\textwidth,height=35mm]{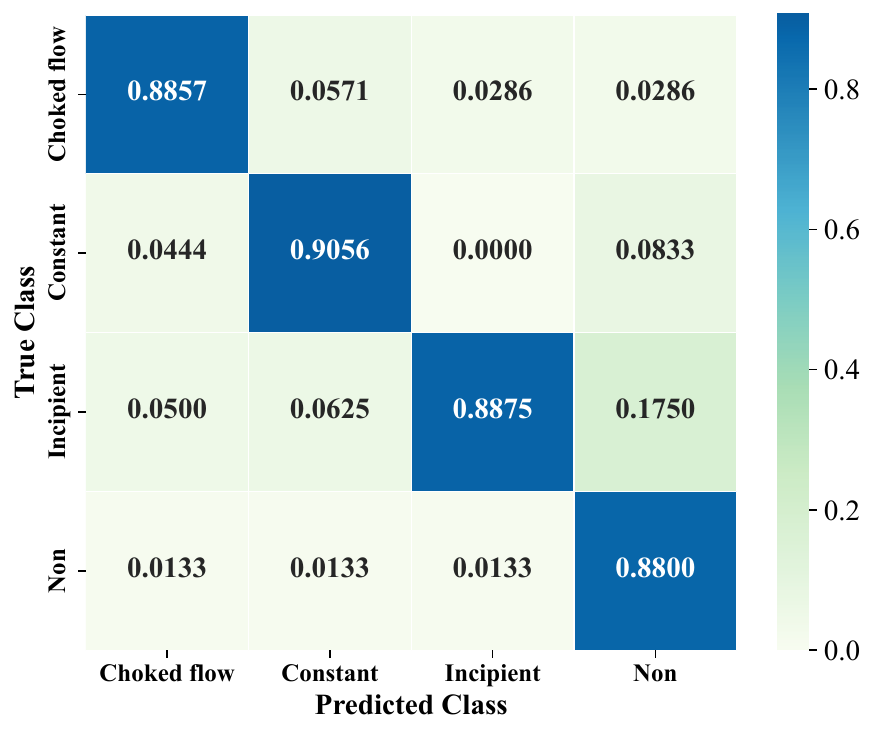}
\end{minipage}%
}%
\subfigure[\tiny UniFormer-B (CL)]{
\begin{minipage}[t]{0.25\linewidth}
\centering
\includegraphics[width=\textwidth,height=35mm]{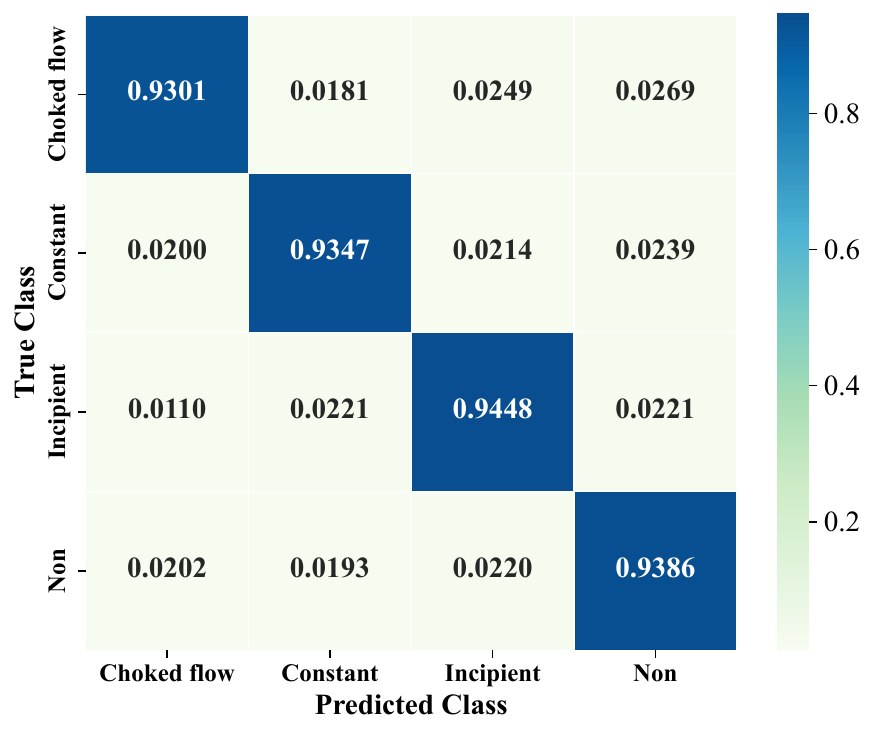}
\end{minipage}%
}%
\subfigure[\tiny UniFormer-B (CN)]{
\begin{minipage}[t]{0.25\linewidth}
\centering
\includegraphics[width=\textwidth,height=35mm]{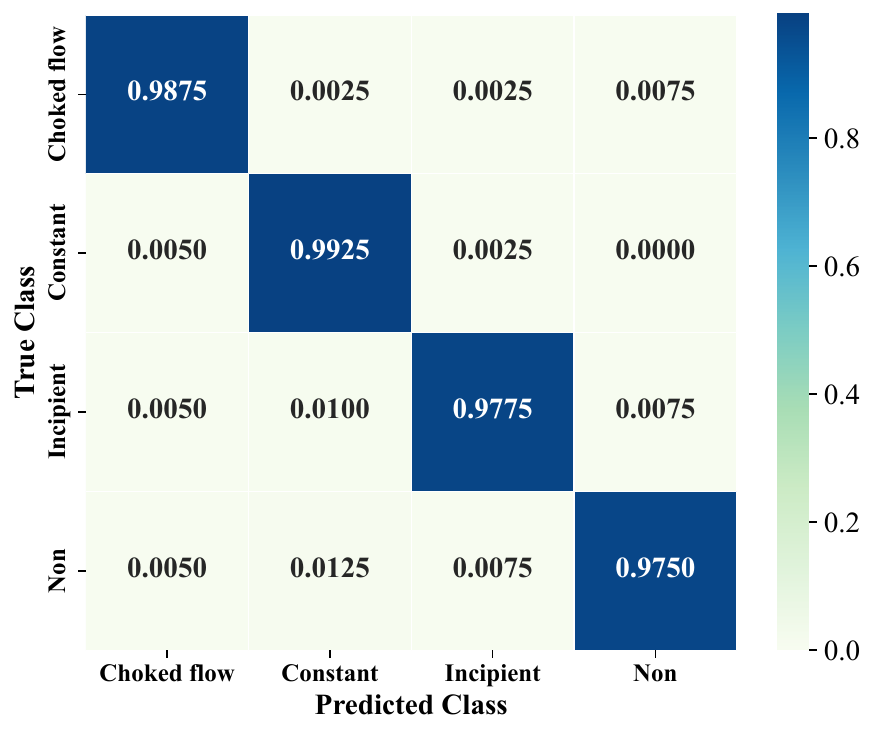}
\end{minipage}%
}%
\subfigure[\tiny HKG-ViT-S (PUB)]{
\begin{minipage}[t]{0.25\linewidth}
\centering
\includegraphics[width=\textwidth,height=35mm]{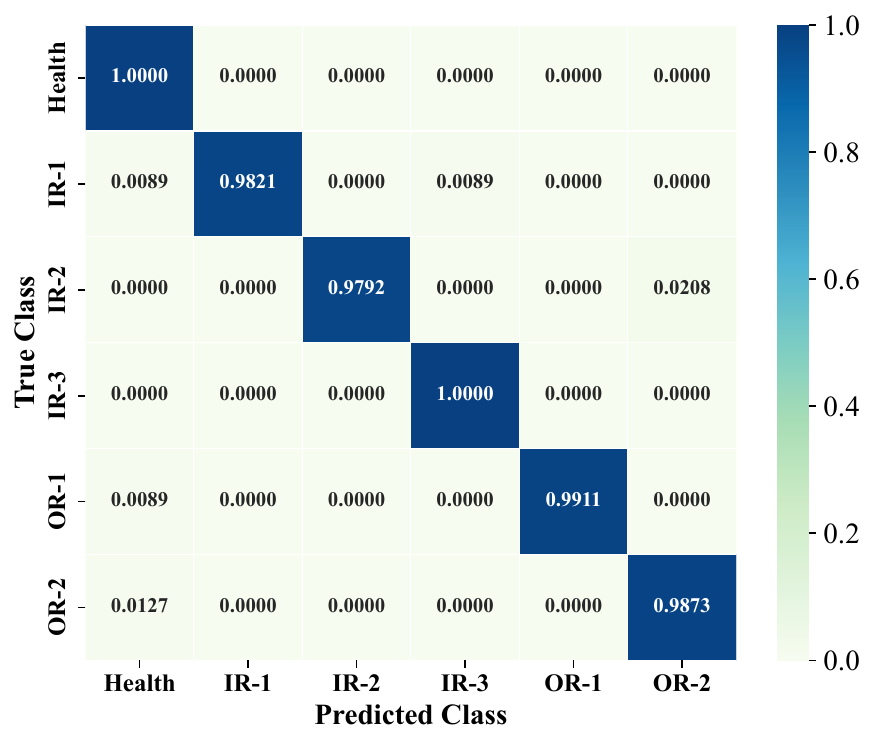}
\end{minipage}%
}%

\subfigure[\tiny DHK+UniFormer-B (CS)]{
\begin{minipage}[t]{0.25\linewidth}
\centering
\includegraphics[width=\textwidth,height=35mm]{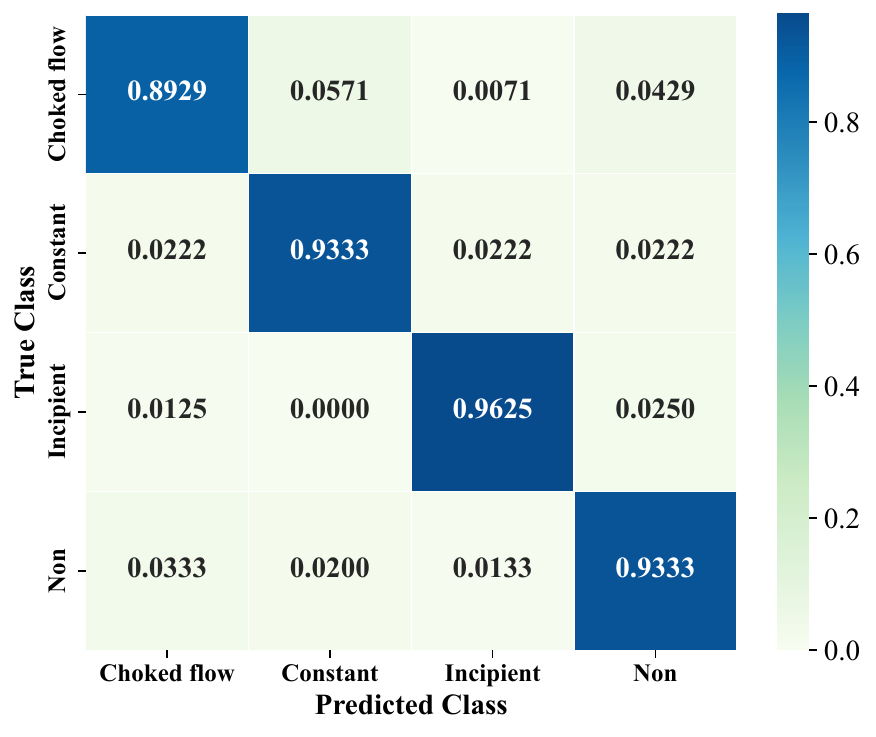}
\end{minipage}%
}%
\subfigure[\tiny DHK+UniFormer-B (CL)]{
\begin{minipage}[t]{0.25\linewidth}
\centering
\includegraphics[width=\textwidth,height=35mm]{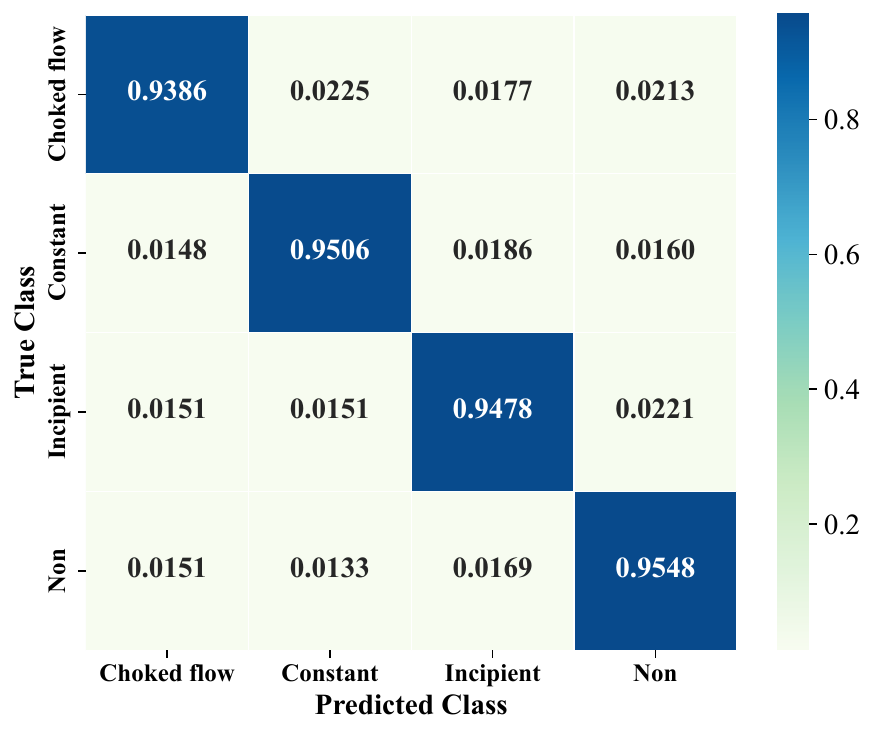}
\end{minipage}%
}%
\subfigure[\tiny DHK+UniFormer-B (CN)]{
\begin{minipage}[t]{0.25\linewidth}
\centering
\includegraphics[width=\textwidth,height=35mm]{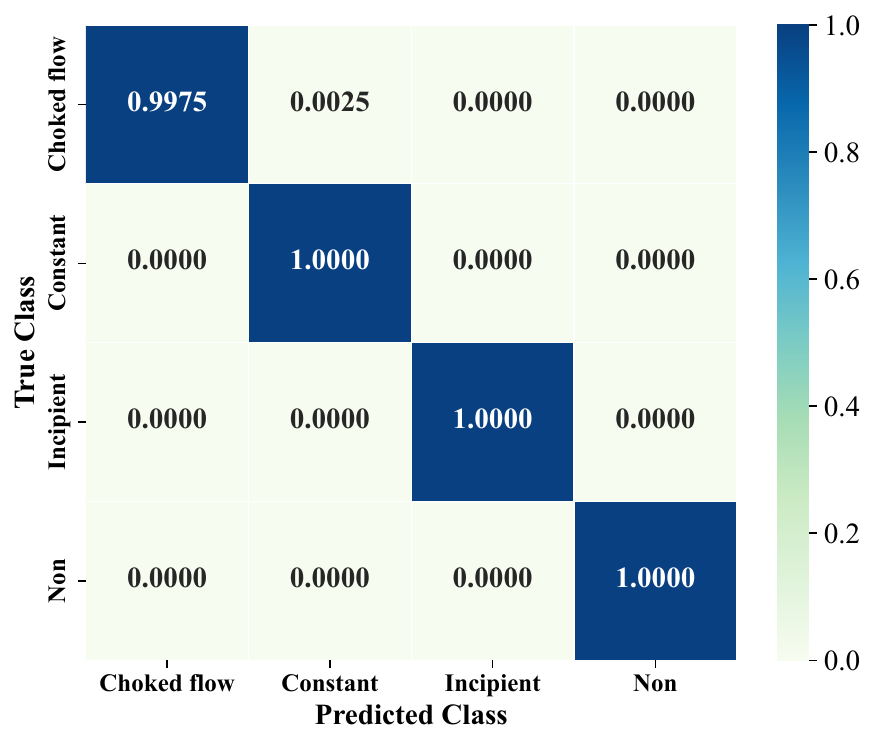}
\end{minipage}%
}%
\subfigure[\tiny DHK+PerViT (PUB)]{
\begin{minipage}[t]{0.25\linewidth}
\centering
\includegraphics[width=\textwidth,height=35mm]{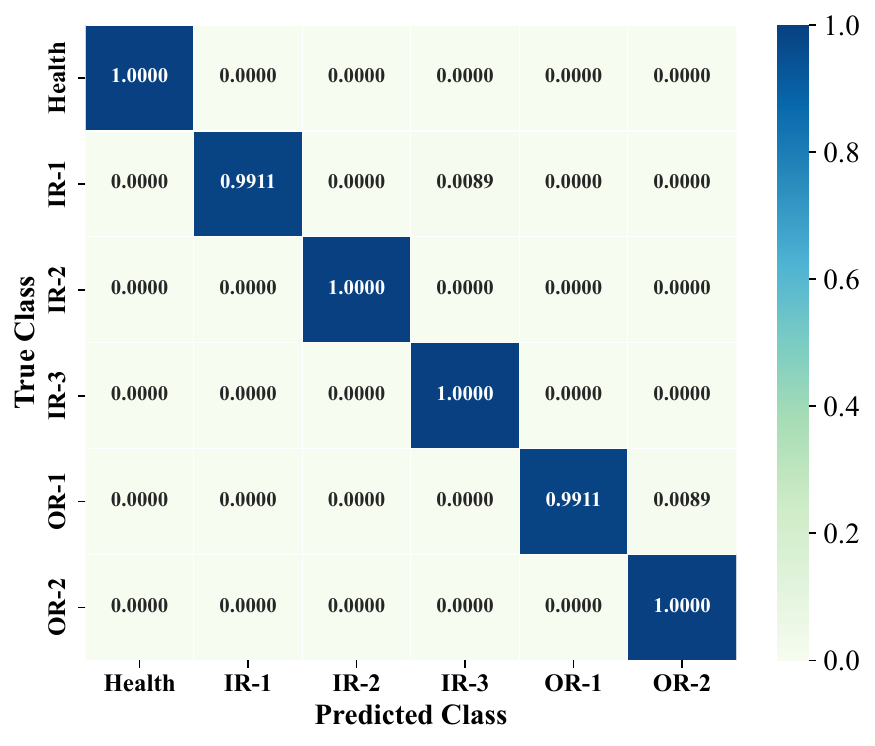}
\end{minipage}%
}%
\centering
\caption{The confusion matrix of DHK with different backbone networks on various datasets. (a)-(c) and (e)-(g) denote the confusion matrix on Cavitation-Short (CS), Cavitation-Long (CL) and Cavitation-Noise (CN), respectively. (d) and (h) represent the confusion matrix on the PUB dataset.} 
\label{app: Appendix Confusion Matrix}
\end{figure*}

We show the confusion matrix of the best performance backbone and our method on four real-world datasets, as shown in \hreffigure{app: Appendix Confusion Matrix}. 

\textbf{Cavitation Datasets:} It can be seen that our method can significantly improve the performance of each cavitation state, especially the incipient cavitation state. Specifically, DHK+Uniformer-B outperforms UniFormer-B by \textbf{7.5}$\%$, \textbf{0.3}$\%$ and \textbf{2.25}$\%$ for the incipient cavitation state on three different cavitation datasets, respectively. 

\textbf{PUB Dataset:} DHK+PerViT-B achieves \textbf{100}$\%$ accuracy in health, IR-2, IR-3 and OR-2. Moreover, DHK+PerViT-B reaches \textbf{99.11}$\%$ accuracy on both IR-1 and OR-1.

\subsubsection{Computational Complexity for DHK} 
\label{subsubsection: Computational Complexity}
Compared to ${\mathcal{L}^{CCE}}$ and ${\mathcal{L}^{Focal}}$, the proposed DHK introduces hierarchical constraints and a group triplet loss, resulting in minimal additional computational complexity. Notably, unlike methods dependent on hierarchical feature extraction modules or specially designed architectures, the proposed DHK significantly reduces implementation and computational costs without requiring additional hierarchical feature extraction modules or knowledge-embedded structures. Therefore, we analyse the computational complexity and efficiency of DHK from theoretical and experimental angles:
\begin{itemize}
    \item ${\mathcal{L}^{CCE}}$/${\mathcal{L}^{Focal}}$: The computational complexity of ${\mathcal{L}^{CCE}}$ is $\mathcal{O}(NC)$ ($N$: number of samples, $C$: number of classes). While ${\mathcal{L}^{Focal}}$ only adds a weighted factor, which cannot increase complexity, i.e. the computational complexity is also $\mathcal{O}(NC)$.
    \item DHK Loss: The DHK loss composed of ${\mathcal{L}^{HT}}$/${\mathcal{L}^{FHT}}$ and ${\mathcal{L}^{GTT}}$. ${\mathcal{L}^{HT}}$ computes hierarchical constraints with computational complexity $\mathcal{O}(NH)$ ($H$: depth of tree). ${\mathcal{L}^{FHT}}$ adds a weighted factor and cannot increase complexity, i.e. the computational complexity is also $\mathcal{O}(NH)$. ${\mathcal{L}^{GTT}}$ computes inter-class distances with computational complexity $\mathcal{O}(T)$ ($T$: number of valid triplets). Based on the above, the overall computational complexity of DHK loss is $\mathcal{O}(NH+T)$.
    \item Experimental Evaluation: We conducted an experimental comparison of the training time for ResNet18 with ${\mathcal{L}^{CCE}}$ and ResNet18 with DHK under the same conditions on Cavitation-Short, as shown in \hreftable{apptab: training time}. It can be seen that DHK loss only introduces a minimal training time cost compared to CCE loss.
\end{itemize}
In summary, although DHK loss has a slightly higher computational complexity compared CCE and Focal loss, we believe that the performance improvement brought by hierarchical optimization strategy far outweigh the minor additional computational cost. 

\subsubsection{Inference Time for \hrefequation{eq: BCE Inference}} 
\label{subsubsection: Inference Time}
During inference, \hrefequation{eq: BCE Inference} calculates each event score from the root to the leaf path. In fact, the essence of \hrefequation{eq: BCE Inference} is a greedy algorithm. Therefore, we analyse \hrefequation{eq: BCE Inference} from both theoretical and experimental aspects, as follows:
\begin{itemize}
    \item Computational Complexity Analysis: The computational cost of \hrefequation{eq: BCE Inference} mainly depends on the depth $H$ of hierarchical label tree $\mathcal{T}$ and the branching factor at each node. Since the inference process only requires computing the accumulated score along a single optimal path, the time complexity of this method is approximately $\mathcal{O}(H)$.
    \item Comparison with Traditional Classification Methods: In typical flat classification, the model typically computes probabilities for all possible classes and uses softmax normalization, resulting in time complexity of $\mathcal{O}(C)$ ($C$: number of classes). In contrast, \hrefequation{eq: BCE Inference} reduces the search space by path constraints and the computational cost is reduced when the $H \ll C$. In general, the depth of $\mathcal{T}$ for fault signal or natural image datasets (e.g. ImageNet, CIFAR, ANIMAL-10N and Pathology) is less than the number of subnodes.
    \item Experimental Evaluation: We calculate the sample inference time on Cavitation-Short, as shown in \hreftable{apptab: inference time}. The server is equipped with 24GB RAM, a 12-core CPU and an NVIDIA RTX 2070. The test dataset is 600 and batch size is 8. From \hreftable{apptab: inference time}, It can clearly be seen that there is almost no difference in the inference time between DHK and CCE.
\end{itemize}


\section{Discussion}
\label{app: Discussion}
In this section, we reflect on the key assumptions underlying our method (DHK), discuss its limitations, extensibility and consider the broader impact of our work. In addition, we also outline potential directions for future improvements to enhance the effectiveness and applicability of our approach.

\subsection{Assumption}
\label{subsection: Assumption}
In this paper, we assume that the labels of the dataset (three real-world cavitation dataset provided by SAMSON AG and one publicly available bearing dataset) are \textbf{clean and accurate}, i.e. all samples are are annotated with \textbf{true and noise-free labels}. Based on this assumption, the hierarchical label tree (\hreffigure{fig: HierarchicalCavitationTree} and \hreffigure{fig: PUBD Hierarchical Tree}) constructed based on dataset labels can truly reflect the \textbf{semantic relationship} and \textbf{hierarchical structure information} among target classes. Therefore, this hierarchical structure information is regarded as reliable prior knowledge throughout the model design and training process.

\subsection{Limitations}
\label{subsection Limitations}
\noindent\textbf{Manually Build Hierarchical Label Tree.} In this study, the hierarchical label trees (\hreffigure{fig: HierarchicalCavitationTree} and \hreffigure{fig: PUBD Hierarchical Tree}) are manually constructed based on semantic relationships among target classes and domain-specific prior knowledge. In some cases, manually building a hierarchical label tree can more accurately capture the hierarchical information between target classes, particularly when dealing with a moderate number classes and clearly defined hierarchical structures. In addition, even for the same dataset, the organization of hierarchical label tree may be different when facing different task objectives. At this point, the manual construction method has obvious advantages in terms of flexibility and task adaptability. However, the manual approach also suffers from a certain degree of subjectivity, high construction costs and limited scalability when applied to large-scale class systems or cross-domain tasks.

\noindent\textbf{Label Noise.} In practical applications, label noise is inevitable and it mainly originates from annotation errors, data ambiguity or sensor inaccuracies. These noises may cause the model to incorrectly learn class relationships, which affects its generalization ability and performance. Although this study assumes that the labels in the dataset are accurate, label noise remains an important challenge for future research. It is worth noting that our method (DHK) relies on the relative hierarchical relationships among classes, which to some extent enhances the robustness for label noise (see \hrefappendix{subsubsection: Label Noise Results}). However, label noise may still interfere with the learning of hierarchical relationships and impact model performance. In most cases, the proportion of label noise among all labels is extremely small and its impact on the overall model is negligible. Nevertheless, from a scientific rigor standpoint, this issue still needs to be addressed.

\subsection{Extensibility}
\label{subsection: Extensibility}
In this study, although the proposed method (DHK) focuses on applications within complex industrial systems (cavitation intensity diagnosis industrial system and bearing strength diagnosis industrial system), the proposed hierarchical tree loss with two adaptive weighting schemes ($cf.$ \hrefequation{eq: HT Loss} / \hrefequation{eq: FHT Loss} w/ \hrefequation{eq: Adaptive Weight Schemes}) and group tree triplet loss with a hierarchical dynamic margin($cf.$ \hrefequation{eq: GTT Loss} w/ \hrefequation{eq: Hierarchical Dynamic Margin}) are novel loss functions designed for general tree structures. They are specifically developed for the general hierarchical classification task and have significant theoretical innovation. Based on this, our proposed method in this study can also be effectively extended to other hierarchical classification scenarios, as follows:
\begin{itemize}
    \item \textbf{Natural Image Classification:} In natural images, classes often exhibit semantic hierarchical relationships (e.g. animal $\to$ mammal $\to$ dog $\to$ golden retriever). The model makes rational use of hierarchical information to improve classification performance and robustness. The hierarchical natural image classification datasets include ImageNet, CIFAR and ANIMAL-10N, etc.
    \item \textbf{Medical Disease Grading:} Medical diagnostic tasks often involve grading systems (e.g. diabetic retinopathy is organized into five grades), where the grades exhibit correlations and progression order. Therefore, hierarchical modeling can more accurately reflect the disease progression path and enhance the model's ability to support clinical decision.
    \item \textbf{Protein Function Prediction:} Protein functions are typically organized into a multi-hierarchy system based on the Gene Ontology structure, where there are significant parent-child relationships among various functions. Therefore, introducing hierarchical information into the model can help enhance its ability to integrate biological priors and improve the biological plausibility of predictions.
    \item \textbf{Scene Recognition:} In scene recognition tasks, classes such as "traffic scene" can be organized into "highway", "city road" and "rural road". By leveraging the hierarchical relationships among scene classes can guide the model to understand scene semantics from coarse to fine levels, improving recognition stability in complex environments.
\end{itemize}
For these tasks, the hierarchical organization and relationship of the target classes have a significant impact on the model performance. Therefore, our proposed method (DHK) provides a hierarchical structure modeling and optimization strategy that can effectively enhance the model's performance.

\subsection{Broader Impact}
\label{subsection: Broader Impact}
Our proposed DHK method whose goal is to advance the field of "AI + Industry". The DHK is designed to enhance the performance of fault intensity diagnosis and cavitation intensity recognition, which are crucial for ensuring the reliability and efficiency of industrial systems. In addition, the DHK enhances model performance and interpretability in industrial fault diagnosis tasks, potentially alleviating the workload of monitors in complex industrial systems. Its strong adaptability to hierarchical classification tasks also makes DHK suitable for deployment in high-risk domains.
We advocate for a cautious and responsible attitude in real-world applications to ensure the method's reliability and compliance in terms of fairness, transparency, and safety.

\subsection{Future Improvements}
Although the DHK performs excellently in industrial fault diagnosis, there is still several aspects that warrant further improvement and exploration in the future, as outlined below:
\begin{itemize}
    \item \textbf{Label Noise Handling:} Although most datasets contain a relatively low proportion of noisy labels, label noise can still affect the training performance of the model. In the future, more robust algorithms can be investigated into our proposed method DHK to better deal with noisy labels (e.g. co-teaching, label smoothing and self-supervised learning, etc.), aiming to achieve the model stability and accuracy even in the presence of inconsistent or incorrect labels.
    \item \textbf{Dynamic Hierarchical Structure Modeling:} The hierarchical label tree in this study assumes that the class relationships are fixed and explicitly defined. However, fault patterns and system states may evolve over time in real-world industrial applications. Future work can explore more flexible hierarchical tree modeling approaches (e.g. graph neural network, structure learning mechanisms and time-aware graph modeling techniques, etc.) to dynamically adjust the hierarchical structure, enabling the model to adapt to evolving system states and fault patterns.
\end{itemize}
\end{document}